\documentclass[vecphys]{svmult}
\usepackage{makeidx}
\usepackage{multicol}
\usepackage{graphicx}
\usepackage{amssymb}
\usepackage{cite}
\input epsf
\newcommand{\aaa}{\mbox{\boldmath$a$}}
\newcommand{\kk}{\mbox{\boldmath$k$}}
\newcommand{\uu}{\mbox{\boldmath$u$}}
\newcommand{\vv}{\mbox{\boldmath$v$}}
\newcommand{\yy}{\mbox{\boldmath$y$}}
\newcommand{\xx}{\mbox{\boldmath$x$}}

\newcommand{\xixi}{\mbox{\boldmath$\xi$}}
\newcommand{\SSS}{\mbox{\boldmath$S$}}
\newcommand{\PiPi}{\mbox{\boldmath$\Pi$}}

\begin{document}
\title*{Basic Types of Coarse-Graining}

\author{ Alexander N. Gorban}
\institute{University of Leicester, Leicester LE1 7RH, UK
\texttt{ag153@le.ac.uk}}

\maketitle

\begin{abstract}

We consider two basic types of coarse-graining: the Ehrenfests'
coarse-graining and its extension to a general principle of
non-equilibrium thermodynamics, and the coarse-graining based on
uncertainty of dynamical models and $\varepsilon$-motions (orbits).
Non-technical discussion of basic notions and main coarse-graining
theorems are presented: the theorem about entropy overproduction for
the Ehrenfests' coarse-graining and its generalizations, both for
conservative and for dissipative systems, and the theorems about
stable properties and the Smale order for $\varepsilon$-motions of
general dynamical systems including structurally unstable systems.
Computational kinetic models of macroscopic dynamics are considered.
We construct a theoretical basis for these kinetic models using
generalizations of the Ehrenfests' coarse-graining. General theory
of reversible regularization and filtering semigroups in kinetics is
presented, both for linear and non-linear filters.  We obtain
explicit expressions and entropic stability conditions for filtered
equations. A brief discussion of coarse-graining by rounding and by
small noise is also presented.
\end{abstract}

\section{Introduction}
Almost a century ago, Paul and Tanya Ehrenfest in their paper for
scientific Encyclopedia \cite{Ehrenfest} introduced a special
operation, the coarse-graining. This operation transforms a
probability density in phase space into a ``coarse-grained" density,
that is a piece-wise constant function, a result of density
averaging in cells. The size of cells is assumed to be small, but
finite, and does not tend to zero. The coarse-graining models
uncontrollable impact of surrounding (of a thermostat, for example)
onto ensemble of mechanical systems.

\begin{figure}[t]
a)\\ \includegraphics[width=110mm, height=65mm]{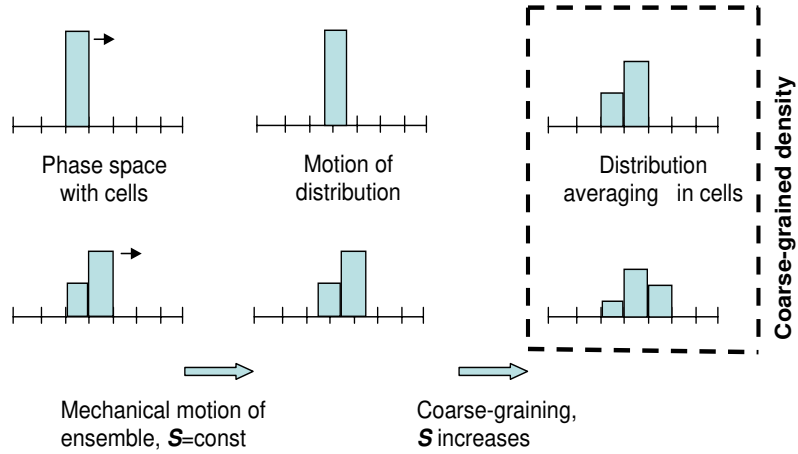}\\
b) \\ \includegraphics[width=110mm, height=35mm]{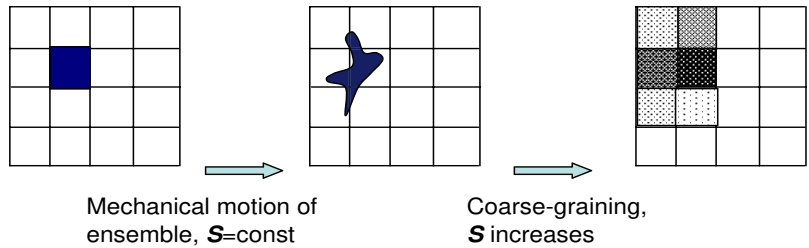}
\caption{\label{FigEhr}The Ehrenfests' coarse-graining: two ``motion
-- coarse-graining" cycles in 1D (a, values of probability density
are presented by the height  of the columns) and one such cycle in
2D (b, values of probability density are presented by hatching
density).}
\end{figure}

To understand reasons for introduction of this new notion, let us
take a phase drop, that is, an ensemble of mechanical systems with
constant probability density localized in a small domain of phase
space. Let us watch  evolution of this drop in time according to the
Liouville equation. After a long time, the shape of the drop may be
very complicated, but the density value remains the same, and this
drop remains ``oil in water." The ensemble can tend to the
equilibrium in the weak sense only: average value of any continuous
function tends to its equilibrium value, but the entropy of the
distribution remains constant. Nevertheless, if we divide the phase
space into cells and supplement the mechanical motion by the
periodical averaging in cells (this is the Ehrenfests' idea of
coarse-graining), then the entropy increases, and the distribution
density tends uniformly to the equilibrium. This periodical
coarse-graining is illustrated by Fig.~\ref{FigEhr} for
one-dimensional (1D)\footnote{Of course, there is no mechanical
system with one-dimensional phase space, but dynamics with
conservation of volume is possible in 1D case too: it is a motion
with constant velocity.} and two-dimensional (2D) phase spaces.

Recently, we can find the idea of coarse-graining everywhere in
statistical physics (both equilibrium and non-equilibrium). For
example, it is the central idea of the Kadanoff transformation, and
can be considered as a background of the Wilson renormalization
group \cite{Wilson} and modern renormalisation group approach to
dissipative systems \cite{RG,Kun3}. \footnote{See also the paper of
A. Degenhard  and J. Javier Rodriguez-Laguna in this volume.} It
gave a simplest realization of the projection operators technique
\cite{Grabert} long before this technic was developed. In the method
of invariant manifold \cite{CMIM,GorKar} the generalized Ehrenfests'
coarse-graining allows to find slow dynamics without a slow manifold
construction. It is also present in the background of the so-called
equation-free methods \cite{KevFree}. Applications of the
Ehrenfests' coarse-graining outside statistical physics include
simple, but effective filtering \cite{Raz}. The Gaussian filtering
of hydrodynamic equations that leads to the Smagorinsky equations
\cite{Smag} is, in its essence, again a version of the Ehrenfests'
coarse-graining. In the first part of this paper we elaborate in
details the Ehrenfests' coarse-graining for dynamical systems.

The central idea of the Ehrenfests' coarse-graining remains the same
in most generalizations: we combine the genuine motion with the
periodic {\it partial equlibration}. The result is the Ehrenfests'
chain. After that, we can find the macroscopic equation that does
not depend on an  initial distribution and describes the Ehrenfests'
chains as results of continuous autonomous motion
\cite{GKIOeNONNEWT2001,GKOeTPRE2001}. Alternatively, we can just
create a computational procedure without explicit equations
\cite{KevFree}. In the sense of entropy production, the resulting
macroscopic motion is ``more dissipative" than initial (microscopic)
one. It is the theorem about entropy overproduction. In its general
form it was proven in \cite{UNIMOLD}.

Kinetic models of fluid dynamics become very popular during the
last decade. Usual way of model simplification leads from kinetics
to fluid dynamics, it is a sort of dimension reduction. But
kinetic models go back, and it is the simplification also. Some of
kinetic equations are very simple and even exactly solvable. The
simplest and most popular example is the free flight kinetics,
$\partial f(\xx,\vv,t)/\partial t=-\sum_i v_i \partial
f(\xx,\vv,t)/\partial x_i$, where $f(\xx,\vv,t)$ is one-particle
distribution function, $\xx$ is  space vector, $\vv$ is velocity.
We can ``lift" a continuum equation to a kinetic model, and than
approximate the solution by a chain, each link of which is a
kinetic curve with a jump from the end of this curve to the
beginning of the next link. In this paper, we describe how to
construct these curves, chains, links and jumps on the base of
Ehrenfests' idea. Kinetic model has more variables than continuum
equation. Sometimes simplification in modeling can be reached by
dimension increase, and it is not a miracle.

In practice, kinetic models in the form of lattice Boltzmann models
are in use \cite{LB2}. The Ehrenfests' coarse-graining provides
theoretical basis for kinetic models. First of all, it is possible
to replace projecting (partial equilibration) by involution (i.e.
reflection with respect to the partial equilibrium). This {\it
entropic involution} was developed for the lattice Boltzmann methods
in \cite{ELB1}. In the original Ehrenfests' chains,
``motion--partial equilibration--motion--...," dissipation is
coupled with time step, but the chains
``motion--involution--motion--..." are conservative. The family of
chains between conservative (with entropic involution) and maximally
dissipative (with projection) ones give us a possibility to model
hydrodynamic systems with various dissipation (viscosity)
coefficients that are decoupled with time steps.

Large eddy simulation, filtering and subgrid modeling are very
popular in fluid dynamics
\cite{Leray,Smag,Germano,Carati,LES2005}. The idea is that small
inhomogeneities should somehow equilibrate, and their statistics
should follow the large scale details of the flow. Our goal is to
restore a link between this approach and initial coarse-raining in
statistical physics. Physically, this type of coarse-graining is
transference  the energy of small scale motion from macroscopic
kinetic energy to microscopic internal energy. The natural
framework for analysis of such transference provides physical
kinetics, where initially exists no difference between kinetic and
internal energy. This difference appears in the continuum mechanic
limit. We proposed this idea several years ago, and an example for
moment equations was published in \cite{AnsKarlFiltr}. Now the
kinetic approach for filtering is presented. The general
commutator expansion for all kind of linear or non-linear filters,
with constant or with variable coefficients is constructed. The
condition for stability of filtered equation is obtained.

The upper boundary for the filter width $\Delta$ that guaranties
stability of the filtered equations is proportional to the square
root of the Knudsen number. $\Delta/L \sim \sqrt{{K\!n}}$ (where
$L$ is the characteristic macroscopic length). This scaling,
$\Delta/L \sim \sqrt{{K\!n}}$, was discussed in
\cite{AnsKarlFiltr} for moment kinetic equations because different
reasons: if $\Delta/L \gg \sqrt{{K\!n}}$ then the Chapman--Enskog
procedure for the way back from kinetics to continuum is not
applicable, and, moreover, the continuum description is probably
not valid, because the filtering term with large coefficient
$\Delta/L$ violates the conditions of hydrodynamic limit. This
important remark  gives the frame for $\eta$ scaling. It is proven
in this paper for the broad class of model kinetic equations. The
entropic stability conditions presented below give the stability
boundaries inside this scale.

Several other notions of coarse-graining were introduced and studied
for dynamical systems during last hundred years. In this paper, we
shall consider one of them, the coarse-graining by
$\varepsilon$-motions ($\varepsilon$-orbits, or pseudo orbits) and
briefly mention two other types: coarse-graining by rounding and by
small random noise.

$\varepsilon$-motions describe dynamics of models with
uncertainty. We never know our models exactly, we never deal with
isolated systems, and the surrounding always uncontrollably affect
dynamics of the system. This dynamics can be presented as a usual
phase flow supplemented by a periodical {\it
$\varepsilon$-fattening}: after time $\tau$, we add a
$\varepsilon$-ball to each point, hence, points are transformed
into sets. This periodical fattening expands all attractors: for
the system with fattening they are larger than for original
dynamics.

Interest to the dynamics of $\varepsilon$-motions was stimulated
by the famous work of S. Smale \cite{Smeil}. This paper destroyed
many naive dreams and expectations. For generic 2D system the
phase portrait is the structure of attractors (sinks), repellers
(sources), and saddles. For generic 2D systems all these
attractors are either fixed point or closed orbits. Generic 2D
systems are structurally stable. It means that they do not change
qualitatively after small perturbations.  Our dream was to find a
similar stable structure in generic systems for higher dimensions,
but S. Smale showed it is impossible: Structurally stable systems
are not dense! Unfortunately, in higher dimensions there are
regions of dynamical systems that can change qualitatively under
arbitrary small perturbations.

One of the reasons to study $\varepsilon$-motions (flow with
fattening) and systems with sustained perturbations was the hope
that even small errors coarsen the picture and can wipe some of
the thin peculiarities off. And this hope was realistic, at least,
partially \cite{[69],Diss,SloRelMono}. The thin peculiarities that
are responsible for appearance of regions of structurally unstable
systems vanish after the coarse-graining via arbitrary small
periodical fattening. All the models have some uncertainty, hence,
the features of dynamics that are unstable under arbitrary small
coarse-graining are unobservable.

\begin{figure}[t]
\begin{centering}
a)\includegraphics[width=55mm, height=43 mm]{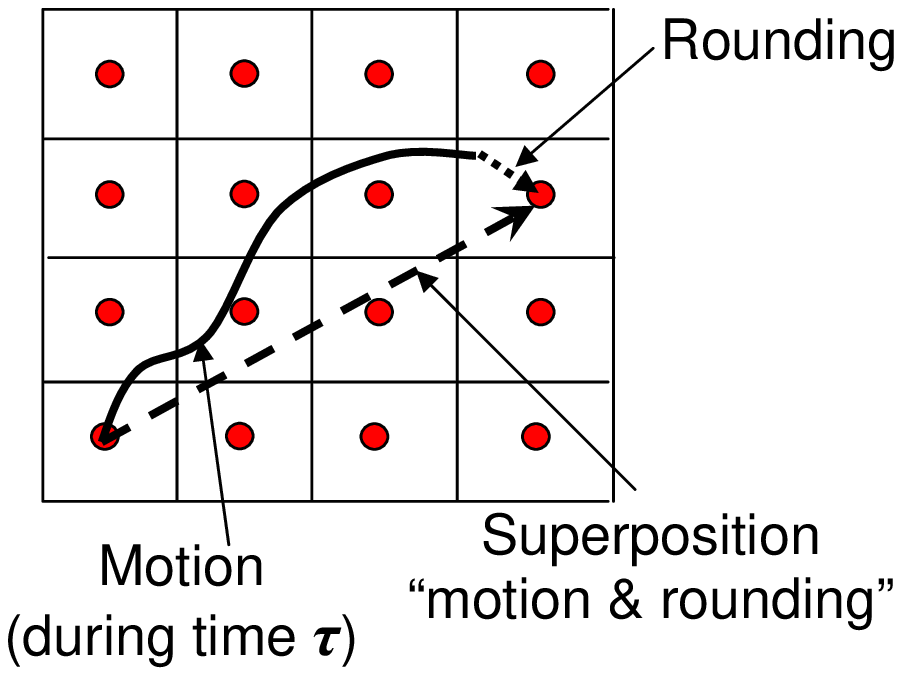}
b)\includegraphics[width=40mm, height=40mm]{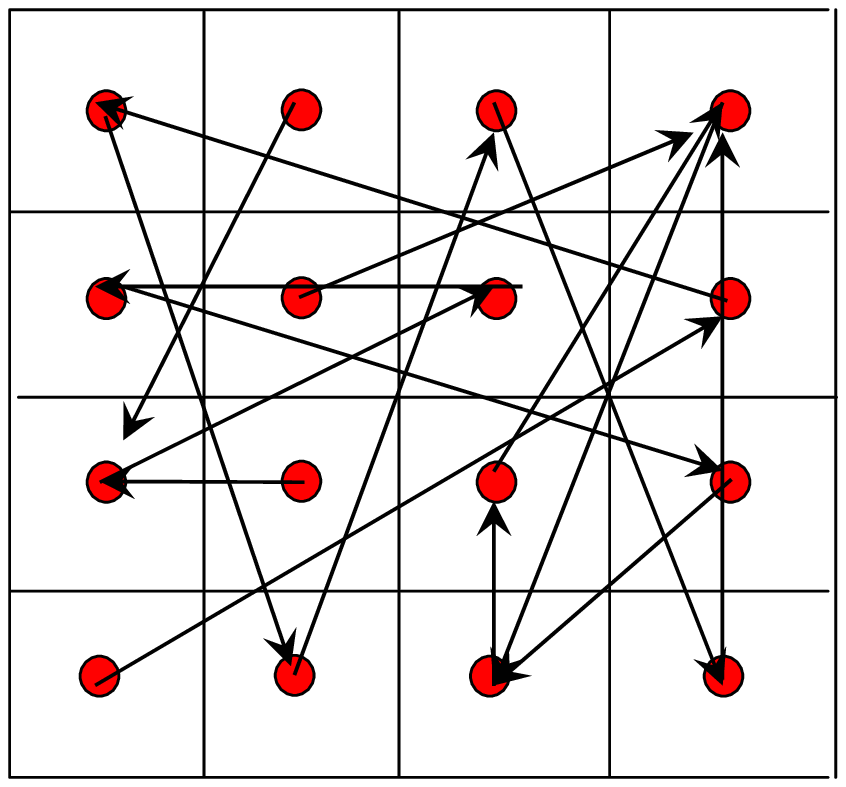}
\caption{\label{FigRound}Motion, rounding and ``motion with
rounding" for a dynamical system (a), and the universal result of
motion with rounding: a random dynamical system (b).}
\end{centering}
\end{figure}

Rounding is a sort of coarse-graining that appears automatically
in computer simulations. It is very natural that in era of
intensive computer simulation of complex dynamics the
coarse-graining by rounding attracted special attention
\cite{Hub,Beck,GrebYor,Diamond,Longa,Binder,Hoower}.  According to
a very idealized popular dynamic model, rounding might be
represented as restriction of shift in given time $\tau$ onto
$\varepsilon$-net in phase space. Of courses, the restriction
includes some perturbation of dynamics (Fig.~\ref{FigRound}). The
formal definition of rounding action includes a tiling: around any
point of the $\varepsilon$-net there is a cell, these cells form a
tiling of the phase space, and rounding maps a cell into
corresponding point of the $\varepsilon$-net. These cells have
equal volumes if there are no special reasons to make their
volumes different. If this volume is dynamically invariant then,
for sufficiently large time of motion between rounding steps, all
the mixing dynamical systems with rounding can be described by an
universal object. This is a random dynamical system, the random
map of a finite set: any point of the $\varepsilon$-net can be the
image of a given point with probability $1/m$ (where $m$ is the
number of points in the $\varepsilon$-net). The combinatorial
theory of such {\it random graphs} is well--developed
\cite{BBoll}.

After rounding, some unexpected properties of dynamics appear. For
example, even for transitive systems with strong mixing
significant part of points of the $\varepsilon$-net becomes
transient after rounding. Initially, attractor of such a
continuous system is the whole phase space, but after rounding
attractor of discrete dynamical system on the $\varepsilon$-net
includes, roughly speaking, a half of its points (or, more
precisely, the expectation of the number of transient points is $m
({\rm e}-1)/{\rm e}$, where $m$ is number of points, ${\rm
e}=2.7...$). In some circumstances, complicated dynamics has a
tendency to collapse to trivial and degenerate behaviour as a
result of discretizations \cite{Diamond}. For systems without
conservation of volume, the number of periodic points after
discretization is linked to the dimension of the attractor $d$.
The simple estimates based on the random map analysis, and
numerical experiments with chaotic attractors give $\sim
\varepsilon^{-d}$ for the number of periodic points, and $\sim
\varepsilon^{-d/2}$ for the scale of the expected period
\cite{GrebYor,Hoower}. The first of them is just the number of
points in  $\varepsilon$-net in $d$-dimensional compact, the
second becomes clear after the following remark. Let us imagine a
random walk in a finite set with $m$ elements (a
$\varepsilon$-net). When the length of the trajectory is of order
$\sqrt{m}$ then the next step returns the point to the trajectory
with probability $\sim 1/\sqrt{m}$, and a loop appears with
expected period $\sim \sqrt{m}$ (a half of the trajectory length).
After $\sim \sqrt{m}$ steps the probability of a loop appearance
is near $1$, hence, for the whole system the expected period is
$\sim \sqrt{m}\sim \varepsilon^{-d/2}$.

It is easy to demonstrate the difference between coarse-graining
by fattening and coarse-graining by rounding. Let us consider a
trivial dynamics on a connected phase space: let the shift in time
be identical transformation. For coarse-graining by fattening the
$\varepsilon$-motion of any point tends to cover the whole phase
space for any positive $\varepsilon$ and time $t\rightarrow
\infty$: periodical $\varepsilon$-fattening with trivial dynamics
transforms, after time $n\tau$, a point into the sum of $n$
$\varepsilon$-balls. For coarse-graining by rounding this trivial
dynamical system generates the same trivial dynamical system on
$\varepsilon$-net: nothing moves.

Coarse-graining by small noise seems to be very natural. We add
small random term to the right hand side of differential equations
that describe dynamics. Instead of the Liouville equation for
probability density the Fokker--Planck equation appears. There is no
fundamental difference between various types of coarse-graining, and
the coarse-graining by $\varepsilon$-fattening includes major
results about the coarse-graining by small noise that are
insensitive to most details of noise distribution. But the knowledge
of noise distribution gives us additional  tools. The {\it action
functional} is such a tool for the description of fluctuations
\cite{[52]}. Let $X^{\varepsilon}(t)$ be a random process ``dynamics
with $\varepsilon$-small fluctuation" on the time interval $[0,T]$.
It is possible to introduce such a functional ${\bf S} [\varphi]$ on
functions $x=\varphi(t)$ ($t \in [0,T]$) that for sufficiently small
$\varepsilon,\delta > 0$ $$ {\bf P}\{ \|X^{\varepsilon} - \varphi \|
< \delta \} \approx \exp(-{\bf S} [\varphi]/{\varepsilon}^2). $$
Action functional is constructed for various types of random
perturbations \cite{[52]}. Introduction to the general theory of
random dynamical systems with invariant measure is presented in
\cite{LArn}.

In following sections, we consider  two types of coarse-graining:
the Ehrenfests' coarse-graining and its extension to a general
principle of non-equilibrium thermodynamics, and the coarse-graining
based on the uncertainty of dynamical models and
$\varepsilon$-motions.

\section{The Ehrenfests' Coarse-graining}

\subsection{Kinetic equation and entropy}

\paragraph{Entropy conservation in systems with conservation of phase
volume}

The Erenfest's coarse-graining was originally defined for
conservative\footnote{In this paper, we use the term
``conservative" as an opposite term to ``dissipative:"
conservative = with entropy conservation. Another use of the term
``conservative system" is connected with energy conservation. For
kinetic systems under consideration conservation of energy is a
simple linear balance, and we shall use the first sense only.}
systems. Usually, Hamiltonian systems are considered as
conservative ones, but in all constructions only one property of
Hamiltonian systems is used, namely, conservation of the phase
volume $\D \Gamma$ (the Liouville theorem). Let $X$ be phase
space, $v(x)$ be a vector field, $\D \Gamma = \D^n x$ be the
differential of phase volume. The flow,
\begin{equation}\label{flow}
{\D x \over \D t }=v(x),
\end{equation}
conserves the phase volume, if ${\rm div} v(x)=0$. The continuity
equation,
\begin{equation}\label{continuity}
\frac{\partial f}{\partial t} = - \sum_i\frac{\partial
(fv_i(x))}{\partial x_i},
\end{equation}
describes the induced dynamics of the probability density $f(x,t)$
on phase space. For incompressible flow (conservation of volume),
the continuity equation can be rewritten in the form
\begin{equation}\label{incomp}
\frac{\partial f}{\partial t} = - \sum_i v_i(x) \frac{\partial
f}{\partial x_i}.
\end{equation}
This means that the probability density is constant along the flow:
$f(x, t+\D t)=f(x-v(x)\D t , t)$. Hence, for any continuous function
$h(f)$ the integral
\begin{equation}\label{integral}
H(f)=\int_X h(f(x))\, \D \Gamma(x)
\end{equation}
does not change in time, provided the probability density satisfies
the continuity equation (\ref{continuity}) and the flow $v(x)$
conserves the phase volume. For $h(f)=-f \ln f$ integral
(\ref{integral}) gives the classical Boltzmann--Gibbs--Shannon (BGS)
entropy functional:
\begin{equation}\label{BGS}
S(f)=-\int_X f(x) \ln (f(x)) \, \D \Gamma(x).
\end{equation}
For flows with conservation of volume, entropy is conserved: $\D S/
\D t \equiv 0$ .

\paragraph{Kullback entropy conservation in systems with regular
invariant distribution}

Suppose the phase volume is not invariant with respect to flow
(\ref{flow}), but a regular invariant density $f^*(x)$ (equilibrium)
exists:
\begin{equation}\label{invariantDen}
\sum_i\frac{\partial (f^*(x)v_i(x))}{\partial x_i}=0.
\end{equation}
In this case, instead of an invariant phase volume $\D \Gamma$, we
have an invariant volume $f^*(x)\, \D \Gamma$. We can use
(\ref{invariantDen}) instead of the incompressibility condition and
rewrite (\ref{continuity}):
\begin{equation}\label{genincomp}
\frac{\partial (f(x,t)/f^*(x)) }{\partial t} = -  \sum_i v_i(x)
\frac{\partial (f(x,t)/f^*(x))}{\partial x_i}.
\end{equation}
The function $f(x,t)/f^*(x)$ is constant along the flow, the
measure $f^*(x)\, \D \Gamma(x)$ is invariant, hence, for any
continuous function $h(f)$ integral
\begin{equation}\label{integralKu}
H(f)=\int_X h(f(x,t)/f^*(x)) f^*(x)\, \D \Gamma(x)
\end{equation}
does not change in time, if the probability density satisfies the
continuity equation. For $h(f)=-f \ln f$ integral
(\ref{integralKu}) gives the Kullback entropy functional
\cite{Kull}:
\begin{equation}\label{Kul}
S_K(f)=-\int_X f(x) \ln \left(\frac{f(x)}{f^*(x)}\right)\, \D
\Gamma(x).
\end{equation}
This situation does not differ significantly from the entropy
conservation in systems with conservation of volume. It is just a
kind of change of variables.

\paragraph{General entropy production formula}

Let us consider the general case without assumptions about phase
volume invariance and existence of a regular invariant density
(\ref{invariantDen}). In this case, let a probability density
$f(x,t)$ be a solution of the continuity equation
(\ref{continuity}). For the BGS entropy functional (\ref{BGS})
\begin{equation}\label{EntProdGen}
\frac{\D S(f)}{\D t}= \int_X f(x,t) {\rm div}v(x)\, \D \Gamma(x),
\end{equation}
if the left hand side exists. This {\it entropy production formula}
can be easily proven for small phase drops with constant density,
and then for finite sums of such distributions with positive
coefficients. After that, we obtain formula (\ref{EntProdGen}) by
limit transition.

For a regular invariant density $f^*(x)$ (equilibrium) entropy
$S(f^*)$ exists, and for this distribution $\D S(f)/ \D t=0$,
hence,
\begin{equation}\label{EntPodEq}
\int_X f^*(x) {\rm div}v(x)\, \D \Gamma(x) =0.
\end{equation}

\paragraph{Entropy production in systems without regular equilibrium}

If there is no regular equilibrium (\ref{invariantDen}), then the
entropy behaviour changes drastically. If volume of phase drops
tends to zero, then the BGS entropy (\ref{BGS}) and any Kullback
entropy (\ref{Kul}) goes to minus infinity. The simplest example
clarifies the situation. Let all the solutions converge to unique
exponentially stable fixed point $x=0$. In linear approximation $\D
x / \D t =Ax$ and $S(t)= S(0)+t\,{\rm tr} A.$ Entropy decreases
linearly in time with the rate ${\rm tr} A$ (${\rm tr} A = {\rm div}
v(x)$, ${\rm tr} A < 0$), time derivative of entropy is ${\rm tr} A$
and does not change in time, and the probability distribution goes
to the $\delta$-function $\delta(x)$. Entropy of this distribution
does not exist (it is ``minus infinity"), and it has no limit when
$f(x,t)\rightarrow \delta(x)$.

Nevertheless, time derivative of entropy is well defined and
constant, it is ${\rm tr} A$. For more complicated singular limit
distributions the essence remains the same: according to
(\ref{EntProdGen}) time derivative of entropy tends to the average
value of ${\rm div} v(x)$ in this limit distribution, and entropy
goes linearly to minus infinity (if this average in not zero, of
course). The order in the system increases. This behaviour could
sometimes be interpreted as follows: the system is open and
produces entropy in its surrounding even in a steady--state. Much
more details are in review \cite{Ruelle}.\footnote{Applications of
this formalism are mainly related to Hamiltonian systems in
so-called force thermostat, or, in particular, isokinetic
thermostat. These thermostats were invented in computational
molecular dynamics for acceleration of computations, as a
technical trick. From the physical point of view, this theory can
be considered as a theory about a friction of particles on the
space, the ``ether friction." For isokinetic thermostats, for
example, this ``friction" decelerates some of particles,
accelerates others, and keeps the kinetic energy constant.}

\paragraph{Starting point: a kinetic equation}

For the formalization of the Ehrenfests' idea of coarse-graining, we
start from a formal kinetic equation
\begin{equation}\label{FormKinEq}
\frac{\D f}{\D t}=J(f)
\end{equation}
with a concave entropy functional $S(f)$ that does not increase in
time. This equation is defined in a convex subset $U$ of a vector
space $E$.

Let us specify some notations:  $E^T$ is the adjoint to the $E$
space.  Adjoint spaces and operators will be indicated by $^T$,
whereas the notation $^*$ is earmarked for equilibria and
quasi-equilibria.

We recall that, for an operator $A:E_1\to E_2$, the adjoint
operator, $A^T:E_1^T\to E_2^T$ is defined by the following
relation:  for any $l\in E_2^T$ and $\varphi\in E_1$,
$l(A\varphi)=(A^Tl)(\varphi)$.

Next, $D_fS(f)\in E^T$ is the differential of the functional
$S(f)$, $D^2_fS(f)$ is the second differential of the functional
$S(f)$.  The quadratic functional $D^2_fS(f)(\varphi,\varphi)$ on
$E$ is defined by the Taylor formula,
\begin{equation}
\label{TaylorP} S(f+\varphi)=S(f)+D_fS(f)(\varphi)+
\frac{1}{2}D_f^2S(f)(\varphi,\varphi)+o(\|\varphi\|^2).
\end{equation}
We keep the same notation for the corresponding symmetric bilinear
form, $D_f^2S(f)(\varphi,\psi)$, and also for the linear operator,
$D_f^2S(f):E\to E^T$, defined by the formula
$(D_f^2S(f)\varphi)(\psi)= D_f^2S(f)(\varphi,\psi)$. In this
formula, on the left hand side $D_f^2S(f)$ is the operator, on the
right hand side it is the bilinear form. Operator $D_f^2S(f)$ is
symmetric on $E$, $D_f^2S(f)^T=D_f^2S(f)$.

In finite dimensions the functional $D_fS(f)$ can be presented
simply as a row vector of partial derivatives of $S$, and the
operator $D_f^2S(f)$ is a matrix of second partial derivatives.
For infinite--dimensional spaces some complications exist because
$S(f)$ is defined only for classical densities and not for all
distributions. In this paper we do not pay attention to these
details.

We assume  strict concavity of $S$, $D_f^2S(f)(\varphi,\varphi)< 0$
if $\varphi\ne0$. This means that for any $f$ the positive definite
quadratic form $-D_f^2S(f)(\varphi,\varphi)$ defines a scalar
product
\begin{equation}\label{eproductP}
\langle \varphi , \psi\rangle_{f}=-(D_f^2S)(\varphi,\psi).
\end{equation}
This {\it entropic scalar product} is an important part of
thermodynamic formalism. For the BGS entropy (\ref{BGS}) as well
as for the Kullback entropy (\ref{Kul})
\begin{equation}\label{eprodBGS}
\langle \varphi , \psi\rangle_{f}=\int\frac{\varphi (x)
\psi(x)}{f(x)} \, \D x.
\end{equation}

The most important assumption about kinetic equation
(\ref{FormKinEq}) is: entropy does not decrease in time:
\begin{equation}\label{dissipativity}
\frac{\D S}{\D t}=(D_fS(f))(J(f))\geq 0.
\end{equation}
A particular case of this assumption is: the system
(\ref{FormKinEq}) is conservative and entropy is constant. The
main example of such conservative equations is the Liouville
equation with linear vector field $J(f)=-Lf=\{H,f\}$, where
$\{H,f\}$ is the Poisson bracket with Hamiltonian $H$.

For the following consideration of the Ehrenfests' coarse-graining
the underlying mechanical motion is not crucial, and it is possible
to start from the formal kinetic equation (\ref{FormKinEq}) without
any mechanical interpretation of vectors $f$. We develop below the
coarse-graining procedure for general kinetic equation
(\ref{FormKinEq}) with non-decreasing entropy (\ref{dissipativity}).
After coarse-graining the entropy production increases: conservative
systems become dissipative ones, and dissipative systems become
``more dissipative."

\subsection{Conditional equilibrium instead of averaging in cells}

\paragraph{Microdescription, macrodescription and quasi-equilibrium state}

Averaging in cells is a particular case of entropy maximization.
Let the phase space be divided into cells. For the $i$th cell the
population $M_i$ is $$M_i = m_i(f)= \int_{{\rm cell}_i} f(x) \, \D
\Gamma (x).$$ The averaging in cells for a given vector of
populations $M=(M_i)$  produces the solution of the optimization
problem for the BGS entropy:
\begin{equation}\label{smax} S(f)\to \max,\ m(f)=M,
\end{equation}
where $m(f)$ is vector $(m_i(f))$. The maximizer is a function
$f^*_M(x)$ defined by the vector of averages $M$.

This operation has a well-known generalization. In the more
general statement, vector $f$ is a microscopic description of the
system, vector $M$ gives a macroscopic description, and a linear
operator $m$ transforms a microscopic description into a
macroscopic one: $M=m(f)$. The standard example is the
transformation of the microscopic density into the hydrodynamic
fields (density--velocity--kinetic temperature) with local
Maxwellian distributions as entropy maximizers (see, for example,
\cite{GorKar}).

\begin{figure}[t]
\begin{centering}
\includegraphics[width=60mm, height=60mm]{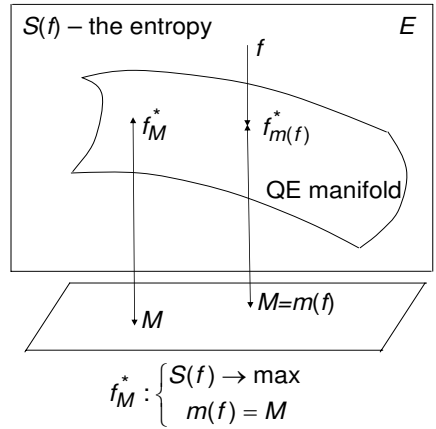}
\caption{\label{FigQEM}Relations between a microscopic state $f$,
a corresponding macroscopic state $M=m(f)$, and a
quasi-equilibrium state $f^*_M$.}
\end{centering}
\end{figure}

For any macroscopic description $M$, let us define the
correspondent $f^*_M$ as a solution to optimization problem
(\ref{smax}) with an appropriate entropy functional $S(f)$
(Fig.~\ref{FigQEM}). This $f^*_M$ has many names in the
literature: MaxEnt distribution, reference distribution (reference
of the macroscopic description to the microscopic one),
generalized canonical ensemble, conditional equilibrium, or {\it
quasi-equilibrium}. We shall use here the last term.

The quasi-equilibrium distribution $f^*_M$ satisfies the obvious,
but important identity of self-consistency:
\begin{equation}\label{SelfConsId}
m(f^*_M)=M,
\end{equation}
or in differential form
\begin{equation}\label{SelfConsIdDif}
m(D_Mf^*_M)=1, \; {\rm i.e.} \; m((D_Mf^*_M)a)\equiv a.
\end{equation}
The last identity means that the infinitesimal change in $M$
calculated through differential of the quasi-equilibrium
distribution $f^*_M$ is simply the infinitesimal change in $M$.
For the second differential we obtain
\begin{equation}\label{SelfConsIdSecDif}
m(D^2_Mf^*_M)=0,\; {\rm i.e.} \; m((D^2_Mf^*_M)(a,b))\equiv 0.
\end{equation}

Following \cite{GorKar} let us mention that most of the works on
nonequilibrium thermodynamics deal with quasi-equilibrium
approximations and corrections to them, or with applications of
these approximations (with or without corrections). This viewpoint
is not the only possible but it proves very efficient for the
construction of a variety of useful models, approximations and
equations, as well as methods to solve them.

From time to time it is discussed in the literature, who was the
first to introduce the quasi-equilibrium approximations, and how
to interpret them. At least a part of the discussion is due to a
different role the quasi-equilibrium plays in the
entropy-conserving and in the dissipative dynamics. The very first
use of the entropy maximization dates back to the classical work
of G.\ W.\ Gibbs \cite{Gibb}, but it was first claimed for a
principle of informational statistical thermodynamics by E.\ T.\
Jaynes \cite{Janes1}. Probably, the first explicit and systematic
use of quasiequilibria on the way from entropy-conserving dynamics
to dissipative kinetics was undertaken by D.\ N.\ Zubarev. Recent
detailed exposition of his approach is given in \cite{Zubarev}.

For dissipative systems, the use of the quasi-equilibrium to
reduce description can be traced to the works of H. Grad on the
Boltzmann equation \cite{Grad}. A review of the informational
statistical thermodynamics was presented in \cite{Garsia1}.  The
connection between entropy maximization and (nonlinear) Onsager
relations was also studied \cite{Nett,Orlov84}. Our viewpoint was
influenced by the papers by L.\ I.\ Rozonoer and co-workers, in
particular, \cite{KoRoz,Ko,Roz}. A detailed exposition of the
quasi-equilibrium approximation for Markov chains is given in the
book \cite{G1} (Chap. 3, {\it Quasi-equilibrium and entropy
maximum}, pp.\ 92-122), and for the BBGKY hierarchy in the paper
\cite{Kark}.

The maximum entropy principle was  applied to the description of the
universal dependence of the three-particle distribution function
$F_3$ on the two-particle distribution function $F_2$ in classical
systems with binary interactions \cite{BGKTMF}. For a discussion of
the quasi-equilibrium moment closure hierarchies for the Boltzmann
equation \cite{Ko} see the papers \cite{MBCh,MBChLANL,Lever}. A very
general discussion of the maximum entropy principle with
applications to dissipative kinetics is given in the review
\cite{Bal}. Recently, the quasi-equilibrium approximation with some
further correction was applied to the description of rheology of
polymer solutions \cite{IKOePhA02,IKOePhA03} and of ferrofluids
\cite{IlKr,IKar2}. Quasi-equilibrium approximations for quantum
systems in the Wigner representation \cite{WIG,CAL} was discussed
very recently \cite{Degon}.

We shall now introduce the quasi-equilibrium approximation in the
most general setting. The coarse-graining procedure will be
developed after that as a method for enhancement of the
quasi-equilibrium approximation  \cite{GKIOeNONNEWT2001}.

\paragraph{Quasi-equilibrium manifold, projector and
approximation}

A {\it quasi-equilibrium manifold} is a set of quasi-equilibrium
states $f^*_M$ parameterized by macroscopic variables $M$. For
microscopic states $f$ the correspondent quasi-equilibrium states
are defined as $f^*_{m(f)}$. Relations between $f$, $M$, $f^*_M$,
and $f^*_{m(f)}$ are presented in Fig.~\ref{FigQEM}.

A {\it quasi-equilibrium approximation} for the kinetic equation
(\ref{FormKinEq}) is an equation for $M(t)$:
\begin{equation}\label{QEapp}
\frac{\D M}{\D t}=m(J(f^*_M)).
\end{equation}
To define $\dot{M}$ in the quasi-equilibrium approximation for
given $M$, we find the correspondent quasi-equilibrium state
$f^*_M$ and the time derivative of $f$ in this state $J(f^*_M)$,
and then return to the macroscopic variables by the operator $m$.
If $M(t)$ satisfies (\ref{QEapp}) then $f^*_{M(t)}$ satisfies the
following equation
\begin{equation}\label{QEonQEM}
\frac{\D f^*_M}{\D t}=(D_Mf^*_{M})\left(\frac{\D M}{\D t}\right)
=(D_Mf^*_{M})(m(J(f^*_M))).
\end{equation}
The right hand side of (\ref{QEonQEM}) is the projection of vector
field $J(f)$ onto the tangent space of the quasi-equilibrium
manifold at the point $f=f^*_M$. After calculating the
differential $D_Mf^*_{M}$ from the definition of quasi-equilibrium
(\ref{smax}), we obtain ${\D f^*_M}/{\D t}=\pi_{f^*_M}J(f^*_M)$,
where $\pi_{f^*_M}$ is the {\it quasi-equilibrium projector}:
\begin{equation}\label{QEP}
\pi_{f^*_{M}}=\left(D_Mf^*_M\right) m=\left(D_f^2S
\right)_{f^*_{M}}^{-1}m^T
\left(m\left(D_f^2S\right)_{f^*_{M}}^{-1}m^T\right)^{-1}m.
\end{equation}
It is straightforward to check the equality
$\pi_{f^*_{M}}^2=\pi_{f^*_{M}}$, and the self-adjointness of
$\pi_{f^*_{M}}$ with respect to entropic scalar product
(\ref{eproductP}). In this scalar product, the quasi-equilibrium
projector is the orthogonal projector onto the tangent space to the
quasi-equilibrium manifold.
\begin{figure}[t]
\begin{center}
\begin{centering}
\includegraphics[width=60mm, height=55mm]{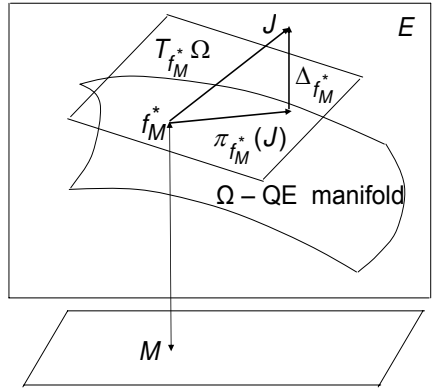}
\caption{\label{QEpro}Quasi-equilibrium manifold $\Omega$, tangent
space $T_{f^*_{M}}\Omega$, quasi-equilibrium projector
$\pi_{f^*_{M}}$, and defect of invariance,
$\Delta_{f^*_{M}}=J-\pi_{f^*_{M}}(J)$.}
\end{centering}
\end{center}
\end{figure}
The quasi-equilibrium projector for a quasi-equilibrium
approximation was first constructed by B. Robertson
\cite{Robertson}.

Thus, we have introduced the basic constructions:
quasi-equilibrium manifold, entropic scalar product, and
quasi-equilibrium projector (Fig. \ref{QEpro}).

\paragraph{Preservation of dissipation}

For the quasi-equilibrium approximation the entropy is
$S(M)=S(f^*_{M})$. For this entropy,
\begin{equation}\label{TypePreser}
\frac{\D S(M)}{\D t}=\left(\frac{\D S(f)}{\D
t}\right)_{f=f^*_{M}},
\end{equation}
Here, on the left hand side stands the macroscopic entropy
production for the quasi-equilibrium approximation (\ref{QEapp}),
and the right hand side is the microscopic entropy production
calculated for the initial kinetic equation (\ref{FormKinEq}). This
equality implies {\it preservation of the type of dynamics}
\cite{G1,Ocherki}:
\begin{itemize}
\item{If for the initial kinetics (\ref{FormKinEq}) the
dissipativity inequality (\ref{dissipativity}) holds then the same
inequality is true for the quasi-equilibrium approximation
(\ref{QEapp});} \item{If the initial kinetics (\ref{FormKinEq}) is
conservative then the quasi-equilibrium approximation
(\ref{QEapp}) is conservative also.}
\end{itemize}
For example, let the initial kinetic equation be the Liouville
equation for a system of many identical particles with binary
interaction. If we choose as macroscopic variables the
one-particle distribution function, then the quasi-equilibrium
approximation is the Vlasov equation. If we choose as macroscopic
variables the hydrodynamic fields, then the quasi-equilibrium
approximation is the compressible Euler equation with
self--interaction of liquid. Both of these equations are
conservative and turn out to be even Hamiltonian systems
\cite{Morrison}.

\paragraph{Measurement of accuracy}

Accuracy of the quasi-equilibrium approximation near a given $M$
can be measured by the {\it defect of invariance}
(Fig.~\ref{QEpro}):
\begin{equation}\label{defect}
\Delta_{f^*_{M}} = J(f^*_M) - \pi_{f^*_{M}} J(f^*_M).
\end{equation}
A dimensionless criterion of accuracy is the ratio $\|
\Delta_{f^*_{M}} \| / \| J(f^*_M)\| $ (a ``sine" of the angle
between $J$ and tangent space). If $\Delta_{f^*_{M}}\equiv 0$ then
the quasi-equilibrium manifold is an invariant manifold, and the
quasi-equilibrium approximation is exact. In applications, the
quasi-equilibrium approximation is usually not exact.

\paragraph{The Gibbs entropy and the
Boltzmann entropy}

For analysis of micro-macro relations some authors
\cite{LebBloEnt,Bentr} call entropy $S(f)$ the {\it Gibbs entropy},
and introduce a notion of the {\it Boltzmann entropy}. Boltzmann
defined the entropy of a macroscopic system in a macrostate $M$ as
the $\log$ of the volume of phase space (number of microstates)
corresponding to $M$.  In the proposed level of generality
\cite{G1,Ocherki}, the Boltzmann entropy of the state $f$ can be
defined as $S_{\rm B}(f)=S(f^*_{m(f)})$. It is entropy of the
projection of $f$ onto quasi-equilibrium manifold (the ``shadow"
entropy). For conservative systems the Gibbs entropy is constant,
but the Boltzmann entropy increases \cite{Ocherki} (during some
time, at least) for motions that start on the quasi-equilibrium
manifold, but not belong to this manifold.

These notions of the Gibbs or Boltzmann entropy are related to
micro-macro transition and may be applied to any convex entropy
functional, not the BGS entropy (\ref{BGS}) only. This may cause
some terminological problems (we hope, not here), and it may be
better just to call $S(f^*_{m(f)})$ the {\it macroscopic entropy}.

\paragraph{Invariance equation and the Chapman--Enskog expansion}

The first method for improvement of the quasi-equilibrium
approximation was the Chapman--Enskog method for the Boltzmann
equation \cite{Chapman}. It uses  the explicit structure of
singularly perturbed systems. Many other methods were invented
later, and not all of them use this explicit structure (see, for
example review in \cite{GorKar}). Here we develop the
Chapman--Enskog method for one important class of {\it model
equations} that were invented to substitute the Boltzmann equation
and other more complicated systems when we don't know the details of
microscopic kinetics. It includes the well-known
Bhatnagar--Gross--Krook (BGK) kinetic equation \cite{BGK} , as well
as wide class of generalized model equations \cite{GKMod}.

As a starting point we take a formal kinetic equation with a small
parameter $\epsilon$
\begin{equation}\label{FormSingPert}
\frac{\D f}{\D t} = J(f)= F(f)+\frac{1}{\epsilon}(f^*_{m(f)}- f).
\end{equation}
The term $(f^*_{m(f)}- f)$ is non-linear because nonlinear
dependency $f^*_{m(f)}$ on $m(f)$.

We would like to find a reduced description valid for macroscopic
variables $M$. It means, at least, that we are looking for an
invariant manifold parameterized by $M$, $f=f_M$, that satisfies
the {\it invariance equation}:
\begin{equation}\label{InvEq}
(D_M f_M)( m(J(f_M)))=J(f_M).
\end{equation}
The invariance equation means that the time derivative of $f$
calculated through the time derivative of $M$ ($\dot{M} =m(J(f_M))$)
by the chain rule coincides with the true time derivative $J(f)$.
This is the central equation for the model reduction theory and
applications. First general results about existence and regularity
of solutions to that equation were obtained by Lyapunov \cite{Lya}
(see review in \cite{CMIM,GorKar}).  For kinetic equation
(\ref{FormSingPert}) the invariance equation has a form
\begin{equation}\label{InvEqSing}
(D_M f_M)( m(F(f_M)))=F(f_M)+\frac{1}{\epsilon}(f^*_{M}- f_M),
\end{equation}
because the self-consistency identity (\ref{SelfConsId}).

Due to presence of small parameter $\epsilon$ in $J(f)$, the zero
approximation is obviously the quasi-equilibrium approximation:
$f^{(0)}_M=f^*_M$. Let us look for $f_M$ in the form of power
series: $f_M =f^{(0)}_M+{\epsilon}f^{(1)}_M+\ldots$;
$m(f^{(k)}_M)=0$ for $k\geq 1$. From (\ref{InvEqSing}) we
immediately find:
\begin{equation}\label{firstChEnBGKmic}
f^{(1)}_M = F(f^{(0)}_M) - (D_M f^{(0)}_M)( m(F(f^{(0)}_M)))=
\Delta_{f^*_M}.
\end{equation}
It is very natural that the first term of the Chapman--Enskog
expansion for model equations (\ref{FormSingPert}) is just the
defect of invariance for the quasi-equilibrium approximation.
Calculation of the following terms is also straightforward.

The correspondent first--order in $\epsilon$ approximation for the
macroscopic equations is:
\begin{equation}\label{firstChEnBGKmac}
\frac{\D M}{\D t}=m(F(f^*_M))+ \epsilon m((D_f
F(f))_{f^*_M}\Delta_{f^*_M}).
\end{equation}
We should remind that $m(\Delta_{f^*_M})=0$. The last term in
(\ref{InvEqSing}) vanishes in macroscopic projection for all orders.

The typical situation for the model equations (\ref{FormSingPert})
is: the vector field $F(f)$ is conservative, $(D_f S(f))F(f)=0$. In
that case, the first term $m(F(f^*_M))$ also conserves the
correspondent Boltzmann (i.e. macroscopic, but not obligatory BGS)
entropy $S(f^*_M)$. But the straightforward calculation of the
Boltzmann entropy $S(f^*_M)$ production for the first-order
Chapman--Enskog term in equation (\ref{firstChEnBGKmac}) gives us
for conservative $F(f)$:
\begin{equation}\label{EntProdChap-Ensk}
\frac{\D S(M)}{\D t} = \epsilon \langle
\Delta_{f^*_M},\Delta_{f^*_M}\rangle_{f^*_M} \geq 0.
\end{equation}
where $\langle \bullet,\bullet \rangle_{f}$ is the entropic scalar
product (\ref{eproductP}). The Boltzmann  entropy production in
the first Chapman--Enskog approximation is zero if and only if
$\Delta_{f^*_M}=0$, i.e. if at  point $M$ the quasi-equilibrium
manifold is locally invariant.

To prove (\ref{EntProdChap-Ensk}) we differentiate the
conservativity identity:
\begin{eqnarray}\label{ConsIdentDiff}
&&(D_f S(f))F(f)\equiv 0 \\
 &&(D^2_f S(F))(F(f),a)+(D_f S(f))((D_f F(f))a) \equiv 0 \nonumber \\
 &&(D_f S(f))((D_f F(f))a)\equiv \langle
 F(f),a\rangle_{f},\nonumber
\end{eqnarray}
use the last equality in the expression of the entropy production,
and take into account that the quasi-equilibrium projector is
orthogonal, hence $$\langle
F({f^*_M}),\Delta_{f^*_M}\rangle_{f^*_M}=\langle
\Delta_{f^*_M},\Delta_{f^*_M}\rangle_{f^*_M}.$$

Below we apply the Chapman--Enskog method to the analysis of
filtered BGK equation.

\subsection{The Ehrenfests' Chain,  Macroscopic Equations and Entropy production}

\paragraph{The Ehrenfests' Chain and entropy growth}

Let $\Theta_t$ be the time shift transformation for the initial
kinetic equation (\ref{FormKinEq}): $$\Theta_t (f(0)) = f(t).$$ The
Ehrenfests' chain (Fig.~\ref{FigCh}) is defined for a given
macroscopic variables $M=m(f)$ and a fixed time of coarse-graining
$\tau$. It is a chain of quasi-equilibrium states $f_0, f_1, \ldots
$:

\begin{figure}[t]
\begin{centering}
\includegraphics[width=60mm, height=55mm]{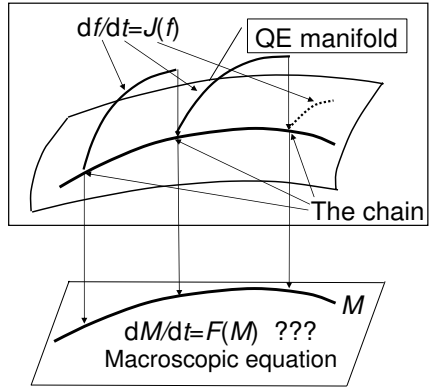}
\caption{\label{FigCh}The Ehrenfests' chain.}
\end{centering}
\end{figure}

\begin{equation}\label{chain}
f_{i+1}=f^*_{m(\Theta_{\tau}(f_i))}.
\end{equation}
To get the next point of the chain, $f_{i+1}$, we take $f_i$, move
it by the time shift $\Theta_{\tau}$, calculate the corresponding
macroscopic state $M_{i+1}=m(\Theta_{\tau}(f_i))$, and find the
quasi-equilibrium state $f^*_{M_{i+1}}=f_{i+1}$.

If the point $\Theta_{\tau}(f_i)$ is not a quasi-equilibrium state,
then $S(\Theta_{\tau}(f_i))<S(f^*_{m(\Theta_{\tau}(f_i))})$ because
of quasi-equilibrium definition (\ref{smax}) and strict concavity of
entropy. Hence, if the motion between $f_i$ and $\Theta_{\tau}(f_i)$
does not belong to the quasi-equilibrium manifold, then
$S(f_{i+1})>S(f_i)$, entropy in the Ehrenfests' chain grows. The
entropy gain consists of two parts: the gain in the motion (from
$f_i$ to $\Theta_{\tau}(f_i)$), and the gain in the projection (from
$\Theta_{\tau}(f_i)$ to $f_{i+1}=f^*_{m(\Theta_{\tau}(f_i))}$). Both
parts are non-negative. For conservative systems the first part is
zero. The second part is strictly positive if the motion leaves the
quasi-equilibrium manifold. Hence, we observe some sort of duality
between entropy production in the Ehrenfests' chain and invariance
of the quasi-equilibrium manifold. The motions that build the
Ehrenfests' chain restart periodically from the quasi-equilibrium
manifold and the entropy growth along this chain is similar to the
Boltzmann entropy growth in the Chapman--Enskog approximation, and
that similarity is very deep, as the exact formulas show below.

\paragraph{The natural projector and macroscopic dynamics}

\begin{figure}[t]
\begin{centering}
\includegraphics[width=62mm, height=73mm]{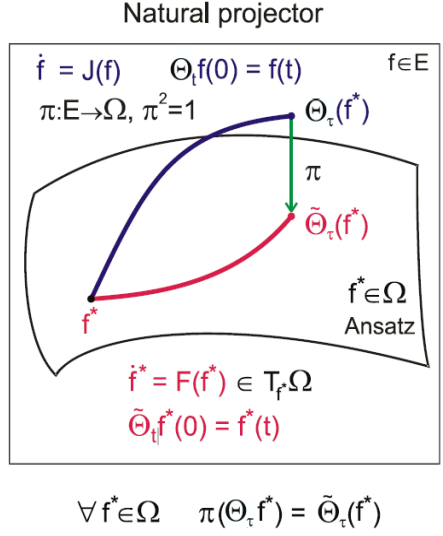}
\caption{\label{FigNatPro}Projection of segments of trajectories:
The microscopic motion above the manifold $\Omega$ and the
macroscopic motion on this manifold. If these motions begin in the
same point on  $\Omega$, then, after time $\tau$, projection of
the microscopic state onto  $\Omega$ should coincide with the
result of the macroscopic motion on $\Omega$. For
quasi-equilibrium $\Omega$, projector $\pi: E \rightarrow \Omega$
acts as $\pi(f)=f^*_{m(f)}$.}
\end{centering}
\end{figure}

How to use the Ehrenfests' chains? First of all, we can try to
define the {\it macroscopic kinetic equations} for $M(t)$ by the
requirement that for any initial point of the chain $f_0$ the
solution of these macroscopic equations with initial conditions
$M(0)=m(f_0)$ goes through all the points $m(f_i)$: $M(n
\tau)=m(f_n)$ ($n=1,2,\ldots$) (Fig.~\ref{FigCh})
\cite{GKIOeNONNEWT2001} (see also \cite{GorKar}). Another way is an
``equation--free approach" \cite{KevFree} to the direct computation
of the Ehrenfests' chain with a combination of microscopic
simulation and macroscopic stepping.

For the definition of the macroscopic equations only the first link
of the Ehrenfests' chain is necessary. In general form, for an
ansatz manifold $\Omega$, projector $\pi: U \rightarrow \Omega$ of
the vicinity of $\Omega$ onto $\Omega$, phase flow of the initial
kinetic equation $\Theta_t$, and macroscopic phase flow
$\tilde{\Theta}_t$ on $\Omega$ the matching condition is
(Fig.~\ref{FigNatPro}):
\begin{equation}\label{match1}
\pi(\Theta_{\tau}(f))=\tilde{\Theta}_{\tau}(f) \; \mbox{for any}
\; f \in \Omega.
\end{equation}
We call this projector of the flow $\Theta$ onto an ansatz
manifold $\Omega$ by fragments of trajectories of given duration
$\tau$ the {\it natural projector} in order to distinguish it from
the standard infinitesimal projector of vector fields on tangent
spaces.

Let us look for the macroscopic equations of the form
\begin{equation}
\frac{\D M}{\D t}=\Psi(M)
\end{equation}
with the phase flow $\Phi_t$: $M(t)=\Phi_t M(0)$. For the
quasi-equilibrium manifold and projector the matching condition
(\ref{match1}) gives
\begin{equation}\label{match2}
m(\Theta_{\tau}(f^*_M))=\Phi_{\tau}(M) \; \mbox{for any
macroscopic state} \; M.
\end{equation}
This condition is the equation for the macroscopic vector field
$\Psi(M)$. The solution of this equation is a function of $\tau$:
$\Psi=\Psi(M,\tau)$. For sufficiently smooth microscopic vector
field $J(f)$ and entropy $S(f)$ it is easy to find the Taylor
expansion of $\Psi(M,\tau)$ in powers of $\tau$. It is a
straightforward exercise in differential calculus. Let us find the
first two terms: $\Psi(M,\tau)=\Psi_0(M)+\tau \Psi_1(M)+o(\tau)$. Up
to the second order in $\tau$ the matching condition (\ref{match2})
is
\begin{eqnarray}\label{match3}
&&m(J(f^*_M))\tau+m((D_f J(f))_{f=f^*_M}
(J(f^*_M)))\frac{\tau^2}{2} \nonumber
\\&& =\Psi_0(M)\tau+ \Psi_1(M)\tau^2+
(D_M \Psi_0(M))(\Psi_0(M))\frac{\tau^2}{2}.
\end{eqnarray}
From this condition immediately follows:
\begin{eqnarray}\label{CoaGr2}
\Psi_0(M)&=&m(J(f^*_M));\\ \Psi_1(M)&=&\frac{1}{2}m[(D_f
J(f))_{f=f^*_M} (J(f^*_M))-(D_M J(f^*_M))(m(J(f^*_M)))]\nonumber
\\ &=&m((D_f J(f))_{f=f^*_M} \Delta_{f^*_M}) \nonumber
\end{eqnarray}
where $\Delta_{f^*_M}$ is the defect of invariance (\ref{defect}).
The macroscopic equation in the first  approximation is:
\begin{eqnarray}\label{MACRO1}
&&\frac{\D M}{\D t}=m(J(f^*_M))+\frac{\tau}{2}m((D_f J(f))_{f=f^*_M}
\Delta_{f^*_M}).
\end{eqnarray}
It is exactly the first Chapman--Enskog approximation
(\ref{firstChEnBGKmac}) for the model kinetics (\ref{FormSingPert})
with $\varepsilon={\tau}/{2}$. The first term $m(J(f^*_M))$ gives
the quasi-equilibrium approximation, the second term increases
dissipation.  The formula for entropy production follows from
(\ref{MACRO1}) \cite{GKOeTPRE2001}. If the initial microscopic
kinetic (\ref{FormKinEq}) is conservative, then for macroscopic
equation (\ref{MACRO1}) we obtain as for the Chapman--Enskog
approximation:
\begin{equation}\label{EntProdCoar}
\frac{\D S(M)}{\D t} = \frac{\tau}{2} \langle
\Delta_{f^*_M},\Delta_{f^*_M}\rangle_{f^*_M},
\end{equation}
where $\langle \bullet,\bullet \rangle_{f}$ is the entropic scalar
product (\ref{eproductP}). From this formula we see again a
duality between the invariance of the quasi-equilibrium manifold
and the dissipativity: entropy production is proportional to the
square of the defect of invariance of the quasi-equilibrium
manifold.

For linear microscopic equations ($J(f)=Lf$) the form of the
macroscopic equations is
\begin{equation}\label{MACROlin}
\frac{\D M}{\D t} =mL\left[1+\frac{\tau}{2}(1- \pi_ {f^*_M}
)L\right]f^*_M,
\end{equation}
where $\pi_ {f^*_M}$ is the quasi-equilibrium projector
(\ref{QEP}).

\paragraph{The Navier--Stokes equation from the free flight dynamics}

The free flight equation describes dynamics of one-particle
distribution function $f(\xx,\vv)$ due to free flight:
\begin{equation}\label{free}
\frac{\partial f(\xx,\vv,t)}{\partial t}=-\sum_i v_i
\frac{\partial f(\xx,\vv,t)}{\partial x_i}.
\end{equation}
The difference from the continuity equation (\ref{continuity}) is
that there is no velocity field $\vv(\xx)$, but the velocity
vector $\vv$ is an independent variable. Equation (\ref{free}) is
conservative and has an explicit general solution
\begin{equation}\label{freeSol}
f(\xx,\vv,t)=f_0(\xx-\vv t,\vv).
\end{equation}
The coarse-graining procedure for (\ref{free}) serves for modeling
kinetics with an unknown  dissipative term $I(f)$
\begin{equation}\label{NonLinKin}
\frac{\partial
f(\xx,\vv,t)}{\partial t}=-\sum_i v_i \frac{\partial
f(\xx,\vv,t)}{\partial x_i} + I(f).
\end{equation}
The Ehrenfests' chain realizes a splitting method for
(\ref{NonLinKin}): first, the free flight step during time $\tau$,
than the complete relaxation to a quasi-equilibrium distribution due
to dissipative term $I(f)$, then again the free flight, and so on.
In this approximation the specific form of $I(f)$ is not in use, and
the only parameter is time $\tau$. It is important that this
hypothetical $I(f)$ preserves all the standard conservation laws
(number of particles, momentum, and energy) and has no additional
conservation laws: everything else relaxes. Following this
assumption, the macroscopic variables are: $M_0=n(\xx,t)=\int f \D
\vv$, $M_i=nu_i=\int v_i f \D \vv$ ($i=1,2,3$), $M_4=\frac{3 n
k_{\rm B }T}{m}+nu^2=\int v^2 f \D \vv$. The zero-order
(quasi-equilibrium) approximation (\ref{QEapp}) gives the classical
Euler equation for compressible non-isothermal gas. In the first
approximation (\ref{MACRO1}) we obtain the Navier--Stokes equations:
\begin{eqnarray}\label{NSEq}
{\partial n\over
\partial t}&=&-\sum_i{\partial (nu_i)\over \partial x_i},\nonumber\\
{\partial (nu_k) \over \partial t}&=&-\sum_i{\partial (nu_ku_i)
\over \partial x_i}-\frac{1}{m}{\partial P \over \partial
x_k}\nonumber\\ && + \underline{{\tau\over 2}\frac{1}{m} \sum_i {\partial
\over \partial x_i}\left[P\left({\partial u_k\over
\partial x_i}+{\partial u_i\over \partial x_k}-{2\over 3}
\delta_{ki}{\rm div} u \right)\right]},\label{NS}\\* {\partial
\mathcal{E}\over
\partial t}&=&-\sum_i{\partial (\mathcal{E} u_i) \over \partial x_i}
- \frac{1}{m} \sum_i{\partial (Pu_i)  \over \partial x_i}
+\underline{\frac{\tau}{2}{5k_{\rm B}\over 2m^2}\sum_i{\partial \over
\partial x_i}\left(P{\partial T\over
\partial x_i}\right)},\nonumber
\end{eqnarray}
where $P= nk_{\rm B}T$ is the ideal gas pressure,
$\mathcal{E}=\frac{1}{2}\int v^2 f \, \D \vv=\frac{3nk_{\rm
B}T}{2m}+\frac{n}{2}u^2$ is the energy density per unite mass
($P=\frac{2m}{3}\mathcal{E}-\frac{mn}{3}u^2$, $T=\frac{2m}{3nk_{\rm
B}}\mathcal{E} - \frac{m}{3k_{\rm B}}u^2$), and the underlined terms
are results of the coarse-graining additional to the
quasi-equilibrium approximation.

The dynamic viscosity in (\ref{NSEq}) is $\mu =\frac{\tau}{2}n
k_{\rm B }T$. It is useful to compare this formula to the
mean--free--path theory that gives $\mu = \tau_{\rm col} n k_{\rm
B }T=\tau_{\rm col}P$, where  $\tau_{\rm col}$ is the collision
time (the time for the mean--free--path). According to these
formulas, we get the following interpretation of the
coarse-graining time $\tau$ for this example: $\tau=2 \tau_{\rm
col}$.

The equations obtained (\ref{NSEq}) coincide with the first--order
terms of the Chapman--Enskog expansion (\ref{firstChEnBGKmac})
applied to the BGK equations with $ \tau_{\rm col}=\tau/2$ and meet
the same problem: the Prandtl number (i.e., the dimensionless ratio
of viscosity and thermal conductivity) is ${\rm Pr}=1$ instead of
the value ${\rm Pr}=\frac{2}{3}$ verified by experiments with
perfect gases and by more detailed theory \cite{Cercignani} (recent
discussion of this problem for the BGK equation with some ways for
its solution is presented in \cite{Struch}).

In the next order in $\tau$ we obtain the stable
post--Navier--Stokes equations instead of the unstable Burnett
equations that appear in the Chapman--Enskog expansion
\cite{GKOeTPRE2001,KTGOePhA2003}. Here we can see the difference
between two approaches.

\paragraph{Persistence of invariance and mistake of differential pursuit}

\begin{figure}[t]
\begin{centering}
a)\includegraphics[width=55mm, height=50mm]{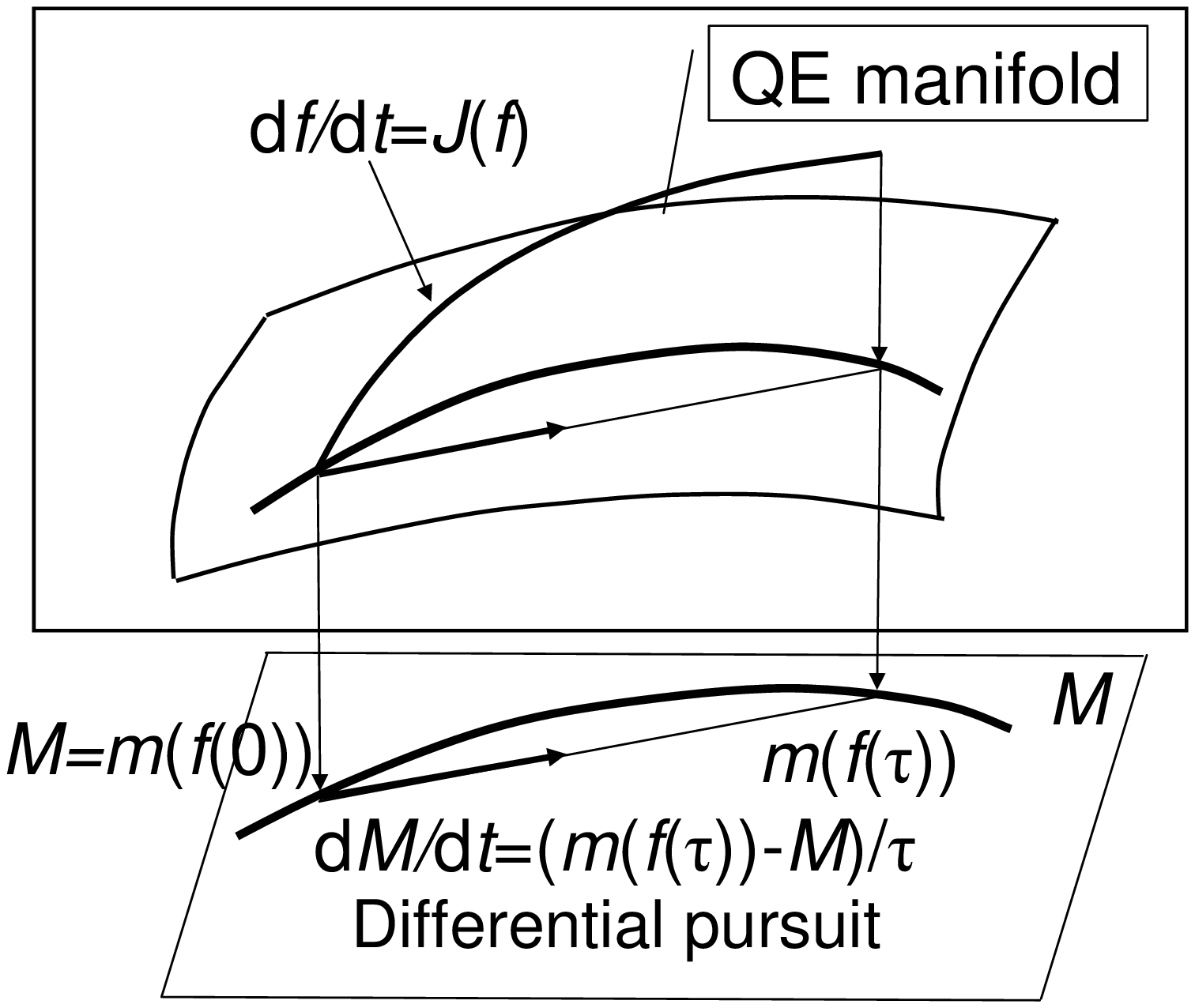}
b)\includegraphics[width=48mm, height=48mm]{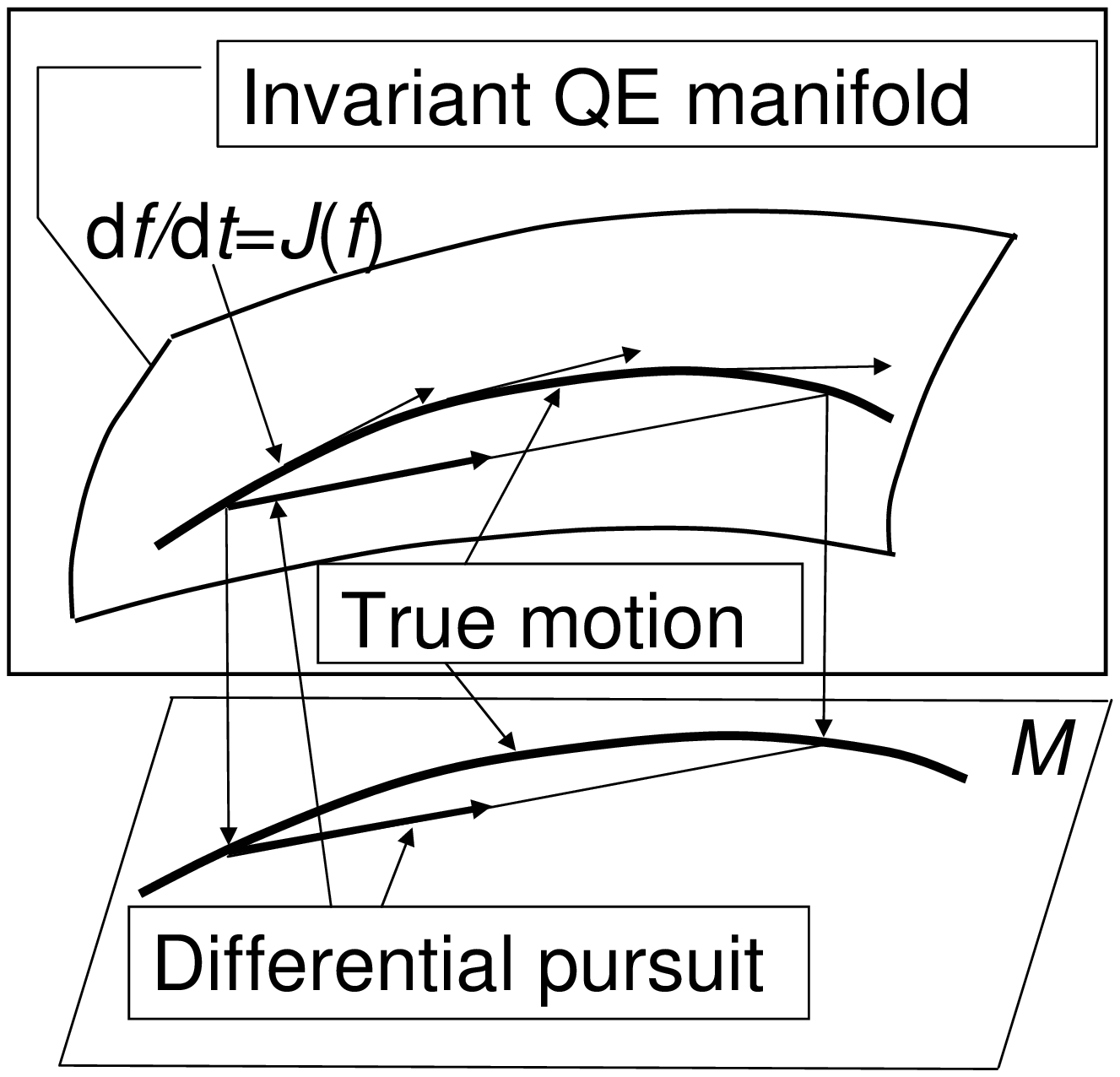}
\caption{\label{FigDifPur}Differential pursuit of the projection
point (a). The mistake of differential pursuit (b): invariant
manifold should preserve its invariance, but it does not!}
\end{centering}
\end{figure}

L.M. Lewis called a generalization of the Ehernfest's approach  a
``unifying principle in statistical mechanics," but he created
other macroscopic equations: he produced the differential pursuit
(Fig.~\ref{FigDifPur}a)
\begin{equation}\label{DifPur}
\frac{\D M}{\D t}=\frac{m(\Theta_{\tau}(f^*_M))-M}{\tau}
\end{equation}
from the full matching condition (\ref{match1}). This means that the
macroscopic motion was taken in the first-order Tailor
approximation, while for the microscopic motion the complete shift
in time (without the Taylor expansion) was used. The basic idea of
this approach is a non-differential time separation: the
infinitesimal shift in macroscopic time is always such a significant
shift for microscopic time that no Taylor approximation for
microscopic motion may be in use. This sort of non-standard analysis
deserves serious attention, but its realization in the form of the
differential pursuit (\ref{DifPur}) does not work properly in many
cases. If the quasi-equilibrium manifold is invariant, then  the
quasi-equilibrium approximation is exact and the Ehrenfests' chain
(Fig.~\ref{FigCh}) just follows the quasi-equilibrium trajectory.
But the differential pursuit does not follow the trajectory
(Fig.~\ref{FigDifPur}b); this motion leaves the invariant
quasi-equilibrium manifolds, and  the differential pursuit does not
approximate the Ehrenfests' chain, even qualitatively.

\paragraph{Ehrenfests' coarse-graining as a method for model
reduction}

\begin{figure}[t]
\begin{centering}
\includegraphics[width=70mm, height=45mm]{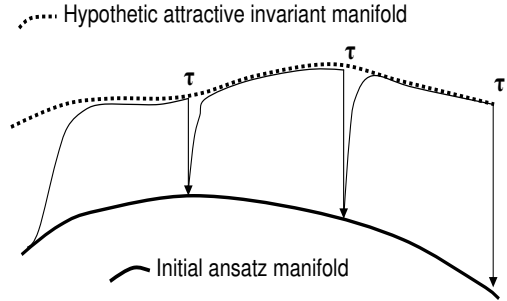}
\caption{\label{NatProInMan}Natural projector and attractive
invariant manifolds. For large $\tau$, the natural projector gives
the approximation of projection of the genuine motion from the
attractive invariant manifold  onto the initial ansatz manifold
$\Omega$. }
\end{centering}
\end{figure}

The problem of model reduction in dissipative kinetics is
recognized as a problem of time separation  and construction  of
slow invariant manifolds. One obstacle on this way is that the
slow invariant manifold is the thing that many people would like
to find, but nobody knows exactly what it is. There is no
conventional definition of {\it slow} invariant manifold without
explicit small parameter that tends to zero. It seems now that the
most reasonable way for such a definition is the analysis of
induced dynamics of manifolds immersed into phase space. Fixed
points of this dynamics are invariant manifolds, and
asymptotically stable (stable and attracting) fixed points are
slow invariant manifolds. This concept was explicitly  developed
very recently \cite{Shnol,CMIM,GorKar}, but the basic idea was
used in earlier applied works \cite{Ocherki,Kev}.

The coarse-graining procedure was developed for {\it erasing} some
details of the dynamics in order to provide entropy growth and
uniform tendency to equilibrium. In this sense, the
coarse-graining is opposite to the model reduction, because for
the model reduction we try to find slow invariant manifolds as
exactly, as we can. But unexpectedly the coarse-graining becomes a
tool for model reduction without any ``erasing."

Let us assume that for dissipative dynamics with entropy growth
there exists an attractive invariant manifold. Let us apply the
Ehrenfests' coarse-graining to this system for sufficiently large
coarse-graining time $\tau$. For the most part of time $\tau$ the
system will spend in a small vicinity of the attractive invariant
manifold. Hence, the macroscopic projection will describe the
projection of dynamics from the  attractive invariant manifold onto
ansatz manifold $\Omega$. As a result, we shall find a shadow of the
proper slow dynamics without looking for the slow invariant
manifold. Of course, the results obtained by the Taylor expansion
(\ref{match3}--\ref{MACRO1}) are not applicable for the case of
large coarse-graining time $\tau$, at least, directly. Some attempts
to utilize the idea of large $\tau$ asymptotic are presented in
\cite{GorKar} (Ch. 12).

One can find a source of this idea in the first work of D.~Hilbert
about the Boltzmann equation solution \cite{Hilbert} (a recent
exposition and development of the Hilbert method is presented in
 \cite{Sone} with many examples of applications). In the
Hilbert method, we start from the local Maxwellian manifold (that
is, quasi-equilibrium one) and iteratively look for ``normal
solutions." The normal solutions $f_{\rm
H}(\vv,\,n(\xx,t),\,\uu(\xx,t),\,T(\xx,t))$ are solutions to the
Boltzmann equation that depend on space and time only through five
hydrodynamic fields. In the Hilbert method no final macroscopic
equation arises. The next attempt to utilize this idea without
macroscopic equations is the ``equation free" approach
\cite{KevFree,GearKap}.

The Ehrenfests' coarse-graining as a tool for extraction of exact
macroscopic dynamics was tested on exactly solvable problems
\cite{GKPRE02}. It gives also a new approach to the
fluctuation--dissipation theorems \cite{GKMex2001}.

\subsection{Kinetic models, entropic involution, and the
second--order ``Euler method"}

\paragraph{Time-step -- dissipation decoupling problem}

Sometimes, the kinetic equation is much simpler than the
coarse-grained dynamics. For example, the free flight kinetics
(\ref{free}) has the obvious exact analytical solution
(\ref{freeSol}), but the Euler or the Navier--Stokes equations
(\ref{NSEq}) seem to be very  far from being exactly solvable. In
this sense, the Ehrenfests' chain (\ref{chain}) (Fig.~\ref{FigCh})
gives a stepwise approximation to a solution of the coarse-grained
(macroscopic) equations by the chain of solutions of the kinetic
equations.

If we use the second-order approximation in the coarse-graining
procedure (\ref{match3}), then the Ehrenfests' chain with step
$\tau$ is the second--order  (in time step $\tau$) approximation to
the solution of macroscopic equation (\ref{MACRO1}). It is very
attractive for hydrodynamics: the second--order in time method with
approximation just by broken line built from intervals of simple
free--flight solutions. But if we use the Ehrenfests' chain for
approximate solution, then the strong connection between the time
step $\tau$ and the coefficient in equations (\ref{MACRO1}) (see
also the entropy production formula (\ref{EntProdCoar})) is strange.
Rate of dissipation is proportional to $\tau$, and it seems to be
too restrictive for computational applications: decoupling of time
step and dissipation rate is necessary. This decoupling problem
leads us to a question that is strange from the Ehrenfests'
coarse-graining point of view: {\it how to construct an analogue to
the Ehrenfests' coarse-graining chain, but without dissipation?} The
{\it entropic involution} is a tool for this construction.

\paragraph{Entropic involution}

The entropic involution was invented for improvement of the
lattice--Boltzmann method \cite{ELB1}. We need to construct a chain
with zero macroscopic entropy production and second order of
accuracy in time step $\tau$. The chain consists of intervals of
solution of kinetic equation (\ref{FormKinEq}) that is conservative.
The time shift for this equation is $\Theta_t$. The macroscopic
variables $M=m(f)$ are chosen, and the time shift for corresponding
quasi-equilibrium equation is (in this section)
$\widetilde{\Theta}_t$. The standard example is: the free flight
kinetics (\ref{free},\ref{freeSol}) as a microscopic conservative
kinetics, hydrodynamic fields (density--velocity--kinetic
temperature) as macroscopic variables, and the Euler equations as a
macroscopic quasi-equilibrium equations for conservative case (see
(\ref{NSEq}), not underlined terms).

Let us start from construction of one link of a chain and  take a
point $f_{1/2}$ on the quasi-equilibrium manifold. (It is not an
initial point of the link, $f_0$, but a ``middle" one.) The
correspondent value of $M$ is $M_{1/2}=m(f_{1/2})$. Let us define
$M_0=m(\Theta_{-\tau /2}(f_{1/2}))$, $M_1=m(\Theta_{\tau /2
}(f_{1/2}))$. The dissipative term in macroscopic equations
(\ref{MACRO1}) is linear in $\tau$, hence, there is a symmetry
between forward and backward motion from any quasiequilibrium
initial condition with the second-order accuracy in the time of
this motion (it became clear long ago \cite{Ocherki}). Dissipative
terms in the shift from $M_0$ to $M_{1/2}$ (that decrease
macroscopic entropy $S(M)$) annihilate with dissipative terms in
the shift from $M_{1/2}$ to $M_{1}$ (that increase macroscopic
entropy $S(M)$). As the result of this symmetry, $M_{1}$ coincides
with  $\widetilde{\Theta}_{\tau}(M_0)$ with the second-order
accuracy. (It is easy to check this statement by direct
calculation too.)

It is necessary to stress that the second-order accuracy is
achieved on the ends of the time interval only:
$\widetilde{\Theta}_{\tau}(M_0)$ coincides with $M_1=
m(\Theta_{\tau}(f_0))$ in the second order in $\tau$
 $$m(\Theta_{\tau}(f_0))-\widetilde{\Theta}_{\tau}(M_0) = o(\tau ^2).$$
On the way $\widetilde{\Theta}_{t}(M_0)$ from $M_0$ to
$\widetilde{\Theta}_{\tau}(M_0)$ for $0< t  < \tau $ we can
guarantee the first-order accuracy only (even for the middle point).
It is essentially the same situation as we had for the Ehrenfests'
chain: the second order accuracy of the matching condition
(\ref{match2}) is postulated for the moment $\tau$, and for $0< t <
\tau $ the projection of the $m(\Theta_{t}(f_0))$ follows a solution
of the macroscopic equation (\ref{MACRO1}) with the first order
accuracy only. In that sense, the method is quite different from the
usual second--order methods with intermediate points, for example,
from the Crank--Nicolson schemes. By the way, the middle
quasi-equilibrium point, $f_{1/2}$ appears for the initiation step
only. After that, we work with the end points of links.

The link is constructed. For the initiation step, we used the middle
point $f_{1/2}$ on the quasi-equilibrium manifold. The end points of
the link, $f_0=\Theta_{-\tau /2}(f_{1/2})$ and $f_1=\Theta_{\tau
/2}(f_{1/2})$ don't belong to the quasi-equilibrium manifold, unless
it is invariant. Where are they located? They belong a surface that
we call a {\it film of non-equilibrium states}
\cite{GKGeoNeo,Plenka,GorKar}. It is a trajectory of the
quasi-equilibrium manifold due to initial microscopic kinetics. In
\cite{GKGeoNeo,Plenka,GorKar} we studied mainly the positive
semi-trajectory (for positive time). Here we need shifts in both
directions.

A point $f$ on the film of non-equilibrium states is naturally
parameterized by $M,\tau$: $f=q_{M,\tau}$, where $M=m(f)$ is the
value of the macroscopic variables, and $\tau (f)$ is the time of
shift from a quasi-equilibrium state: $ \Theta_{-\tau }(f)$ is a
quasi-equilibrium state. In the first order in $\tau$,
\begin{equation}\label{filmFirOr}
q_{M, \tau}=f^*_M+ \tau \Delta_{f^*_M},
\end{equation}
and the first-order Chapman--Enskog approximation
(\ref{firstChEnBGKmic}) for the model BGK equations is also here
with $\tau=\epsilon$. (The two--times difference between kinetic
coefficients for the Ehrenfests' chain and the first-order
Chapman--Enskog approximation appears because for the Ehrenfests'
chain the distribution walks linearly between $q_{M, 0}$ and
$q_{M, \tau}$, and for the first-order Chapman--Enskog
approximation it is exactly $q_{M, \tau}$.)

For each $M$ and positive $s$ from some interval $]0,\varsigma[$
there exist two such $\tau_{\pm}(M,s)$ ($\tau_+(M,s) >0$,
$\tau_-(M,s) <0$) that
\begin{equation}\label{Entropy-time}
S(q_{M,\tau_{\pm}(M,s)})=S(M)-s.
\end{equation}
Up to the second order in $\tau_{\pm}$
\begin{equation}\label{Entr-time-Teylor}
s=\frac{\tau_{\pm}^2}{2}\langle
\Delta_{f^*_M},\Delta_{f^*_M}\rangle_{f^*_M} +  o(\tau_{\pm}^2)
\end{equation}
(compare to (\ref{EntProdCoar})), and
\begin{equation}
\tau_+=-\tau_- + o(\tau_-); \ |\tau_{\pm}|=\sqrt{\frac{s}{\langle
\Delta_{f^*_M},\Delta_{f^*_M}\rangle_{f^*_M}}}(1+ o(1)).
\end{equation}
Equation (\ref{Entropy-time}) describes connection between entropy
change $s$ and time coordinate $\tau$ on the film of non-equilibrium
states, and (\ref{Entr-time-Teylor}) presents the first non-trivial
term of the Taylor expansion of (\ref{Entropy-time}).

The {\it entropic involution} $I_S$ is the transformation of the
film of non-equilibrium states:
\begin{equation}\label{EntrInv}
I_S(q_{M,\tau_{\pm}})=q_{M,\tau_{\mp}}.
\end{equation}
This involution transforms $\tau_+$ into $\tau_-$, and back. For a
given macroscopic state $M$, the entropic involution $I_S$
transforms the curve of non-equilibrium states $q_{M,\tau}$ into
itself.

In the first order in $\tau$ it is just reflection $q_{M,\tau} \to
q_{M,-\tau}$. A partial linearization is also in use. For this
approximation, we define nonlinear involutions of straight lines
parameterized by $\alpha$, not of curves:
\begin{equation}\label{EntrInvLin}
I^0_S(f)=f^*_{m(f)}- \alpha (f-f^*_{m(f)}), \ \alpha>0,
\end{equation}
with condition of entropy conservation
\begin{equation}\label{EntrConsInv}
S(I^0_S(f))=S(f).
\end{equation}
The last condition serves as equation for $\alpha$. The positive
solution is unique and exists for $f$ from some vicinity of the
quasi-equilibrium manifold. It follows from the strong concavity
of entropy. The transformation $I^0_S$ (\ref{EntrConsInv}) is
defined not only on the film of non-equilibrium states, but on all
distributions (microscopic) $f$ that are sufficiently closed to
the quasi-equilibrium manifold.

\begin{figure}[t]
\begin{centering}
a)\includegraphics[width=55mm, height=38mm]{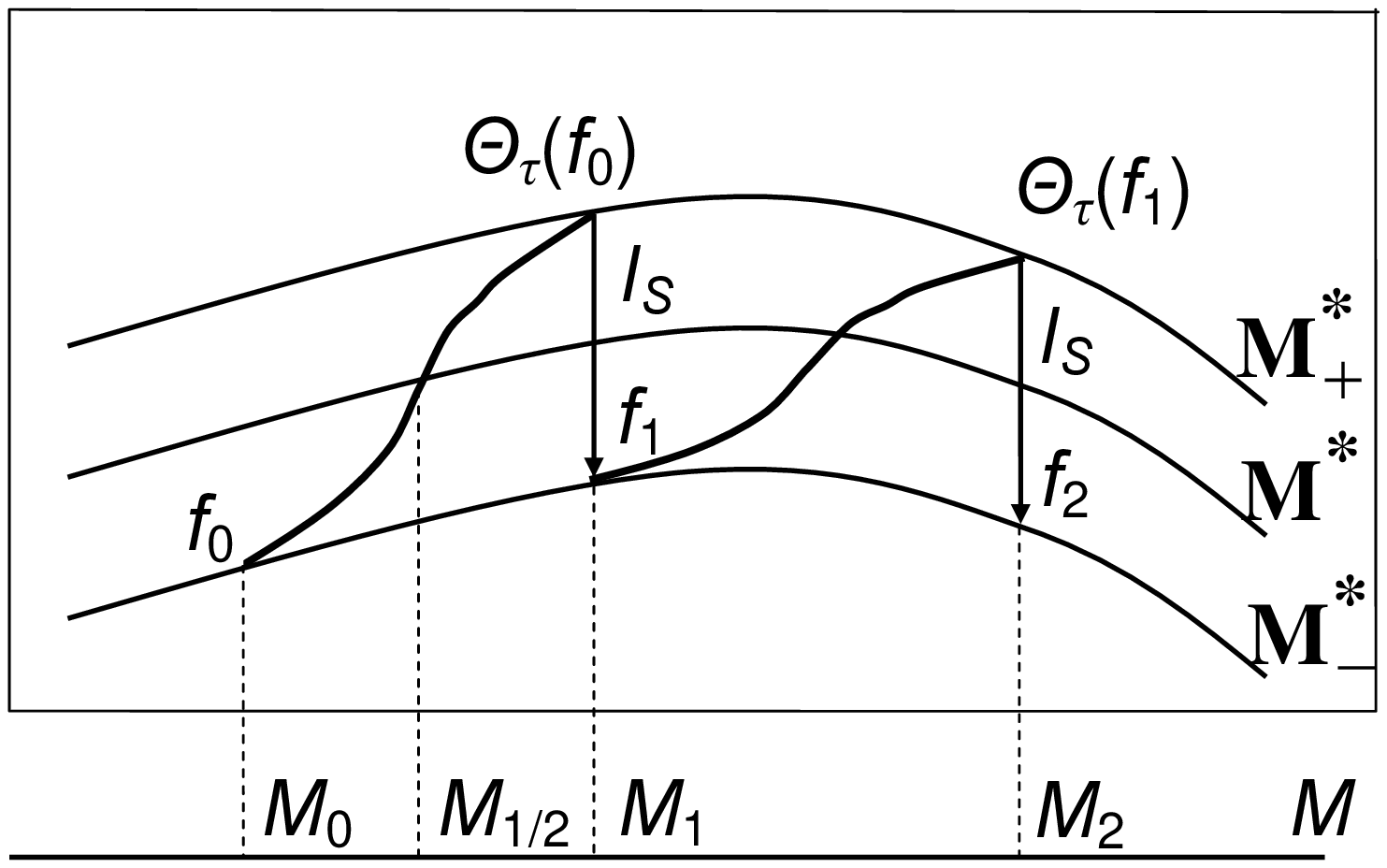}
b)\includegraphics[width=55mm, height=38mm]{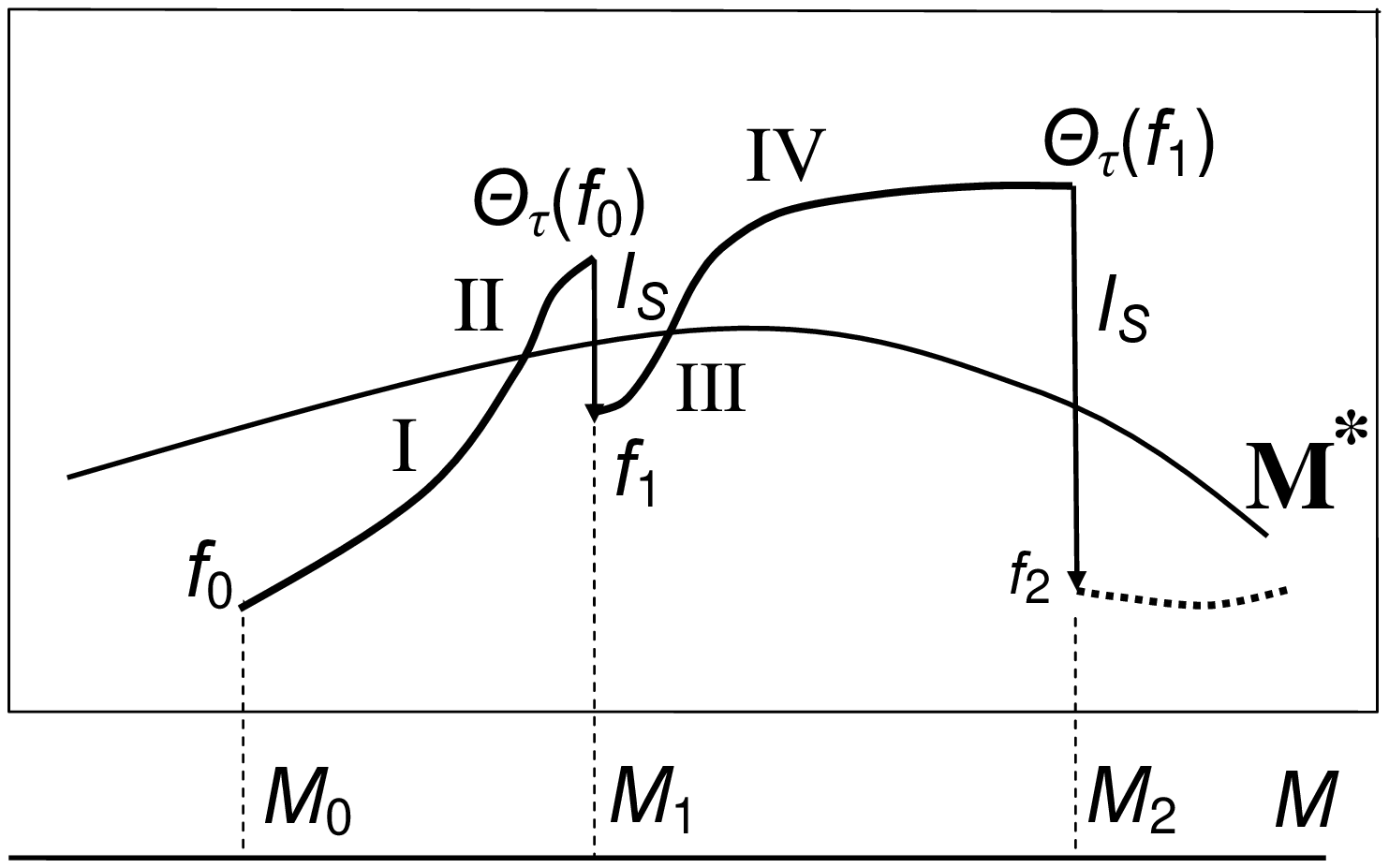}
\caption{\label{FigConsChain}The regular (a) and irregular (b)
conservative chain. Dissipative terms for the regular chain give
zero balance inside each step. For the irregular chain,
dissipative term of part I (the first step) annihilates with
dissipative term of part IV (the second step), as well, as
annihilate dissipartive terms for parts II and III.}
\end{centering}
\end{figure}

In order to avoid the stepwise accumulation of errors in entropy
production, we can choose a constant step in a conservative chain
not in time, but in entropy. Let an initial point in
macro-variables $M_0$ be given, and some $s>0$ be fixed. We start
from the point $f_0=q_{M,\tau_-(M_0,s)}$. At this point, for
$t=0$, $S(m(\Theta_{0}(f_0))))-s=S((\Theta_{0}(f_0)))$
($\Theta_{0}={\rm id}$). Let the motion $\Theta_{t}(f_0)$ evolve
until the equality $S(m(\Theta_{t}(f_0)))-s=S(\Theta_{t}(f_0))$ is
satisfied next time. This time will be the time step $\tau$, and
the next point of the chain is:
\begin{equation}\label{ConsChain}
f_1=I_S(\Theta_{\tau}(f_0)).
\end{equation}
We can present this construction geometrically
(Fig.~\ref{FigConsChain}a). The quasi-equilibrium manifold, ${\bf
M^*}= \{q_{M,0} \}$, is accompanied by two other manifolds, ${\bf
M^*_{\pm}(s)}= \{q_{M,\tau_{\pm}(M,s)} \}$. These manifolds are
connected by the entropic involution: $I_S{\bf M^*_{\pm}(s)}={\bf
M^*_{\mp}(s)}$. For all points $f \in {\bf M^*_{\pm}(s)}$
$$S(f)=S(f^*_{m(f)})-s.$$ The conservative chain starts at a point
on  $f_0 \in {\bf M^*_{-}(s)}$, than the solution of initial
kinetic equations, $\Theta_t(f_0)$, goes to its intersection with
${\bf M^*_{+}(s)}$, the moment of intersection is $\tau$. After
that, the entropic involution transfers $\Theta_{\tau}(f_0)$ into
a second point of the chain, $f_1=I_S(\Theta_{\tau}(f_0)) \in {\bf
M^*_{-}(s)}$.

\paragraph{Irregular conservative chain}

The regular geometric picture is nice, but for some
generalizations we need less rigid structure. Let us combine two
operations: the shift in time $\Theta_{\tau}$ and the entropic
involution $I_S$. Suppose, the motions starts on a point $f_0$ on
the film of non-equilibrium states, and
\begin{equation}\label{IrregConsChain}
f_{n+1} = I_S( \Theta_{\tau} (f_n)).
\end{equation}
This chain we call an {\it irregular conservative chain}, and the
chain that moves from ${\bf M^*_{-}(s)}$ to ${\bf M^*_{+}(s)}$ and
back, the regular one. For the regular chain the dissipative term
is zero (in the main order in $\tau$) already for one link because
this link is symmetric, and the macroscopic entropy ($S(M)$) loose
for a motion from ${\bf M^*_{-}(s)}$ to ${\bf M^*}$ compensate the
macroscopic entropy production on a way from ${\bf M^*}$ to ${\bf
M^*_{+}(s)}$. For the irregular chain (\ref{IrregConsChain}) with
given $\tau$ such a compensation occurs in two successive links
(Fig.~\ref{FigConsChain}b) in main order in $\tau $.

\paragraph{Kinetic modeling for non-zero dissipation. 1. Extension of regular chains}

\begin{figure}[t]
\begin{centering}
a)\includegraphics[width=55mm, height=38mm]{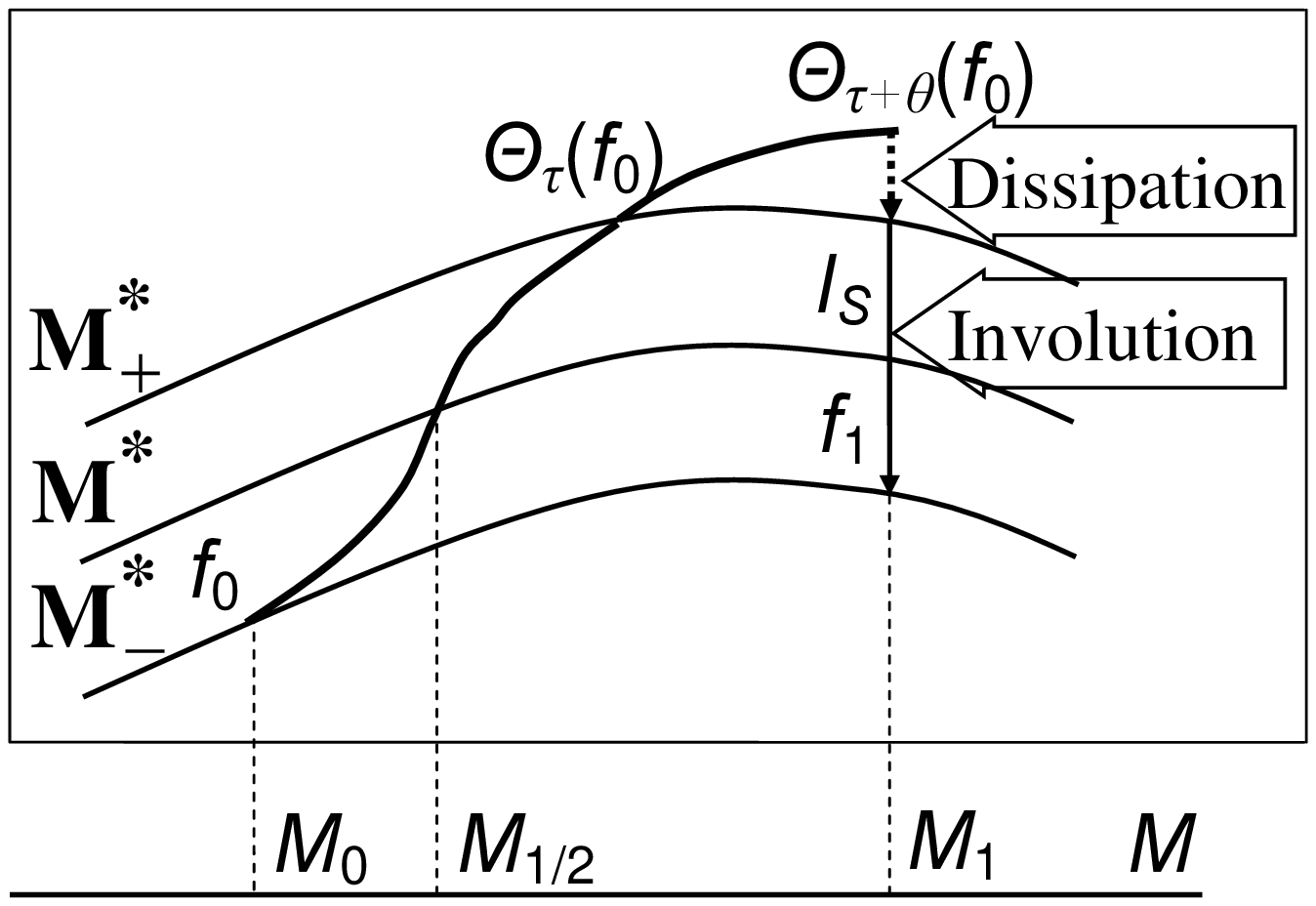}
b)\includegraphics[width=55mm, height=38mm]{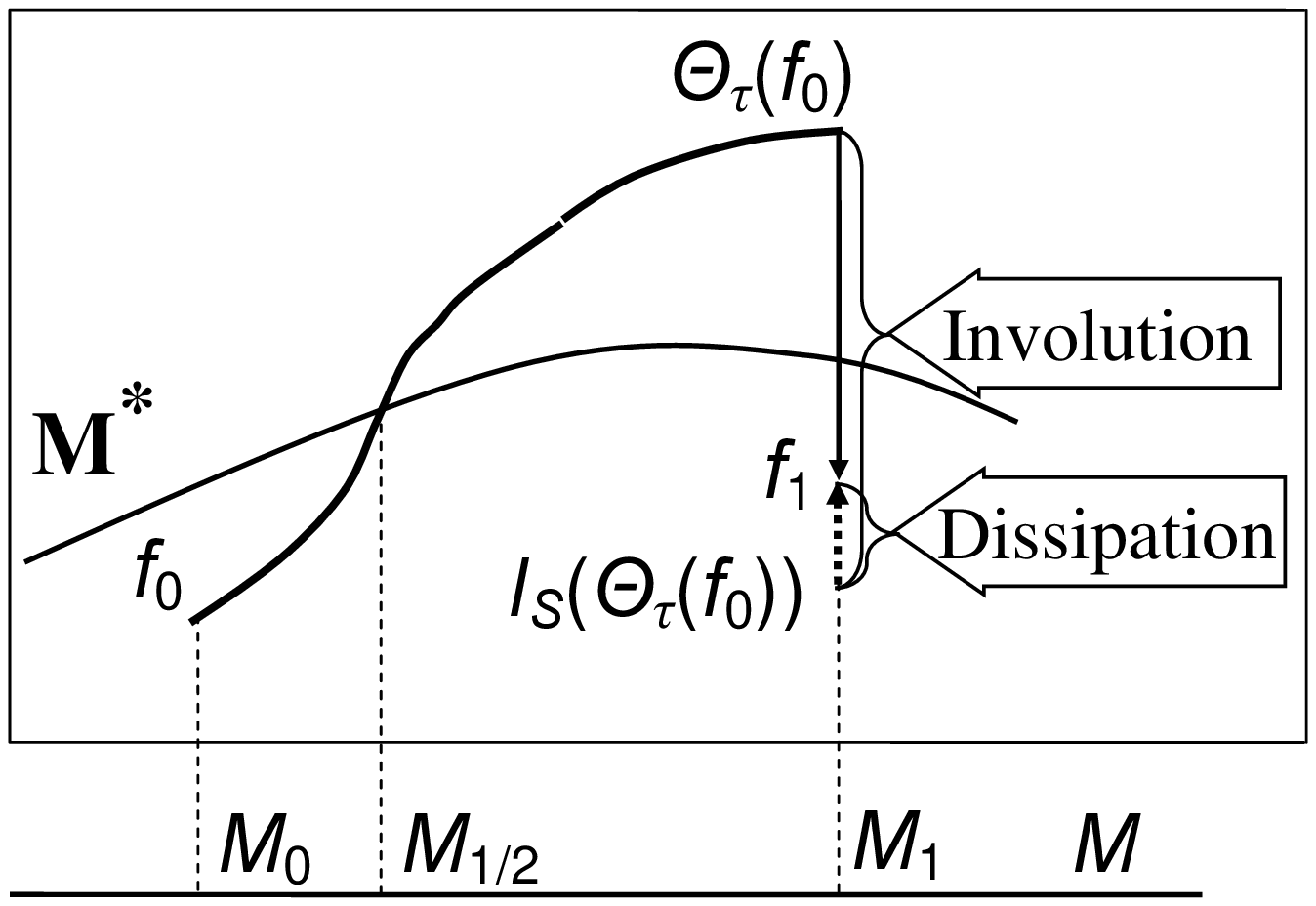}
\caption{\label{FigDissChain} Realization of dissipative chain by
the extra time $\vartheta$ on the base of a regular conservative
chain (a), and by the incomplete involution on the base of an
irregular conservative chain (b).}
\end{centering}
\end{figure}

The conservative  chain of kinetic curves approximates the
quasi-equilibrium dynamics. A typical example of quasi-equilibrium
equations (\ref{QEapp}) is the Euler equation in fluid dynamics.
Now, we combine conservative chains construction with the idea of
the dissipative Ehrenfests' chain in order to create a method for
kinetic modeling of dissipative hydrodynamics (``macrodynamics")
(\ref{MACRO1}) with arbitrary kinetic coefficient that is decoupled
from the chain step $\tau$:
\begin{eqnarray}\label{MACROreg}
&&\frac{\D M}{\D t}=m(J(f^*_M))+\kappa (M) m[(D_f J(f))_{f=f^*_M}
\Delta_{f^*_M}].
\end{eqnarray}
Here, a kinetic coefficient $\kappa (M)\geq 0$ is a non-negative
function of $M$. The entropy production for (\ref{FigDissChain})
is:
\begin{equation}\label{EntProdCoarKappa}
\frac{\D S(M)}{\D t} = \kappa (M) \langle
\Delta_{f^*_M},\Delta_{f^*_M}\rangle_{f^*_M}.
\end{equation}

Let us start from a regular conservative chain and deform it. A
chain that approximates solutions of (\ref{MACROreg}) can be
constructed as follows (Fig.~\ref{FigDissChain}a). The motion
starts from $f_0 \in {\bf M^*_{-}(s)}$, goes by a kinetic curve to
intersection with ${\bf M^*_{+}(s)}$, as for a regular
conservative chain, and, after that, follows the same kinetic
curve an extra time $\vartheta$. This motion stops at the moment
$\tau +\vartheta$ at the point $\Theta_{\tau+\vartheta}(f_0)$
(Fig.~\ref{FigDissChain}a). The second point of the chain, $f_1$
is the unique solution of equation
\begin{equation}\label{DissChain}
m(f_1)=m(\Theta_{\tau+\vartheta}(f_0)), \ f_1 \in {\bf M^*_{-}(s)}.
\end{equation}
The time step is linked with the kinetic coefficient:
\begin{equation}
\kappa=\frac{\vartheta}{2}+o(\tau+ \vartheta).
\end{equation}
For entropy production we obtain the analogue of (\ref{EntProdCoar})
\begin{equation}\label{EntProdDisChain}
\frac{\D S(M)}{\D t} = \frac{\vartheta}{2} \langle
\Delta_{f^*_M},\Delta_{f^*_M}\rangle_{f^*_M} + o(\tau+ \vartheta).
\end{equation}
All these formulas follow from the first--order picture. In the
first order of the time step,
\begin{eqnarray}
q_{M, \tau}&=&f^*_M+ \tau \Delta_{f^*_M}; \nonumber \\
    I_S(f^*_M+ \tau \Delta_{f^*_M})&=&f^*_M- \tau \Delta_{f^*_M}; \nonumber \\
  f_0&=&f^*_{M_0}-\frac{\tau}{2} \Delta_{f^*_{M_0}}; \nonumber \\
  \Theta_{t}(f_0)&=&f^*_{M(t)}+\left(t-\frac{\tau}{2}\right)
  \Delta_{f^*_{M_0}},
\end{eqnarray}
and up to the second order of accuracy (that is, again, the first
non-trivial term)
\begin{equation}
S(q_{M, \tau})=S(M)+\frac{\tau ^2}{2} \langle \Delta_{f^*_M},
  \Delta_{f^*_M} \rangle_{f^*_M}.
\end{equation}
For a regular conservative chains, in the first order
\begin{equation}
f_1=f^*_{M(\tau)}-\frac{\tau}{2}
  \Delta_{f^*_{M_0}}.
\end{equation}
For chains (\ref{DissChain}), in the first order
\begin{equation}
f_1=f^*_{M(\tau + \vartheta)}-\frac{\tau}{2}
  \Delta_{f^*_{M_0}}.
\end{equation}

\paragraph{Kinetic modeling for non-zero dissipation. 2. Deformed involution in irregular chains}

For irregular chains, we introduce dissipation without change of
the time step $\tau$. Let us, after entropic involution, shift the
point to the quasi-equilibrium state (Fig.~\ref{FigDissChain})
with some entropy increase $\sigma(M)$. Because of entropy
production formula (\ref{EntProdCoarKappa}),
\begin{equation}\label{entrgain}
\sigma(M)= \tau \kappa(M)\langle
\Delta_{f^*_M},\Delta_{f^*_M}\rangle_{f^*_M}.
\end{equation}
This formula works, if there is sufficient amount of non-equilibrium
entropy, the difference $S(M_n)-S(f_n)$  should not be too small. In
average, for several (two) successive steps it should not be less
than $\sigma(M)$. The Ehrenfests' chain gives a limit for possible
value of $\kappa(M)$ that we can realize using irregular chains with
overrelaxation:
\begin{equation}\label{EhrLim}
\kappa(M) < \frac{\tau}{2}.
\end{equation}

\begin{figure}[t]
\begin{centering}
\includegraphics[width=80mm, height=40mm]{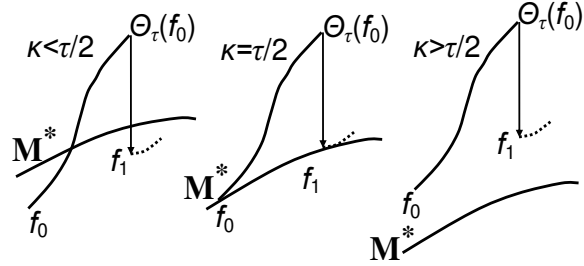}
\caption{\label{StepsBound} The Ehrenfests' limit of dissipation:
three possible links of a dissipative chain: overrelaxation,
$\kappa(M) < \frac{\tau}{2}$ ($\langle\sigma \rangle= s_{\tau} - 2
\sqrt{s_{\tau} \langle s_0 \rangle}$), Ehrenfests' chain, $\kappa(M)
= \frac{\tau}{2}$ ($\sigma= s_{\tau}$), and underrelaxation,
$\kappa(M) > \frac{\tau}{2}$ ($\langle\sigma \rangle= s_{\tau} + 2
\sqrt{s_{\tau} \langle s_0 \rangle}$). }
\end{centering}
\end{figure}

Let  us call the value $\kappa(M) = \frac{\tau}{2}$  the  {\it
Ehrenfests' limit}. Formally, it is possible to realize a chain of
kinetic curves with time step $\tau$ for $\kappa(M) >
\frac{\tau}{2}$ on the other side of the Ehrenfests' limit, without
overrelaxation (Fig.~\ref{StepsBound}).

 Let us choose the
following notation for non-equilibrium entropy: $s_0 =
S(M_0)-S(f_0)$, $s_1 = S(M_1)-S(f_1)$, $s_{\tau}(M)=\frac{\tau
^2}{2}{\langle \Delta_{f^*_M},\Delta_{f^*_M}\rangle_{f^*_M}}$. For
the three versions of steps (Fig.~\ref{StepsBound}) the entropy
gain $\sigma= s(f_1)-S(I_S(\Theta_{\tau}(f_0)))$ in the main order
in $\tau$ is:
\begin{itemize}
\item{For overrelaxation ($\kappa(M) < \frac{\tau}{2}$) $\sigma=
s_{\tau} + s_0-s_1 - 2\sqrt{s_{\tau} s_0}$;}
\item{For the Ehrenfests' chain (full relaxation, $\kappa(M) =
\frac{\tau}{2}$) $s_0=s_1=0$ and $\sigma= s_{\tau}$;}
\item{For underrelaxation ($\kappa(M) > \frac{\tau}{2}$) $\sigma=
s_{\tau} + s_0-s_1 + 2\sqrt{s_{\tau} s_0}$.} \end{itemize}

After averaging in successive steps, the term $s_0-s_1$ tends to
zero, and we can write the estimate of the average entropy gain
$\langle\sigma \rangle$: for overrelaxation $\langle\sigma
\rangle= s_{\tau} - 2 \sqrt{s_{\tau} \langle s_0 \rangle}$ and for
underelaxation $\langle\sigma \rangle= s_{\tau} + 2 \sqrt{s_{\tau}
\langle s_0 \rangle}$.

In the really interesting physical problems the kinetic
coefficient $\kappa(M)$ is non-constant in space. Macroscopic
variables $M$ are functions of space, $\kappa(M)$ is also a
function, and it is natural to take a space-dependent step of
macroscopic entropy production $\sigma(M)$. It is possible to
organize the involution (incomplete involution) step at different
points with different density of entropy production step $\sigma$.

\paragraph{Which entropy rules the kinetic model?}

For linear kinetic equations, for example, for the free flight
equation (\ref{free}) there exist many concave Lyapunov
functionals (for dissipative systems) or integrals of motion (for
conservative systems), see, for example, (\ref{integral}).

There are two reasonable conditions for entropy choice: additivity
with respect to joining of independent systems, and trace form
(sum or integral of some function $h(f,f^*)$). These conditions
select a one-parametric family \cite{ENTR1,ENTR3}, a linear
combination of the classical Boltzmann--Gibbs--Shannon entropy
with $h(f)=-f \ln f$ and the Burg Entropy with $h(f)=\ln f $, both
in the Kullback form: $$S_{\alpha}=-\alpha \int f \ln
\frac{f}{f^*} \, \D \Gamma(x) + (1-\alpha)\int f^* \ln
\frac{f}{f^*} \, \D \Gamma(x),$$
 where $1 \geq \alpha \geq 0$, and $f^* \D
\Gamma$ is invariant measure. Singularity of the Burg term for $f
\to 0$ provides the positivity preservation in all entropic
involutions.

If we weaken these conditions and require that there exists such a
monotonic (nonlinear) transformation of entropy scale that in one
scale entropy is additive, and in transformed one it has a trace
form, then we get additionally a family of Renyi--Tsallis
entropies with $h(f)=\frac{1-f^q}{1-q}$ \cite{ENTR3} (these
entropies and their applications are discussed in details in
\cite{Abe}).

Both the Renyi--Tsallis entropy and the Burge entropy are in use
in the entropic lattice Boltzmann methods from the very beginning
\cite{ELB1,Boghos}. The connection of this entropy choice with
Galilei invariance is demonstrated in \cite{Boghos}.

\paragraph{Elementary examples}

In the most popular and simple example, the conservative formal
kinetic equations (\ref{FormKinEq}) is the free flight equation
(\ref{free}). Macroscopic variables $M$ are the hydrodynamic
fields: $n(\xx)=\int f(\xx,\vv)\, \D \vv$, $n(\xx)\uu(\xx)=\int
\vv f(\xx,\vv)\, \D \vv$, $3n(\xx)k_{\rm B}T/2m  =\frac{1}{2} \int
\vv^2 f(\xx,\vv)\, \D \vv - \frac{1}{2}n(\xx) \uu^2(\xx)$, where
$m$ is particle mass. In 3D at any space point we have five
independent variables.

For a given value of five macroscopic variables $M=\{n,\uu, T\}$
(3D), the quasi-equilibrium distribution is the classical local
Maxwellian:
\begin{equation}\label{LM}
f^*_M(\xx,\vv)=n\left({2\pi k_{\rm B}T\over m}\right)^{-3/2} \exp
\left(- {m(\vv-\uu)^{2}\over 2k_{\rm B}T}\right),
\end{equation}

The standard choice of entropy for this example is the classical
Boltzmann--Gibbs--Shannon entropy (\ref{BGS}) with entropy density
$s(\xx)$. All the involution operations are performed pointwise: at
each point $\xx$ we calculate hydrodynamic moments $M$, the
correspondent local Maxwellian (\ref{LM}) $f^*_M$, and find the
entropic inversion at this point with the standard entropy. For
dissipative chains, it is useful to take the dissipation (the
entropy density gain in one step) proportional to the $S(M)-S(f)$,
and not with fixed value.

The special variation of the discussed example is the free flight
with finite number of velocities: $f(\xx,\vv)= \sum_i
f_i(\xx)\delta(\vv-\vv _i)$. Free flight does not change the set
of velocities $\{\vv_1, \ldots\ldots \vv_n\}$. If we define
entropy, then we can define an equilibrium distribution for this
set of velocity too. For the entropy definition let us substitute
$\delta$-functions in expression for $f(\xx,\vv)$ by some ``drops"
with unite volume, small diameter, and fixed density that may
depend on $i$. After that, the classical entropy formula
unambiguously leads to expression:
\begin{equation}\label{entfin}
s(\xx)= - \sum_i f_i(\xx) \left(\ln \frac{f_i(\xx)}{f^0_i}-1
\right).
\end{equation}
This formula is widely known in chemical kinetics (see elsewhere,
for example \cite{G1,Ocherki,YBGE}). After classical work of
Zeldovich \cite{Zeld} (1938), this function is recognized as a
useful instrument for analysis of chemical kinetic equations. Vector
of values $f^0=f^0_i$ gives us a ``particular equilibrium:" for
$M=m(f^0)$ the conditional equilibrium ($s \to \max$, $M=m(f^0)$) is
$f^0$. With entropy (\ref{entfin}) we can construct all types of
conservative and dissipative chains for discrete set of velocities.
If we need to approximate the continuous local equilibria and
involutions by our discrete equilibria and involutions, then we
should choose a particular equilibrium distribution $\sum_i f^0_i
\delta(\vv-\vv _i)$ in velocity space as an approximation to the
Maxwellian $f^{*0}(\vv)$ with correspondent value of macroscopic
variables $M^0$ calculated for the discrete distribution $f^0$:
$n=\sum_i f^0_i$, ... This approximation of distributions should be
taken in the weak sense. It means that $\vv _i$ are nodes, and
$f^0_i$ are weights for a cubature formula in 3D space with weight
$f^{*0}(\vv)$:
\begin{equation}\label{entrcub}
\int p(\vv) f^{*0}(\vv) \, \D \vv \approx \sum_i p(\vv_i) f^0_i.
\end{equation}
There exist a huge population of cubature formulas in 3D with
Gaussian weight that are optimal in various senses \cite{cubature}.
Each of them contains a hint for a choice of nodes $\vv _i$ and
weights $f^0_i$ for the best discrete approximation of continuous
dynamics. Applications of this entropy (\ref{entfin}) to the lattice
Boltzmann models are developed in \cite{LBperfect}.

There is one more opportunity to use entropy (\ref{entfin}) and
related involutions for discrete velocity systems. If for some of
components $f_i=0$, then we can find the correspondent {\it
positive} equilibrium, and perform the involution in the whole
space. But there is another way: if for some of velocities $f_i=0$,
then we can reduce the space, and find an equilibrium for non-zero
components only, for the shortened list of velocities. These {\it
boundary equilibria} play important role in the chemical
thermodynamic estimations \cite{Kagan}.

This approach allows us to construct systems with variable in space
set of velocities. There could be ``soft particles" with given
velocities, and the density distribution in these particles changes
only when several particles collide. In 3D for the possibility of a
non-trivial equilibrium that does not obligatory coincide with the
current distribution we need more than 5 different velocity vectors,
hence, a non-trivial collision ($\approx$ entropic inversion) is
possible only for 6 one-velocity particles. If in a collision
participate 5 one-velocity particles or less, then they are just
transparent and don't interact at all. For more moments, if we add
some additional fields (stress tensor, for example), the number of
velocity vectors that is necessary for non-trivial involution
increases.

\paragraph{Lattice Boltzmann models: lattice is not a tool for discretization}

In this section, we presented the theoretical backgrounds of kinetic
modeling. These problems were discussed previously for development
of lattice Boltzmann methods in computational fluid dynamics. The
``overrelaxation" appeared in \cite{Succi89}.  In papers
\cite{LBGK1,LBGK2} the overrelaxation based method for the
Navier--Stokes equations was further developed, and the entropic
involution was invented in \cite{ELB1}. Due to historical reasons,
we propose to call it the {\it Karlin--Succi} involution. The
problem of computational stability of entropic lattice Boltzmann
methods was systematically analyzed in \cite{LBperfect,LBentr}.
$H$-theorem for lattice Boltzmann schemes was presented with details
and applications in \cite{LB3}. For further discussion and
references we address to \cite{LB2}.

In order to understand links from the Ehrenfests' chains to the
lattice Boltzmann models, let us take the model with finite number
of velocity vectors and entropy (\ref{entfin}). Let the velocities
from the set $\{\vv_1, \ldots\ldots \vv_n\}$ be automorphisms of
some lattice $\mathbf{L}$: $\mathbf{L}+\vv_i=\mathbf{L}$. Then the
restriction of free flight in time $\tau$ on the functions on the
lattice $\tau \mathbf{L}$ is exact. It means that the free flight
shift in time $\tau$, $f(x,v) \mapsto f(x-v\tau,v)$ is defined on
functions on the lattice, because $\vv_i \tau$ are automorphisms of
$\tau \mathbf{L}$. The entropic involution (complete or incomplete
one) acts pointwise, hence, the restriction of the chains on the
lattice $\tau \mathbf{L}$ is exact too. In that sense, the role of
lattice here is essentially different from the role of grid in
numerical methods for PDE. All the discretization contains in the
velocity set $\{\vv_1, \ldots\ldots \vv_n\}$, and the accuracy of
discretization is the accuracy of cubature formulas (\ref{entrcub}).

The lattice $\tau \mathbf{L}$ is a tool for presentation of
velocity set as a subset of $\mathbf{L}$ automorphism group. At
the same time, it is a perfect screen for presentation of the
chain dynamics, because restriction of that dynamics on this
lattice is an autonomous dynamic of lattice distribution. (Here we
meet a rather rare case of exact model reduction.)

The boundary conditions for the lattice Boltzmann models deserve
special attention. There were many trials of non-physical conditions
until the proper (and absolutely natural) discretization of
well-known classical kinetic boundary conditions (see, for example,
\cite{Cercignani}) were proposed \cite{AK4}. It is necessary and
sufficient just to describe scattering of particles on the boundary
with maximal possible respect to the basic physics (and given
proportion between elastic collisions and thermalization).

\section{Coarse-graining by filtering}

The most popular area for filtering applications in mathematical
physics is the Large Eddy Simulation (LES) in fluid dynamics
\cite{LES2005}. Perhaps, the first attempt to turbulence modeling
was done by Boussinesq in 1887. After that, Taylor (1921, 1935,
1938) and Kolmogorov (1941) have provided the bases of the
statistical theory of turbulence. The Kolmogorov theory of
turbulence self similarity inspired many attempts of so-called
subgrid-scale modeling (SGS model): only the large scale motions of
the flow are solved by filtering out the small and universal eddies.
For the dynamic subgrid-scale models a filtering step is required to
compute the SGS stress tensor. The filtering a hydrodynamic field is
defined as convoluting the field functions with a filtering kernel,
as it is done in electrical engineering:
\begin{equation}\label{ConvFilter}
\overline{\{n, n \uu, n T \}}(\xx) = \int G(\xx-\yy){\{n, n \uu, n T
\}}(\yy)\, \D \yy.
\end{equation}

Various filter kernels are in use. Most popular of them are:
\begin{enumerate}
\item{The box filter $G(x)=H(\Delta/2 - |x|)/\Delta$;}
\item{The Gaussian filter $G(x)=\frac{1}{\Delta}\sqrt{6/ \pi } \exp {(-6 x^2 /
\Delta ^2)}$,}
\end{enumerate}
where $\Delta$ is the filter width (for the Gaussian filter,
$\Delta/2 =\sqrt{3} \sigma$, this convention corresponds to 91.6\%
of probability in the interval $[-\Delta/2,+\Delta/2]$ for the
Gaussian distribution) , $H$ is the Heaviside function, $G(\xx)=
\prod_i G(x_i)$.

In practical applications, implicit filtering is sometimes done by
the grid itself. This filtering by grids should be discussed in
context of the Whittaker--Nyquist--Kotelnikov--Shannon  sampling
theory \cite{Higgins1,Higgins2}. Bandlimited  functions (that is,
functions which Fourier transform has compact support) can be
exactly reconstructed from their values on a sufficiently fine grid
by the Nyquist-Shannon interpolation formula and its
multidimensional analogues. If, in 1D, the Fourier spectrum of
$f(x)$ belongs to the interval $[-k_{\max}, k_{\max}]$, and the grid
step $h$ is less than $\pi / k_{\max}$ (it is, twice less than the
minimal wave length), then this formula gives the exact
representation of $f(x)$ for all points $x$:
\begin{equation}
f(x)= \sum_{n=-\infty}^{+\infty} f(n h) {\rm sinc} \left( \pi
\left[ \frac{x}{h} - n\right]\right),
\end{equation}
where ${\rm sinc} (x)= \frac{\sin x}{x}$. That interpolation
formula implies an exact differentiation formula in the nodes:
\begin{equation}
\left.\frac{\D f(x)}{\D x}\right|_{x=nh}= 2 \pi
\sum_{k=1}^{\infty} (-1)^{k+1}\frac{f((n+k)h)-f((n-k)h)}{2kh}.
\end{equation}
Such ``long tail" exact differentiation formulas are useful under
assumption about bounded Fourier spectrum.

As a background for SGS modeling, the {\it Boussinesq hypothesis}
is widely used. This hypothesis is that the turbulent terms can be
modeled as directly analogues to the molecular viscosity terms
using a ``turbulent viscosity." Strictly speaking, no hypothesis
are needed for equation filtering, and below a sketch of {\it
exact filtering theory} for kinetic equations is presented. The
idea of {\it reversible} regularization without apriory closure
assumptions in fluid dynamics was proposed by Leray \cite{Leray}.
Now it becomes popular again \cite{Germano2,Holm,toyLeonard}.

\subsection{Filtering as auxiliary kinetics}

\paragraph{Idea of filtering in kinetics}

The variety of possible filters is too large, and we need some
fundamental conditions that allow to select physically reasonable
approach.

Let us start again from the formal kinetic equation
(\ref{FormKinEq}) ${\D f}/{\D t}=J(f)$ with concave entropy
functional $S(f)$ that does not increase in time and is defined in
a convex subset $U$ of a vector space $E$.

The filter transformation $\Phi_{\Delta}: U \to U$, where $\Delta$
is the filter width, should satisfy the following conditions:

\begin{enumerate}
\item{Preservation of conservation laws: for any basic
conservation law of the form $C[f]={\rm const}$ filtering does not
change the value $C[f]$: $C[\Phi_{\Delta}(f)]=C[f]$. This
condition should be satisfied for the whole probability or for
number of particles (in most of classical situations), momentum,
energy, and filtering should not change the center of mass, this
is not so widely known condition, but physically obvious
consequence of Galilei invariance.}
\item{The Second Law (entropy growth): $S(\Phi_{\Delta}(f)) \geq S(f)$.}
\end{enumerate}
It is easy to check the conservation laws for convoluting filters
(\ref{ConvFilter}), and here we find the first benefit from the
kinetic equation filtering: for usual kinetic equations and all
mentioned conservation laws functionals $C[f]$ are linear, and the
conservation preservation conditions are very simple linear
restrictions on the kernel $G$ (at least, far from the boundary).
For example, for the Boltzmann ideal gas distribution function
$f(\xx,\vv)$, the number of particles, momentum, and energy conserve
in filtering $\overline{f(\xx,\vv)} = \int G(\xx-\yy)f(\yy,\vv)\, \D
\yy$, if $\int G(\xx) \, \D \xx =1$; for the center of mass
conservation we need also a symmetry condition $\int \xx G(\xx) \,
\D \xx =0$. It is necessary to mention that usual filters extend the
support of distribution, hence, near the boundary the filters should
be modified, and boundary can violate the Galilei invariance, as
well, as momentum conservation. We return to these problems in this
paper later.

For continuum mechanics equations, energy is not a linear
functional, and operations with filters require some accuracy and
additional efforts, for example, introduction of spatially variable
filters \cite{Vreman}. Perhaps, the best way is to lift the
continuum mechanics to kinetics, to filter the kinetic equation, and
then to return back to filtered continuum mechanics. On kinetic
level, it becomes obvious how filtering causes the redistribution of
energy between internal energy and mechanical energy: energy of
small eddies and of other small-scale inhomogeneities partially
migrates into internal energy.

\paragraph{Filtering semigroup}

If we apply the filtering twice, it should lead just just to
increase of the filter width. This natural semigroup condition
reduces the set of allowed filters significantly. The approach
based on filters superposition was analyzed by Germano
\cite{Germano} and developed by many successors.  Let us formalize
it in a form
\begin{equation}\label{filtSemi}
\Phi_{\Delta '} (\Phi_{\Delta}(f)) = \Phi_{\Delta ''} (f),
\end{equation}
where $\Delta ''(\Delta ',\Delta)$ is a monotonic function, $\Delta
'' \geq \Delta '$ and $\Delta '' \geq \Delta $.

The semigroup condition (\ref{filtSemi}) holds for the Gaussian
filter with $\Delta ''^2=\Delta '^2+\Delta^2$, and does not hold
for the box filter. It is convenient to parameterize the semigroup
$\{\Phi_{\Delta} | \Delta \geq 0\}$ by an additive parameter $\eta
\geq 0$ (``auxiliary time"): $\Delta=\Delta(\eta)$,
$\Phi_{\eta}\circ \Phi_{\eta'} =\Phi_{\eta + \eta '}$, $\Phi_0 =
{\rm id}$. Further we use this parameterization.

\paragraph{Auxiliary kinetic equation}

The filtered distribution $f(\eta)=\Phi_{\eta}(f_0)$ satisfies
differential equation
\begin{equation}\label{FiltAuxKin}
\frac{\D f(\eta)}{\D \eta} = \phi (f(\eta)), \; \mbox{where} \;
\phi (f)= \lim_{\eta \to 0} \frac{\Phi_{\eta}(f)-f}{\eta}.
\end{equation}
For Gaussian filters this equation is the simplest diffusion
equation ${\D f(\eta)}/{\D \eta}={\rm \Delta} f$ (here ${\rm
\Delta}$ is the Laplace operator).

Due to physical restrictions on possible filters, auxiliary
equation (\ref{FiltAuxKin}) has main properties of kinetic
equations: it respects conservation laws and the Second Law. It is
also easy to check that in the whole space (without boundary
effects) diffusion, for example, does not change the center of
mass.

So, when we discuss filtering of kinetics, we deal with two
kinetic equations in the same space, but  in two times $t$  and
$\eta$: initial kinetics (\ref{FormKinEq}) and filtering equation
(\ref{FiltAuxKin}). Both have the same conservation laws and the
same entropy.

\subsection{Filtered kinetics}

\paragraph{Filtered kinetic semigroup}

Let $\Theta_t$ be the semigroup of the initial kinetic phase flow.
We are looking for kinetic equation that describes dynamic of
filtered distribution $\Phi_{\eta}f$ for given $\eta$. Let us call
this equation with correspondent dynamics the {\it filtered
kinetics}. It is the third kinetic equation in our consideration,
in addition to the initial kinetics (\ref{FormKinEq}) and the
auxiliary filtering kinetics (\ref{FiltAuxKin}). The natural phase
space for this filtered kinetics is the set of filtered
distributions $\Phi_{\eta} (U)$. For the phase flow of the
filtered kinetics we use notation $\Psi_{(\eta)\,t}$ This filtered
kinetics should be the exact shadow of the true kinetics. It means
that the motion $\Psi_{(\eta)\,t}(\Phi_{\eta}f_0)$ is the result
of filtering of the true motion $\Theta_t(f_0)$: for any $f_0 \in
U$ and $t > 0$
\begin{equation}\label{filtMatch}
\Psi_{(\eta)\,t}(\Phi_{\eta}f_0)=\Phi_{\eta}(\Theta_t(f_0)).
\end{equation}
This equality means that
\begin{equation}\label{filtKinSem}
\Psi_{(\eta)\,t}=\Phi_{\eta} \circ\Theta_t \circ\Phi_{-\eta}
\end{equation}
The transformation $\Phi_{-\eta}$ is defined on the set of
filtered distributions $\Phi_{\eta} (U)$, as well as
$\Psi_{(\eta)\,t}$ is. Now it is necessary to find the vector
field $$\psi_{(\eta)} (f) = \left. \frac{\D \Psi_{(\eta)\,t}
(f)}{\D t }\right|_{t=0} $$ on the base of conditions
(\ref{filtMatch}), (\ref{filtKinSem}). This vector field is the
right-hand side of the filtered kinetic equations
\begin{equation}\label{filtKinEquat}
\frac{\D f}{\D t}= \psi_{(\eta)} (f).
\end{equation}
From (\ref{filtKinSem}) immediately follows:
\begin{equation}\label{EqForEq}
\frac{\D \psi_{(\eta)}(f)}{\D \eta}=(D_f \phi(f))\psi_{(\eta)}(f)-
(D_f \psi_{(\eta)}(f)) \phi(f)=[\psi _{(\eta)},\phi](f),
\end{equation}
where $[\psi,\phi]$ is the Lie bracket of vector fields.

In the first approximation in $\eta$
\begin{equation}\label{GenSmag}
\psi_{(\eta)}(f)=J(f) + \eta ((D_f \phi(f))J(f)- (D_f J(f))
\phi(f))=J(f) + \eta [J,\phi](f),
\end{equation}
the Taylor series expansion for $\psi_{(\eta)}(f)$ is
\begin{equation}\label{GenSmagNonlin}
\psi_{(\eta)}(f)=J(f) + \eta [J,\phi](f)+
\frac{\eta^2}{2}[[J,\phi],\phi](f)+\ldots +
\frac{\eta^n}{n!}[\ldots[J,\underbrace{\phi],...\phi]}_n(f)+\ldots
\end{equation}
We should stress again that filtered  equations
(\ref{filtKinEquat}) with vector field $\psi_{(\eta)}(f)$ that
satisfies (\ref{EqForEq}) is exact and presents just a shadow of
the original kinetics. Some problems may appear (or not) after
truncating the Taylor series (\ref{GenSmagNonlin}), or after any
other approximation.

So, we have two times: physical time $t$ and auxiliary filtering
time $\eta$, and  four different equations of motion in these
times:
\begin{itemize}
\item{initial equation (\ref{FormKinEq}) (motion in time $t$)},
\item{filtering equation (\ref{FiltAuxKin}) (motion in time $\eta$)},
\item{filtered equation (\ref{filtKinEquat}) (motion in time $t$),}  \item{and
equation for the right hand side of filtered equation
(\ref{EqForEq}) (motion in time $\eta$)}.
\end{itemize}

\paragraph{Toy example: advection + diffusion}

Let us consider kinetics of system that is presented by one scalar
density in space (concentration), with only one linear
conservation law, the total number of particles.

In the following example the filtering equation (\ref{FiltAuxKin})
is
\begin{equation}\label{Heat}
\frac{\partial f(\xx, \eta)}{\partial \eta}= {\rm \Delta} f(\xx,
\eta) \: (=\phi(f)).
\end{equation}
The differential of $\phi(f)$ is simply the Laplace operator ${\rm
\Delta}$. The correspondent 3D heat kernel (the fundamental
solution of (\ref{Heat})) is
\begin{equation}\label{HeatKernel}
K(\eta, \xx - \xixi)=\frac{1}{(4\pi \eta)^{3/2}}
\exp\left(-\frac{(\xx - \xixi)^2}{4 \eta}\right).
\end{equation}
After comparing this kernel with the Gaussian filter we find the
filter width $\Delta = \sqrt{24 \eta}$.

Here we consider the diffusion equation (\ref{Heat}) in the whole
space with zero conditions at infinity. For other domains and
boundary conditions the filtering kernel is the correspondent
fundamental solution.

The equation for the right hand side of filtered equation
(\ref{EqForEq}) is
\begin{equation}\label{EqForEqLapl}
\frac{\D \psi_{(\eta)}(f)}{\D \eta}= {\rm
\Delta}(\psi_{(\eta)}(f))- (D_f \psi_{(\eta)}(f)) ({\rm \Delta}f)
\: (=[\psi,\phi](f) ).
\end{equation}

For the toy example we select the advection + diffusion equation
\begin{equation}\label{AdDif}
\frac{\partial f(\xx, t)}{\partial t}=  \kappa {\rm \Delta} f(\xx,
t) -  {\rm div}( \vv(\xx) f(\xx, t)) \: (=J(f)).
\end{equation}
where $\kappa >0$ is a given diffusion coefficient, $\vv(\xx)$ is
a given velocity field. The differential $D_f J(f)$ is simply the
differential operator from the right hand side of (\ref{AdDif}),
because this vector field is linear. After simple straightforward
calculation we obtain the first approximation (\ref{GenSmag}) to
the filtered equation:
\begin{eqnarray}\label{comutatorA}
[J,\phi](f)&=&{\rm \Delta}(J(f))- (D_f J(f)) ({\rm \Delta}f) = -
{\rm \Delta}[{\rm div}(\vv f)]+{\rm div}(\vv {\rm \Delta} f)
 \\&=&{\rm div}[\vv{\rm \Delta} f- {\rm \Delta}(\vv
f)]=-{\rm div}\left[ f{\rm \Delta}\vv+ 2 \sum_r \frac{\partial
\vv}{\partial x_r} \frac{\partial f}{\partial x_r}
\right]\nonumber \\&=&-{\rm div}(f{\rm \Delta}\vv)-\sum_i
\frac{\partial }{\partial x_i} \left[\sum_r\left(\frac{\partial
v_i}{\partial x_r} + \frac{\partial v_r}{\partial x_i}
\right)\frac{\partial f}{\partial x_r} \right. \nonumber \\&&
\left. - \sum_r \frac{\partial f}{\partial x_r}
\left(\frac{\partial v_i}{\partial x_r} - \frac{\partial
v_r}{\partial x_i} \right)\right]\nonumber \\&=& -\sum_i
\frac{\partial }{\partial x_i} \left[\sum_r\left(\frac{\partial
v_i}{\partial x_r} + \frac{\partial v_r}{\partial x_i}
\right)\frac{\partial f}{\partial x_r} \right] - \sum_r
\frac{\partial }{\partial x_r} \left(f \frac{\partial {\rm
div}\vv}{\partial x_r}\right). \nonumber
\end{eqnarray}
The resulting equations in divergence form are
\begin{eqnarray}\label{AdDifFiltered}
\frac{\partial f(\xx, t)}{\partial t}&=&J(f)+\eta [J,\phi](f) \\
&=&  - {\rm div} \left(-\kappa \nabla f+ (\vv+ \eta{\rm
\Delta}\vv)f + 2 \eta \sum_r \frac{\partial \vv}{\partial x_r}
\frac{\partial f}{\partial x_r} \right)\nonumber \\ &=& {\rm div}
( (\kappa - 2 \eta \SSS (\xx)) \nabla f(\xx, t) ) - {\rm
div}((\vv(\xx) + \eta \nabla {\rm div}\vv (\xx)) f(\xx, t))
\nonumber ,
\end{eqnarray}
where $\SSS (\xx) = (S_{ij})=\frac{1}{2}\left(\frac{\partial
v_i}{\partial x_j} + \frac{\partial v_j}{\partial x_i}\right)$ is
the strain tensor. In filtered equations (\ref{AdDifFiltered}) the
additional diffusivity tensor $- 2 \eta \SSS (\xx)$ and the
additional velocity $\eta \nabla {\rm div}\vv (\xx)$ are present.
The additional diffusivity tensor $- 2 \eta \SSS (\xx)$ may be not
positive definite. The positive definiteness of the diffusivity
tensor  $\kappa - 2 \eta \SSS (\xx)$ may be also violated. For
arbitrary initial condition $f_0(\xx)$ it may cause some
instability problems, but we should take into account that the
filtered equations (\ref{AdDifFiltered}) are defined on the space
of filtered functions for given filtering time $\eta$. On this
space the negative diffusion ($\partial_t f = - {\rm \Delta} f$)
is possible during time $\eta$. Nevertheless, the approximation of
exponent (\ref{GenSmagNonlin}) by the linear term (\ref{GenSmag})
can violate the balance between smoothed initial conditions and
possible negative diffusion and can cause some instabilities.

Some numerical experiments with this model (\ref{AdDifFiltered})
for incompressible flows (${\rm div} \vv =0$) are presented in
\cite{toyLeonard}.

Let us discuss equation (\ref{EqForEqLapl}) in more details. We
shall represent it as the dynamics of the filtered advection flux
vector $\PiPi$. The filtered equation for any $\eta$ should have
the form: ${\partial f}/{\partial t} = - {\rm div}(-\kappa \nabla
f + \PiPi(f))$, where
\begin{equation}\label{flux}
\PiPi(f)=\left(\sum_{j_1,j_2,j_3 \geq 0} \aaa _{j_1j_2j_3}(\xx
,\eta) \partial^{j_1j_2j_3}_x \right) f(\xx),
\end{equation}
where
\begin{equation}
\partial^{j_1j_2j_3}_x = \left(\frac{\partial }{\partial
x_1}\right)^{j_1}\left(\frac{\partial }{\partial
x_2}\right)^{j_2}\left(\frac{\partial }{\partial x_3}\right)^{j_3}
\end{equation}

For coefficients $\aaa _{j_1j_2j_3}(\xx ,\eta)$ equation
(\ref{EqForEqLapl}) is
\begin{eqnarray}\label{OperFluxEq}
\frac{\partial \aaa _{j_1j_2j_3}(\xx ,\eta)}{\partial \eta}&=&
{\rm \Delta} \aaa _{j_1j_2j_3}(\xx ,\eta) \\ &+&2\frac{\partial
\aaa _{j_1-1\,j_2j_3}(\xx ,\eta)}{\partial x_1} +2 \frac{\partial
\aaa _{j_1j_2-1 \,j_3}(\xx ,\eta)}{\partial x_2}+ 2\frac{\partial
\aaa _{j_1j_2 j_3-1}(\xx ,\eta)}{\partial x_3} \nonumber.
\end{eqnarray}
The initial conditions are: $\aaa _{000}(\xx ,0) = \vv(\xx)$,
$\aaa _{j_1j_2j_3}(\xx ,0)=0$ if at least one of $j_k > 0$. Let us
define formally $\aaa _{j_1j_2j_3}(\xx ,\eta)\equiv 0$  if at
least one of $j_k$ is negative.

We shall consider (\ref{OperFluxEq}) in the whole space with
appropriate conditions at infinity. There are many representation
of solution to this system. Let us use the Fourier transformation:
\begin{eqnarray}\label{FOperFluxEq}
&&\frac{\partial \hat{\aaa} _{j_1j_2j_3}(\kk ,\eta)}{\partial
\eta} = -k^2 \hat{\aaa} _{j_1j_2j_3}(\kk ,\eta) \\&&+ 2i(k_1
\hat{\aaa} _{j_1-1\,j_2j_3}(\kk ,\eta) + k_2 \hat{\aaa} _{j_1j_2-1
\,j_3}(\kk ,\eta)+ k_3 \hat{\aaa} _{j_1j_2 j_3-1}(\kk ,\eta))
\nonumber.
\end{eqnarray}
Elementary straightforward calculations give us:
\begin{equation}\label{FFilteredCoefficient}
\hat{\aaa} _{j_1j_2j_3}(\kk ,\eta) = (2i\eta) ^{|j|} {\rm e}^{-k^2
\eta}\frac{k_1^{j_1}k_2^{j_2}k_3^{j_3}}{j_1!j_2!j_3!}
\hat{\vv}(\kk),
\end{equation}
where $|j|=j_1+j_2+j_3$. To find this answer, we consider all
monotonic paths on the integer lattice from the point $(0,0,0)$ to
the point $(j_1,j_2,j_3)$. In concordance with
(\ref{FOperFluxEq}), every such a path adds a term
$$\frac{(2i\eta) ^{|j|}}{|j|!} {\rm e}^{-k^2
\eta}k_1^{j_1}k_2^{j_2}k_3^{j_3} \hat{\vv}(\kk)$$ to $\hat{\aaa}
_{j_1j_2j_3}(\kk ,\eta)$. The number of these paths is
$|j|!/(j_1!j_2!j_3!)$.

The inverse Fourier transform gives
\begin{equation}\label{FilteredCoefficient}
\aaa _{j_1j_2j_3}(\xx ,\eta) =(2\eta)
^{|j|-3/2}\frac{\partial^{j_1j_2j_3}_x}{{j_1!j_2!j_3!}} \int \exp-
\frac{(\xx - \yy)^2}{4 \eta} \vv (\yy) \, \D \yy.
\end{equation}
Finally, for $\PiPi$ we obtain
\begin{eqnarray}\label{Finflux}
&&\PiPi(f)\\&&=\sum_{j_1,j_2,j_3 \geq 0} \frac{(2\eta)
^{|j|-3/2}}{j_1!j_2!j_3!}\left(\partial^{j_1j_2j_3}_x \int \exp-
\frac{(\xx - \yy)^2}{4 \eta} \vv (\yy) \, \D \yy \,\right)
\partial^{j_1j_2j_3}_x  f(\xx). \nonumber
\end{eqnarray}
By the way, together with (\ref{Finflux}) we received the following
formula for the Gaussian filtering of products \cite{toyLeonard}. If
the semigroup $\Phi_{\eta}$ is generated by the diffusion equation
(\ref{Heat}), then for two functions $f(\xx), g(\xx)$ in $R^n$ (if
all parts of the formula exist):
\begin{equation}\label{FilteredProduct}
\Phi_{\eta}(fg)= \sum_{j_1,j_2,\ldots j_n \geq 0} \frac{(2\eta)
^{|j|-n/2}}{j_1!j_2! \ldots j_n!}(\partial^{j_1j_2\ldots j_n}_x
\Phi_{\eta}(f)) (\partial^{j_1j_2\ldots j_n}_x  \Phi_{\eta}(g)).
\end{equation}
Generalization of this formula for a broader class of filtering
kernels for convolution filters is described in \cite{Carati}. This
is simply the Taylor expansion of the Fourier transformation of the
convolution equality $\Psi_{t}=\Phi \circ\Theta_t \circ\Phi^{-1}$,
where $\Phi$ is the filtering transformation (see
(\ref{filtKinSem})).

For filtering semigroups all such formulas are particular cases of
the commutator expansion (\ref{GenSmagNonlin}), and calculation of
all orders requires differentiation only. This case includes
non-convolution filtering semigroups also (for example, solutions of
the heat equations in a domain with given boundary conditions, it is
important for filtering of systems with boundary conditions), as
well as semigroups of non-linear kinetic equation.

\paragraph{Nonlinear filtering toy example}

Let us continue with filtering of advection + diffusion equation
(\ref{AdDif}) and accept the standard assumption about
incompressibility of advection flow $\vv$: ${\rm div} \vv =0$. The
value of density $f$ does not change in motion with the advection
flow, and for diffusion the maximum principle exists, hence, it
makes sense to study bounded solutions of (\ref{AdDif}) with
appropriate boundary conditions, or in the whole space. Let us
take $\max f < A$. This time we use the filtering semigroup
\begin{equation}\label{NlinHeat}
\frac{\partial f(\xx, \eta)}{\partial \eta}= -{\rm div}(-(A-f)
\nabla f)=  (A-f) {\rm \Delta} f(\xx, \eta) - (\nabla f)^2 \:
(=\phi(f)).
\end{equation}
This semigroup has slightly better properties of reverse filtering
(at least, no infinity in values of $f$). The first-order filtered
equation (\ref{GenSmag}) for this filter is (compare to
(\ref{comutatorA}):
\begin{eqnarray}\label{AdDifFilteredNlin}
\frac{\partial f(\xx, t)}{\partial t}&=&J(f)+\eta [J,\phi](f) \\
&=& - {\rm div} [-\kappa \nabla (f+ \eta (\nabla f)^2)+2 \eta (A-f)
\SSS \nabla f + \vv f] .\nonumber
\end{eqnarray}
Here,  $\SSS$ is the strain tensor, the term $-2 \eta (A-f) \SSS$ is
the additional (nonlinear) tensor diffusivity, and the term  $ \eta
\kappa \nabla (\nabla f)^2$ describes  the flux from areas with high
$f$ gradient. Because this flux vanishes near critical points of
$f$, it contributes to creation of a patch structure.

In the same order in $\eta$, it is convenient to write:
\begin{eqnarray}\label{AdDifFilteredNlin2}
\frac{\partial f(\xx, t)}{\partial t}= - {\rm div} [-(\kappa - 2
\eta (A-f) \SSS ) \nabla (f+ \eta (\nabla f)^2)+ \vv f] .\nonumber
\end{eqnarray}

The nonlinear filter changes not only the diffusion coefficient, but
the correspondent thermodynamic force also: instead of $-\nabla f$
we obtain $-\nabla (f+ \eta (\nabla f)^2)$. This thermodynamic force
depends on $f$ gradient and can participate in the pattern
formation.

\paragraph{LES + POD filters}

In the title, LES stands for Large Eddy Simulation, as it is before,
and POD stands for Proper Orthogonal Decomposition. POD \cite{POD}
is an application of principal component analysis \cite{princ} for
extraction of main components from the flow dynamics. The basic
procedure is quite simple. The input for POD is a finite set of flow
images (a sample) $\{f_1, \ldots , f_n \}$. These images are
functions in space, usually we have the values of these function on
a grid. In the space of functions an inner product is given. The
first choice gives the $L_2$ inner product $\int fg \, \D x$, or
energetic one, or one of the Sobolev's space inner products. The
mean point $\psi_0=\sum_i f_i /n$ minimizes the sum of distance
squares $\sum_i (f_i - \psi_0)^2$. The first principal component
$\psi_1$ minimizes the sum of distance squares from points $f_i$ to
a straight line $\{ \psi_0 + \alpha \psi_1 \ | \ \alpha \in R \}$,
the second principal component, $\psi_2$, is orthogonal  to $\psi_1$
and minimizes the sum of distance squares from points $f_i$ to a
plain $\{ \psi_0 + \alpha_1 \psi_1 + \alpha_2 \psi_2 \ | \
\alpha_{1,2} \in R \}$, and so on. Vectors of principal components
$\psi_i$ are the eigenvectors of the sample covariance matrix
$\Sigma$, sorted by decreasing eigenvalue $\lambda_i$, where
\begin{equation}
\Sigma = \frac{1}{n}\sum_i (f_i-\psi_0)\otimes(f_i-\psi_0)^T =
\frac{1}{n}\sum_i |f_i-\psi_0\rangle \langle f_i-\psi_0|.
\end{equation}
The projection of a field $f$ on the plane of the $k$ first
principal components is $\psi_0+P_k (f)$, where $P_k$ is the
orthogonal projector on the space spanned by the first $k$
components:
\begin{equation}
P_k(\phi)= \sum_{1 \leq j\leq k} \psi_j (\psi_j,\phi).
\end{equation}
The average square distance from the sample points $f_i$ to the
plane of the $k$ first principal components is
\begin{equation}
\sigma_j^2 = \sum_{j > k} \lambda_j = {\rm tr} \Sigma - \sum_{1 \leq
j \leq k} \lambda_j \;\; \left({\rm tr} \Sigma= \frac{1}{n}\sum_i
(f_i-\psi_0)^2\right) .
\end{equation}
This number, $\sigma_j$, measures the accuracy of substitution of
the typical (in this sample) field $f$ by its projection on the
plane of the $k$ first principal components.

Among many applications of POD in fluid dynamics at least two have
direct relations to the coarse-graining: \begin{itemize}\item{
Postprocessing, that is, analysis of an experimentally observed or
numerically computed flow regime in projection on the
finite-dimensional space of the first principal components;}
\item{Creation of ``optimal" Galerkin approximations (Galerkin POD, \cite{PODGal}). In this
approach, after finding principal components from sampled images of
flow, we project the equations on the first principal components,
and receive a reduced model.}
\end{itemize}
In addition to radical and irreversible step from initial equations
to Galerkin POD, we can use POD filtering semigroup. It suppresses
the component of field orthogonal to selected $k$ first principal
components, but makes this reversibly. The filtering semigroup is
generated by auxiliary equation
\begin{equation}\label{FiltAuxKinPOD}
\frac{\D f(\eta)}{\D \eta} = \phi (f(\eta))= -(1-P_k)(f-\psi_0).
\end{equation}
The filter transformation in explicit form is
\begin{equation}\label{PODfilter}
U_{\eta}(f)=\psi_0+(P_k+{\rm e}^{-\eta}(1-P_k))(f-\psi_0).
\end{equation}
with explicit reverse transformation $U_{-\eta}$.

For equations of the form (\ref{FormKinEq}) $\dot{f}=J(f)$, the POD
filtered equations are
\begin{equation}\label{PODfiltered}
\frac{\D f}{\D t} =(D_fU_{\eta}(f))_{U_{-\eta}(f)}(J(U_{-\eta}(f)))
=(P_k+{\rm e}^{-\eta}(1-P_k))(J(U_{-\eta}f)).
\end{equation}
These equations have nonconstant in space coefficients, because
$P_k$ is combined from functions $\psi_i$. They are also non-local,
because $P_k$ includes integration, but this non-locality appears in
the form of several inner products (moments) only. Of course, this
approach can be combined with usual filtering, nonlinear Galerkin
approximations \cite{GalTem}, and non-linear principal manifold
approaches \cite{GZComp2005}.

\paragraph{Main example: the BGK model kinetic equation}

The famous BGK model equation substitutes the Boltzmann equation
in all cases when we don't care about exact collision integral
(and it is rather often, because usually it is difficult to
distinguish our knowledge about exact collision kernel from the
full ignorance).

For the one-particle distribution function $f(\xx,\vv,t)$ the BGK
equation reads:
\begin{equation}\label{BGK}
\frac{\partial f(\xx,\vv,t)}{\partial t} + \sum_i v_i \frac{\partial
f(\xx,\vv,t)}{\partial x_i} = \frac{1}{\tau_{\rm col}} (f^*_{m(f)}
(\xx,\vv)- f(\xx,\vv,t)),
\end{equation}
where $m(f)=M(t)$ is the cortege of the hydrodynamic fields that
corresponds to $f(\xx,\vv,t)$, and $f^*_{m(f)}$ is the correspondent
local Maxwellian. Let us rescale variables $x,v,t$: we shall measure
$x$ in some characteristic macroscopic units $L$, $v$ in units of
thermal velocity $v_T$ for a characteristic temperature, $t$ in
units $L/v_T$. Of course, there is no exact definition of the
``characteristic time" or length, but usually it works if not take
it too serious. After rescaling, the BGK equation remains the same,
only the parameter becomes dimensionless:
\begin{equation}\label{BGKKn}
\frac{\partial f(\xx,\vv,t)}{\partial t} + \sum_i v_i
\frac{\partial f(\xx,\vv,t)}{\partial x_i} =
\frac{1}{K\!n}(f^*_{m(f)} (\xx,\vv)- f(\xx,\vv,t)),
\end{equation}
where $K\!n=l/L$ is the dimensionless Knudsen number (and $l$ is the
mean--free--path). It is the small parameter in the kinetics --
fluid dynamics transition. If the $K\!n \gtrsim 1$ then the
continuum assumption of fluid mechanics is no longer a good
approximation and kinetic equations must be used.

It is worth to mention that the BGK equation is {\it non-linear}.
The term $f^*_{m(f)}$ depends non-linearly on moments $m(f)$, and,
hence, on the distribution density $f$ too. And $f^*_{m(f)}$ is the
only term in (\ref{BGK}) that don't commute with the Laplace
operator from the filtering equation (\ref{Heat}). All other terms
do not change after filtering.

According to (\ref{GenSmag}), in the first order in $\eta$ the
filtered BGK equation is
\begin{eqnarray}\label{BGKKnF}
&&\frac{\partial f}{\partial t} + \sum_i v_i \frac{\partial
f}{\partial x_i}\\&&  = \frac{1}{K\!n}(f^*_{m(f)} - f)+
 \frac{\eta}{K\!n} (D^2_M f^*_M)_{M=m(f)}(\nabla M,\nabla
M)_{M=m(f)}. \nonumber
\end{eqnarray}
The last notation may require some explanations: $(D^2_M f^*_M)$
is the second differential of  $f^*_M$, for the BGK model equation
it is a quadratic form in $R^5$ that parametrically depends on
moment value $M=\{M_0,M_1,M_2,M_3,M_4\}$. In the matrix form, the
last expression is
\begin{eqnarray}\label{chlen}
&&(D^2_M f^*_M)_{M=m(f)}(\nabla M,\nabla M)_{M=m(f)} \\ &&\qquad
\qquad\qquad\qquad \qquad= \sum_{r=1}^3 \sum_{i,j=0}^4
\left(\frac{\partial^2 f^*_M}{\partial M_i M_j}
\right)_{M=m(f)}\frac{\partial M_i}{\partial x_r}\frac{\partial
M_j}{\partial x_r}. \nonumber
\end{eqnarray}
This expression depends on the macroscopic fields $M$ only. From
identity (\ref{SelfConsIdSecDif}) it follows that the filtering
term gives no inputs in the quasi-equilibrium approximation,
because $m(D^2_Mf^*_M)=0$.

This fact is a particular case of the general {\it commutation
relations} for general quasi-equilibrium distributions. Let a
linear operator $\mathbf{B}$ acts in the space of distributions
$f$, and there exists such a linear operator $\mathbf{b}$ which
acts in the space of macroscopic states $M$ that
$m\mathbf{B}=\mathbf{b}m$. Then
\begin{equation}\label{CommRel}
m(\mathbf{B} f^*_{m(f)} - (D_f f^*_{m(f)}) (\mathbf{B}f))=0.
\end{equation}
This means that the macroscopic projection of the Lie bracket for
the vector fields of equations $\partial_{\eta} f = \mathbf{B}f$
(a field $\phi$) and $\partial_t f=f^*_{m(f)}-f$ (a field
$\theta$) is zero: $m([\theta,\phi])=0$.\footnote{The term $-f$
gives zero input in these Lie brackets for any linear operator
$\mathbf{B}$.} These commutation relations follow immediately from
the self-consistency identities (\ref{SelfConsId}),
(\ref{SelfConsIdDif}), if we use relations
$m\mathbf{B}=\mathbf{b}m$ to carry $m$ through $\mathbf{B}$. In
the case of BGK equation, relations (\ref{CommRel}) hold for any
linear differential or pseudodifferential  operator
$\mathbf{B}=Q(\xx,\partial/\partial \xx)$ that acts on functions
of $\xx$. In this case, $\mathbf{b}=\mathbf{B}$, if we use the
same notation for differentiation of functions and of
vector-functions.

Relations (\ref{CommRel}) imply a result that deserves special
efforts for physical understanding: the filtered kinetic equations
in zero order in the Knudsen number produce the classical Euler
equations for filtered hydrodynamic fields without any trace of the
filter terms. At the same time, direct filtering of the Euler
equation adds new terms.

To obtain the next approximation we need the Chapman--Enskog
method for equation (\ref{BGKKnF}). We developed a general method
for all equations of this type (\ref{firstChEnBGKmic}), and now
apply this method to the filtered BGK equation. Let us take in
(\ref{FormSingPert}) $\epsilon ={K\!n}$, $F(f)=F_{0}(f)+F_{\rm
filt}(f)$, where $F_0 = - \vv \partial /\partial \xx$ is the free
flight operator and
\begin{equation}\label{filtNotTerm}
F_{\rm filt}(f)= \frac{\eta}{K\!n} (D^2_M f^*_M)_{M=m(f)}(\nabla
M,\nabla M)_{M=m(f)}.
\end{equation}
In these notations, for the zero term in the Chapman--Enskog
expansion we have $f^{(0)}_M=f^*_M$, and for the first term
\begin{eqnarray}
&&f^{(1)}_M =f^{\rm NS}_M + f^{\rm filt}_M= \Delta^{\rm
NS}_{f^*_M}+\Delta^{\rm filt}_{f^*_M},
\\&& f^{\rm NS}_M=\Delta^{\rm NS}_{f^*_M}= F_0(f^*_M) - (D_M
f^{*}_M)( m(F_0(f^{*}_M))) \nonumber \\&& f^{\rm filt}_M=
\Delta^{\rm filt}_{f^*_M}=F_{\rm filt}(f)\; \;({\rm because}  \;
m(F_{\rm filt}(f))=0), \nonumber
\end{eqnarray}
where NS stands for Navier--Stokes. The correspondent continuum
equations (\ref{firstChEnBGKmac}) are
\begin{equation}\label{firstChEnBGKmacFiltr}
\frac{\D M}{\D t}=m(F_0(f^*_M))+ {K\!n}\, m(F_0 (\Delta^{\rm
NS}_{f^*_M}+\Delta^{\rm filt}_{f^*_M})).
\end{equation}
Here, the first term includes non-dissipative terms (the Euler ones)
of the Navier--Stokes equations, and the second term includes both
the dissipative terms of the Navier--Stokes equations and the
filtering terms. Let us collect all the classical hydrodynamic terms
together:
\begin{eqnarray}\label{chlenVfiltre}
\frac{\partial M(\xx,t)}{\partial
t}&=&\underbrace{\ldots\ldots\ldots \ldots}_{\rm NS \, terms} +
{K\!n} \, m \! \left(\vv \frac{\partial}{\partial \xx} F_{\rm
filt}(f) \right) \\
 &=&\underbrace{\ldots\ldots\ldots \ldots}_{\rm
NS \, terms} + \eta m \! \left(\vv \frac{\partial}{\partial \xx}
\sum_{r=1}^3 \sum_{i,j=0}^4 \frac{\partial^2 f^*_M}{\partial M_i
M_j}\frac{\partial M_i}{\partial x_r}\frac{\partial M_j}{\partial
x_r}\right),
 \nonumber
\end{eqnarray}
The  NS terms here are the right hand sides of the Navier--Stokes
equations for the BGK kinetics (\ref{NSEq}) (with $\tau=2\tau_{\rm
col}=2{K\!n}$). Of course, (\ref{chlenVfiltre}) is one of the
tensor viscosity -- tensor diffusivity models. Its explicit form
for the BGK equation and various similar model equations requires
several quadratures:
\begin{equation}
C_{ij}=m \! \left(\vv \frac{\partial}{\partial \xx} \sum_{r=1}^3
\sum_{i,j=0}^4 \frac{\partial^2 f^*_M}{\partial M_i M_j}\right)
\end{equation}
(for the Maxwell distributions $f^*_M$ that are just Gaussian
integrals).

\paragraph{Entropic stability condition for the filtered kinetic equations}

Instability of filtered equations is a well-known problem. It arises
because the reverse filtering is an ill-posed operation, the balance
between filter and reverse filter in (\ref{filtKinSem}) may be
destroyed by any approximation, as well as a perturbation may move
the hydrodynamic field out of space of pre-filtered fields. (And the
general filtered equations are applicable for sure in that space
only.)

Analysis of entropy production is the first tool for stability
check. This is a main thermodynamic realization of the Lyapunov
functions method (invented in physics before Lyapunov).

The filtration term $F_{\rm filt}(f)$ (\ref{filtNotTerm}) in the
filtered BGK equation (\ref{BGKKnF}) does not produce the
Boltzmann (i.e. macroscopic) entropy $S(f^*_{m(f)})$, but is not
conservative. In more details:
\begin{enumerate}
\item{$(D_M S(f^*_M))(m(F_{\rm
filt}(f^*_{m(f)})))\equiv 0$, because $m(F_{\rm filt}(f))\equiv
0$;}
 \item{$(D_f S(f))_{f^*_{m(f)}}(F_{\rm
filt}(f^*_{m(f)}))=(D_M S(f^*_M))(m(F_{\rm
filt}(f^*_{m(f)})))\equiv 0$;}
 \item{$(D_f S(f))_{f^*_{m(f)}}(F_{\rm filt}(f))=(D_f S(f))_{f^*_{m(f)}}(F_{\rm
filt}(f^*_{m(f)}))\equiv 0$, because $F_{\rm filt}(f)$ depends on
$f^*_{m(f)}$ only;}
 \item{But for any field $F_{\rm filt}(f)$ that
depends on $f^*_{m(f)}$ only, if the conservativity identity
(\ref{ConsIdentDiff}) $(D_f S(f))_{f}(F_{\rm filt}(f))\equiv 0$ is
true even in a small vicinity of quasi-equilibria, then $F_{\rm
filt}(f)\equiv 0$. Hence, the non-trivial filter term $F_{\rm
filt}(f)$ cannot be conservative, the whole field
$F(f)=F_{0}(f)+F_{\rm filt}(f)$ is not conservative, and we cannot
use the entropy production formula (\ref{EntProdChap-Ensk}).}
\end{enumerate}

Instead of (\ref{EntProdChap-Ensk}) we obtain

\begin{equation}\label{EntProdChap-EnskFil}
\frac{\D S(M)}{\D t} = {K\!n} \langle \Delta^{\rm
NS}_{f^*_M},\Delta^{\rm NS}_{f^*_M}\rangle_{f^*_M} + \eta \langle
\Delta^{\rm NS}_{f^*_M},\Delta^{\rm filt}_{f^*_M}\rangle_{f^*_M} .
\end{equation}

The {\it entropic stability condition} for the filtered kinetic
equations is:
\begin{equation}\label{EntStabCond}
{\D S(M)}/{\D t} \geq 0, \;\; {\rm i.e.} \;\; {K\!n} \langle
\Delta^{\rm NS}_{f^*_M},\Delta^{\rm NS}_{f^*_M}\rangle_{f^*_M} +
\eta \langle \Delta^{\rm NS}_{f^*_M},\Delta^{\rm
filt}_{f^*_M}\rangle_{f^*_M} \geq 0 .
\end{equation}

There exists a plenty of convenient sufficient conditions, for
example,
\begin{eqnarray}\label{EntStabCondSuf}
\eta \leq {K\!n} \frac{|\langle\Delta^{\rm NS}_{f^*_M},\Delta^{\rm
filt}_{f^*_M}\rangle_{f^*_M}|}{\langle\Delta^{\rm
NS}_{f^*_M},\Delta^{\rm NS}_{f^*_M}\rangle_{f^*_M}}; \quad {\rm
or}\quad \eta \leq {K\!n} \sqrt{\frac{\langle\Delta^{\rm
filt}_{f^*_M},\Delta^{\rm
filt}_{f^*_M}\rangle_{f^*_M}}{\langle\Delta^{\rm
NS}_{f^*_M},\Delta^{\rm NS}_{f^*_M}\rangle_{f^*_M}}}.
\end{eqnarray}

The upper boundary for $\eta$ that guaranties  stability of the
filtered equations is proportional to ${K\!n}$. For the Gaussian
filter width $\Delta$ this means $\Delta =L \sqrt{24 \eta} \sim
\sqrt{{K\!n}}$ (where $L$ is the characteristic macroscopic length).
This scaling, $\Delta/L \sim \sqrt{{K\!n}}$, was discussed in
\cite{AnsKarlFiltr} for moment kinetic equations because different
reasons: if $\Delta/L \gg \sqrt{{K\!n}}$ then the Chapman--Enskog
procedure is not applicable, and, moreover, the continuum
description is probably not valid, because the filtering term with
large coefficient $\eta$ violates the conditions of hydrodynamic
limit. This important remark gives the frame for $\eta$ scaling, and
(\ref{EntStabCond}), (\ref{EntStabCondSuf}) give the stability
boundaries inside this scale.

\section[Stable Properties of Structurally Unstable Systems]
{Errors of Models, $\varepsilon$-trajectories and Stable Properties of Structurally Unstable
Systems}

\subsection{Phase flow, attractors and repellers}

\paragraph{Phase flow}

In this section, we return from kinetic systems to general dynamical
systems, and lose such specific tools as entropy and
quasi-equilibrium. Topological dynamics gives us a natural language
for general discussion of limit behavior and relaxation of general
dynamical systems  \cite{[1]}. We discuss a general dynamical system
as a semigroup of homeomorphisms (phase flow transformations):
$\Theta (t,x)$ is the result of shifting point $x$ in time $t$.

Let the phase space $X$ be a compact metric space with the metrics
$\rho$,
\begin{equation}\label{e1}
\Theta: [0, \infty[ \times X  \to X
\end{equation}
be a continuous mapping for any $t \geq 0$; let mapping $\Theta
(t, \cdot): X \to X$ be homeomorphism of $X$  into subset of $X$
and   let these homeomorphisms form monoparametric semigroup:
\begin{equation}\label{e2}
\Theta(0, \cdot) = {\rm id},\ \Theta(t,\Theta(t',x)) =
\Theta(t+t', x)
\end{equation}
for any $t, t' \geq 0, x \in X$.

Below we  call the semigroup of mappings $\Theta(t,\cdot)$   a
{\it semiflow of homeomorphisms} (or,  for  short, semiflow),  or
simply system (\ref{e1}). We assume that the continuous map
$\Theta(t,x)$ is continued to negative time $t$ as far as it is
possible with preservation of the semigroup property (\ref{e2}).
For phase flow we use also notations $\Theta_t$ and $\Theta_t(x)$.
For any given $x \in X$, $x$-{\it motion} is a function of time
$\Theta(t,x)$, $x$-motion is the {\it whole motion} if the
function is defined on the whole axis $t \in ]-\infty,\infty[$.
The image of $x$-motion is the $x$-{\it trajectory}.

\paragraph{Attractors and repellers}

First of all, for the description of limit behaviour we need a
notion of an $\omega$-limit set.

A point $p \in X$ is called $\omega$- ($\alpha$-)-{\it limit
point} of the $x$-motion (correspondingly of the whole
$x$-motion), if there is such a sequence $t_n \to \infty$  ($t_n
\to -\infty$) that $\Theta(t_n, x) \to p$  as $n \to \infty$.  The
totality of all $\omega$- ($\alpha$-)-limit points of $x$-motion
is called its $\omega$- ($\alpha$-)-{\it limit set} and is denoted
by $\omega(x)$  ($\alpha(x)$).

A set $W \subset X$ is called {\it invariant} set, if,  for any $
x \in W$, the $x$-motion is whole and the whole $x$-trajectory
belongs to $W$.

The sets $\omega (x)$, $\alpha (x)$  (the  last  in  the case when
$x$-motion  is  whole)  are  nonempty,  closed, connected, and
invariant.

The set of all $\omega$-limit points of the system
$\omega_{\Theta}=\bigcup_{x \in X} \omega (x)$ is nonempty and
invariant, but may be disconnected and not closed. The sets
$\omega (x)$ might be considered as attractors, and the sets
$\alpha (x)$ as repellers (attractors for $t \to -\infty$). The
system of these sets represents all limit behaviours of the phase
flow.

Perhaps, the most constructive idea of attractor definition
combines pure topological (metric) and measure points of view. A
{\it weak attractor} \cite{Ashwin} is a closed (invariant) set $A$
such that the set $\mathcal{B}(A)=\{x \ | \ \omega(x)\subset A \}$
(a basin of attraction) has strictly positive measure. A {\it
Milnor attractor} \cite{Milnor} is such a weak attractor that
there is no strictly smaller closed $A'\subsetneqq A$ so that
$\mathcal{B}(A)$ coincides with $\mathcal{B}(A')$ up to a set of
measure zero. If $A$ is a Milnor attractor and for any closed
invariant proper subset $A'\subsetneqq A$ the set
$\mathcal{B}(A')$ has zero measure, then we say that $A$ is a
{\it minimal Milnor attractor.}

Below in this section we follow a purely topological (metric)
point of view, but keep in mind that its combination with
measure--based ideas create a richer theory.

\paragraph{The dream of applied dynamics}

Now we can formulate the ``dream of applied dynamics." There is
such a finite number of invariant sets $A_1,\ldots A_n$ that:
\begin{itemize}
\item{Any attractor or repeller is one of the $A_i$;}
\item{The following relation between sets $A_1,\ldots
A_n$ is acyclic: $A_i \succeq A_j$ if there exists such $x$ that
$\alpha (x)=A_i$ and $\omega (x)=A_j$;}
\item{The system $A_1,\ldots
A_n$ with the preorder $A_i \succeq A_j$ does not change
qualitatively under sufficiently small perturbations of the
dynamical system: all the picture can be restored by a map that is
close to $\rm id$.}
\end{itemize}
For generic two-dimensional systems this dream is the reality:
there is a finite number of fixed points and closed orbits such
that any motion goes to one of them at $t \to \infty$, and to
another one at $t \to -\infty$ for a whole motion.

The multidimensional analogues of generic two-dimensional systems
are the Morse--Smale systems. For them all attractors and
repellers are fixed points or closed orbits. The relation $A_i
\succeq A_j$ for them is the {\it Smale order}. But the class of
the Morse--Smale systems is too narrow: there are many systems
with more complicated attractors, and some of these  systems are
structurally stable and do not change qualitatively after
sufficiently small perturbations.\footnote{Review of  modern
dynamics is presented in \cite{Katok,Katok2}} It is necessary to
take into account that typically some of motions have smaller
attractors (for example, in $A_i$ exists a dense set of closed
orbits), and $\omega (x)=A_j$ not for all, but for almost all $x$.
Finally, the ``dream of applied dynamics" was destroyed by S.
Smale \cite{Smeil}. He demonstrated that ``structurally stable
systems are not dense." It means that even the last item of this
dream contradicts the multidimensional reality.

\subsection{Metric coarse-graining by $\varepsilon$-motions}

\paragraph{$\varepsilon$-motions}

The observable picture must be structurally stable. Any real
system exists under the permanent perturbing influence of the
external world. It is hardly possible to construct a model taking
into account all such perturbations. Besides  that, the model
describes  the  internal properties of the system  only
approximately. The discrepancy between the real system and the
model arising from these two circumstances is different for
different models. So, for the systems of  celestial mechanics it
can be done  very  small.  Quite the contrary,  for chemical
engineering this discrepancy can be if not too large but not such
small to be neglected. Structurally unstable features or phase
portrait should be destroyed by  such an unpredictable divergence
of the model and reality. The perturbations ``conceal" some fine
details of dynamics, therefore these details become irrelevant to
analysis of real systems.

There are two traditional approaches to the  consideration of
perturbed motions. One of them is to investigate the motion in the
presence of small sustained  perturbations \cite{[41],[42],[45]},
the other is the study of fluctuations under the influence of
small stochastic perturbations \cite{[52],LArn}. In this section,
we join mainly the first direction.

A small unpredictable discrepancy between the real system and the
dynamical model can be simulated by periodical ``fattening." For a
set $A \subset X$ its $\varepsilon$-fattening is the set
\begin{equation}\label{fatt}
A_{\varepsilon}=\{x \ | \ \rho(x,y)<\varepsilon \; \mbox{for all} \;
y \in A \}.
\end{equation}
Instead of one $x$-motion we consider motion of a set,
$A(t)=\Theta_t A$, and combine this motion with periodical
$\varepsilon$-fattening for a given period $\tau$. For
superposition of $\Theta_{\tau}$ with $\varepsilon$-fattening we
use the notation ${\bf \Theta}_{\tau}^{\varepsilon}$:
\begin{equation}
{\bf \Theta}_{\tau}^{\varepsilon} A= (\Theta_{\tau} A)_{\varepsilon}
\end{equation}
For $t \in [n \tau, (n+1) \tau[$ We need to generalize this
definition for $t \in [n \tau, (n+1) \tau[$:
\begin{equation}\label{SetMotion}
{\bf \Theta}_{t}^{\varepsilon} A=\Theta_{t-n\tau}(({\bf
\Theta}_{\tau}^{\varepsilon})^n A).
\end{equation}
Analysis of these motions of sets gives us the information about
dynamics with $\varepsilon$-uncertainty in model. Single-point
sets are natural initial conditions for such motions.

One can call this coarse-graining the {\it metric
coarse-graining}, and the Erenfest's coarse-graining for dynamics
of distribution function might be called the {\it measure
coarse-graining}. The concept of {\it metric--measure spaces}
({\it mm}-spaces \cite{Gromov99}) gives the natural framework for
analysis of various sorts of coarse-graining.

\begin{figure}[t]
\begin{centering}
a)\includegraphics[width=45mm, height=35mm]{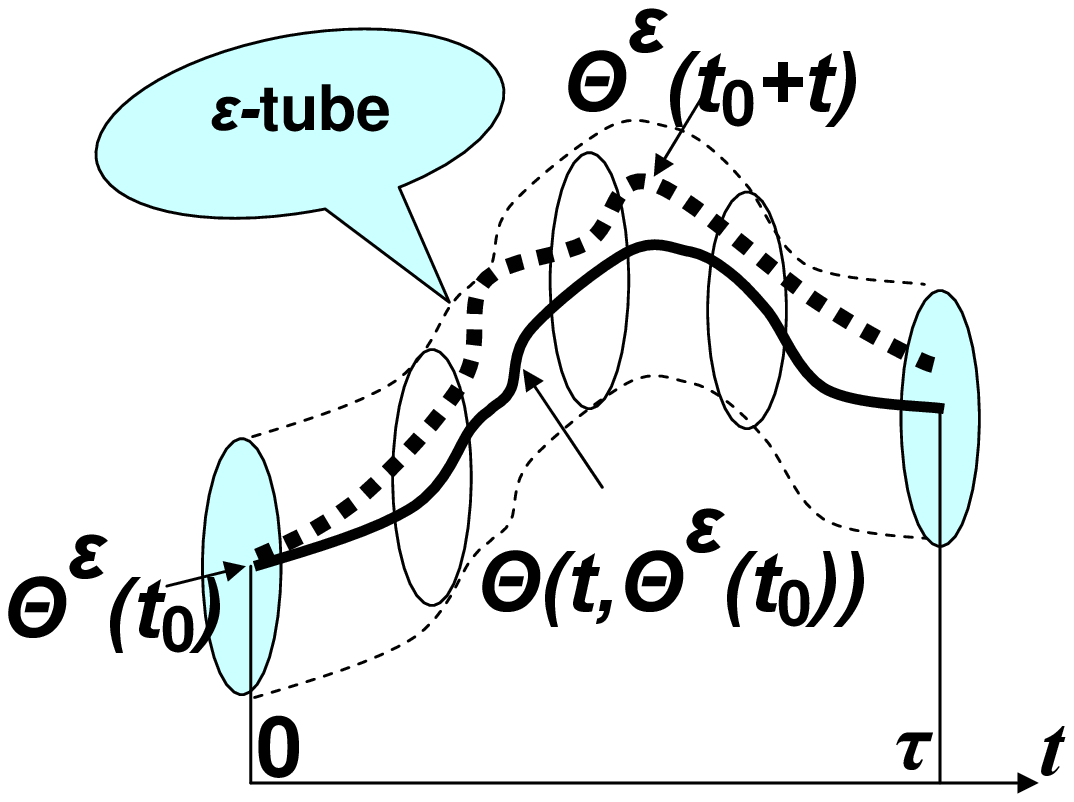}
b)\includegraphics[width=65mm, height=45mm]{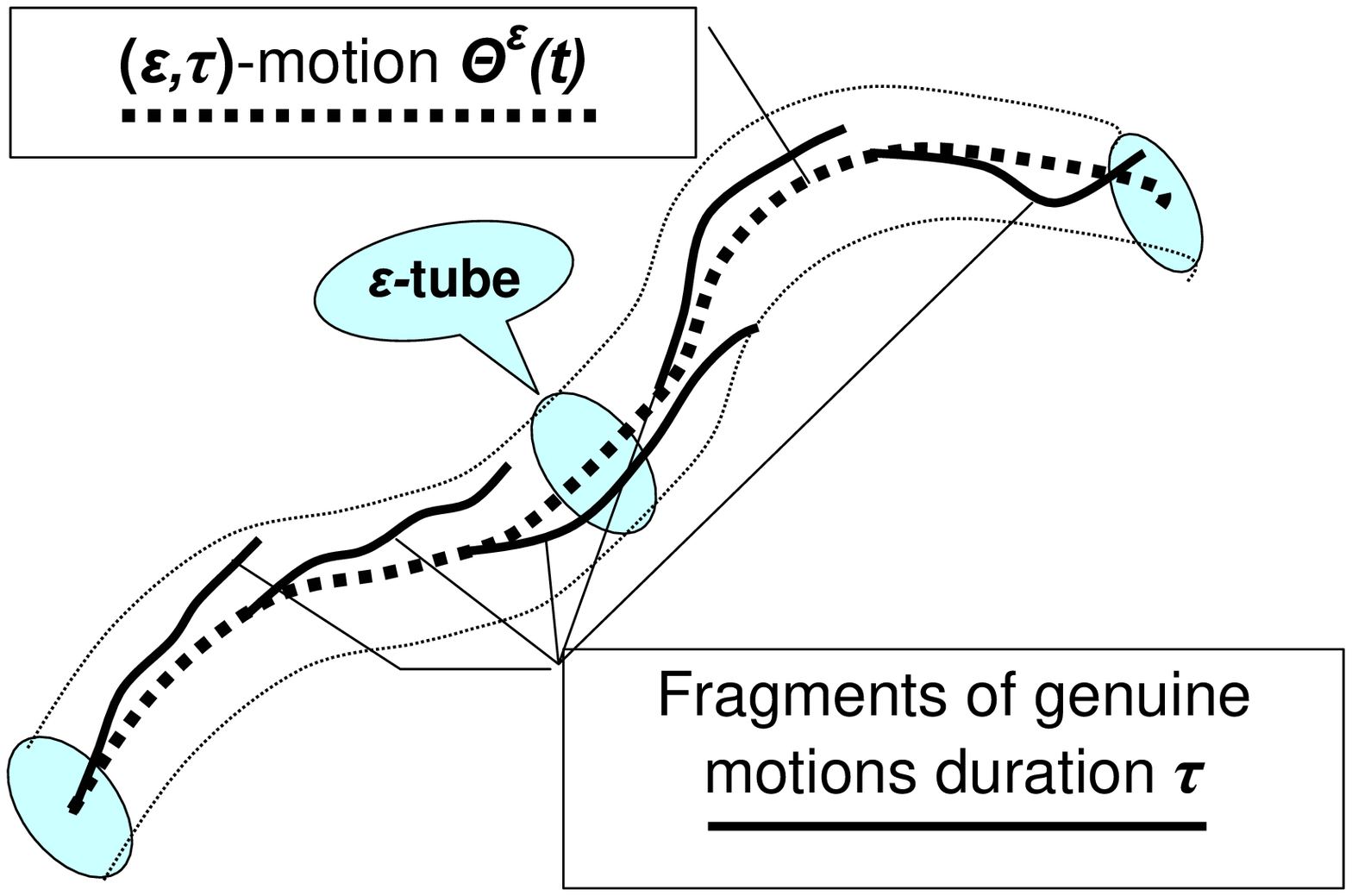}
\caption{\label{TubeT}An $(\varepsilon,\tau)$-motion
$\Theta^{\varepsilon}(t_0+t)$ ($t \in [0,\tau]$) in the
$\varepsilon$-tube near a genuine motion $\Theta(t,
\Theta^{\varepsilon}(t_0))$ ($t \in  [0,\tau]$) duration $\tau$
(a), and an $(\varepsilon,\tau)$-motion $\Theta^{\varepsilon}(t)$
with fragments of genuine motions duration $\tau$ in the
$\varepsilon$-tube near $\Theta^{\varepsilon}(t)$ (b).}
\end{centering}
\end{figure}

It is convenient to introduce individual $\varepsilon$-motions. A
function  of  time $\Theta^{\varepsilon} (t)$ with values in $X$,
defined at $t \geq 0$, is  called $(\varepsilon,x)$-motion
$(\varepsilon > 0)$, if $\Theta^{\varepsilon} (0)=x$ and for any
$t_0 \geq 0$, $t \in [0, \tau]$ the inequality
$\rho(\Theta^{\varepsilon} (t_0+t),\Theta(t,
\Theta^{\varepsilon}(t_0))) < \varepsilon$ holds. In other words,
if for an arbitrary point $\Theta^{\varepsilon} (t_0)$ one
considers its motion due to phase flow of dynamical system, this
motion will diverge $\Theta^{\varepsilon}(t_0+t)$ from no more
than at $\varepsilon$ for $t \in [0, \tau]$. Here $[0,\tau]$ is a
certain interval of time, its length $\tau$ is not very important
(it is important that it  is fixed), because later we shall
consider the case $\varepsilon \rightarrow 0$. For a given $\tau$
we shall call the $(\varepsilon,x)$-motion
$(\varepsilon,x,\tau)$-motion when reference to $\tau$ is
necessary. On any interval $[t_0,t_0+\tau]$ an
$(\varepsilon,x,\tau)$-motion deviates from a genuine motion not
further than on distance $\varepsilon$ if these motions coincide
at time moment $t_0$ (Fig.~\ref{TubeT}a). If a genuine motion
starts from a point of an $(\varepsilon,x,\tau)$-trajectory, it
remains in the $\varepsilon$-tube near that
$(\varepsilon,\tau)$-motion during time $\tau$
(Fig.~\ref{TubeT}b).

\paragraph{Limit sets of $\varepsilon$-motions}

Let us study the limit behaviour of the coarse-grained
trajectories ${\bf \Theta}_{t}^{\varepsilon} A$, and than take the
limit $\varepsilon \to 0$. For systems with complicated dynamics,
this limit may differ significantly from the limit behaviour of
the original system for $\varepsilon=0$. This effect of the
perturbation influence in the zero limit is a ``smile of a
Cheshire cat:" the cat tends to disappear, leaving only its smile
hanging in the air.

For any $\Theta^{\varepsilon} (t)$ the $\omega$-limit set
$\omega(\Theta^{\varepsilon})$ is the set of all limit points of
$\Theta^{\varepsilon} (t)$ at $t \to \infty$. For any $x \in X$ a
set $\omega^{\varepsilon} (x)$ is a totality of all $\omega$-limit
points of all $(\varepsilon,x)$-motions: $$\omega^{\varepsilon}
(x) = \bigcup_{\Theta^{\varepsilon} (0)=x}
\omega(\Theta^{\varepsilon}).$$

For $\varepsilon \to 0$ we obtain the set $$ \omega^0 (x)=
\bigcap_{\varepsilon>0} \omega^{\varepsilon} (x). $$ Firstly, it
is necessary to notice that $\omega^{\varepsilon}(x)$ does not
always tend to $\omega(x)$ as $\varepsilon \to 0$: the set
$\omega^0(x)$ may not coincide with $\omega(x,k)$.

The sets $ \omega^0 (x)$ are closed and invariant. Let $x \in
\omega^0 (x)$. Then for any $\varepsilon > 0$ there exists
periodical $(\varepsilon,x)$-motion (This is a version of Anosov's
$C^0$-closing lemma \cite{[53],Katok2}).

The function $ \omega^0 (x)$ is {\it upper semicontinuous}. It
means that for any sequence $x_i \to x$ all limit points of all
sequences $y_i \in \omega^0 (x_i)$ belong to $\omega^0 (x)$.

In order to study the limit behaviour for all initial conditions,
let us join all $ \omega^0 (x)$:
\begin{equation}\label{omega^0}
\omega^0 = \bigcup_{x \in X} \omega^0 (x) = \bigcup_{x \in X}
\bigcap_{\varepsilon>0} \omega^{\varepsilon} (x) =
\bigcap_{\varepsilon>0} \bigcup_{x \in X} \omega^{\varepsilon}
(x).
\end{equation}
The  set $\omega^0$  is  closed  and invariant. If $y \in
\omega^0$ then $y \in \omega^0 (y)$. If $Q \subset \omega^0$ and
$Q$ is  connected, then $Q \subset \omega^0 (y)$ for any $y \in
Q$.\footnote{For all proofs here and below in this section we
address to \cite{Diss,SloRelMono}.}

The $\varepsilon$-motions were studied earlier in differential
dynamics,  in connection with the theory of Anosov about
$\varepsilon$-trajectories and its applications
\cite{[53],[54],[55],[57],sinai}. For systems with hyperbolic
attractors an important {\it $\varepsilon$-motion shadowing
property} was discovered: for a given $\eta >0$ and sufficiently
small $\varepsilon > 0$ for any $\varepsilon$-motion
$\Theta^{\varepsilon}(t)$ there exists a motion of the
non-perturbed system $\Theta(t,x)$  that belongs to $\eta$ -
neighborhood of $\Theta^{\varepsilon}(t)$:
$$\rho(\Theta^{\varepsilon}(\phi(t)), \Theta(t,x)) < \eta,$$ for
$t>0$ and some monotonous transformation of time $\phi(t)$
($t-\phi(t)= O(\varepsilon t)$). The sufficiently small
coarse-graining changes nothing in dynamics of systems with this
shadowing property, because any $\varepsilon$-motion could be
approximated uniformly by genuine motions  on the whole semiaxis
$t \in [0,\infty[$.

\paragraph{Preorder and equivalence generated by dynamics}

Let $x_1, x_2 \in X$. Let us say $x_1 \succsim_{\Theta} x_2$ if
for any $\varepsilon>0$ there exists such a
$(\varepsilon,x_1)$-motion $\Theta^{\varepsilon}(t)$
($\Theta^{\varepsilon}(0)=x_1$) that $\Theta^{\varepsilon}(t_0) =
x_2$ for some $t_0 \geq 0$.

Let $x_1, x_2 \in X$. Say that points $x_1$ and $x_2$    are
$\Theta$-{\it equivalent} (denotation  $x_1 \sim _{\Theta} x_2$),
if $x_1 \succsim_{\Theta} x_2$ and $x_2 \succsim_{\Theta} x_1$.

The  relation $\succsim_{\Theta}$  is  a  closed
$\Theta$-invariant  preorder relation on $X$:
\begin{itemize}
\item{It is reflexive: $x \succsim_{\Theta} x$ for all $x\in X$;}
\item{It is transitive: $x_1 \succsim_{\Theta} x_2$ and $x_2
\succsim_{\Theta} x_3$  implies $x_1 \succsim_{\Theta} x_3$;}
\item{The set of pairs $(x_1, x_2)$, for which $x_1 \sim_{\Theta}
x_2$ is closed in $X$;}
\item{If $x_1 \succsim_{\Theta} x_2$ then
$\Theta(t,x_1) \succsim_{\Theta} \Theta(t,x_2)$ for any $t>0$.}
\end{itemize}
The necessary and sufficient conditions for the preorder
$\succsim_{\Theta}$ relation are as follows: $x_1
\succsim_{\Theta} x_2$ if and only if either $x_2 \in \omega^0
(x_1)$ or $x_2=\Theta(t,x_1)$ for some $t\geq 0$. Therefore,
\begin{equation}\label{omegaorder}
\omega^0 (x)=\{y\in \omega^0 \ | \ x \succsim_{\Theta} y \}
\end{equation}

The relation $\sim_{\Theta}$  is  a  closed  $\Theta$-invariant
equivalence relation:
\begin{itemize}
\item{The set of pairs $(x_1, x_2)$, for which $x_1 \sim_{\Theta}
x_2$ is closed in $X$;} \item{If $x_1 \sim x_2$ and $x_1 \neq
x_2$, then $x_1$-  and   $x_2$-motions  are whole and
$\sim_{\Theta} \Theta(t, x_2)$ for any $t \in ]-\infty, \infty[\
\Theta(t,x_1)$.}
\end{itemize}
If $x_1 \neq x_2$, then $x_1 \sim_{\Theta} x_2$ if and only if
$\omega^0 (x_1) = \omega^0 (x_2)$ , $x_1 \in \omega^0 (x_1)$, and
$x_2 \in \omega^0 (x_2)$.

Compare  with \cite{[52]}, where analogous theorems  are proved
for relations defined by action functional for randomly perturbed
dynamics.

\paragraph{The coarsened phase portrait}

We present the results about the coarsened phase portrait as a
series of theorems.

Let us remind, that topological space is called {\it totally
disconnected} if there exist a base of topology, consisting of
sets which are simultaneously open and closed. Simple examples of
such spaces are discrete space and Cantor's discontinuum. In a
totally disconnected space all subsets with more than one element
are disconnected. Due to the following theorem, in the coarsened
phase portrait we have a totally disconnected space instead of
finite set of attractors mentioned in the naive dream of applied
dynamics.

\begin{theorem}\label{P4.18}
The quotient space $\omega^0 / \sim_{\Theta} $ is  compact and
totally disconnected.
\end{theorem}

The space $\omega^0 / \sim_{\Theta} $ with the factor-relation
$\succsim_{\Theta}$ on it is the {\it generalized Smale diagram}
 with the  {\it generalized Smale order} on it \cite{Diss,SloRelMono}.

Attractors and basins of attraction are the most important parts
of a phase portrait. Because of (\ref{omegaorder}), all attractors
are {\it saturated downwards}. The set $Y \subset \omega^0$ is
saturated downwards, if for any $y \in Y$, $$ \{ x \in \omega^0 \
| \ y \succsim_{\Theta} x \} \subset Y. $$ Every saturated
downwards subset in $\omega^0$ is saturated also for the
equivalence relation $\sim_{\Theta}$ and includes with any $x$ all
equivalent points. The following theorem states that coarsened
attractors $Y$ (open in $\omega^0$ saturated downwards subsets of
$\omega^0$) have open coarsened basins of attraction
$\mathcal{B}^0 (Y)$.

\begin{theorem}\label{P4.21}
Let $Y \subset \omega^0$  be  open  (in $\omega^0$) saturated
downwards set. Then the set $\mathcal{B}^0 (Y) = \{ x \in X \ | \
\omega^0 (x) \subset Y \}$ is open in $X$.
\end{theorem}

There is a natural expectation that $\omega$-limit sets can change
by jumps on boundaries of basins of attraction only. For the
coarsened phase portrait it is true.

\begin{theorem}\label{P4.13}
The set $B$ of all points of discontinuity  of the function
$\omega^0 (x)$ is the subset of first  category in $X$.\footnote{A
set of first category, or a meagre set is  a countable union of
nowhere dense sets. In a complete metric space a complement of a
meagre set is dense (the Baire theorem).} If $x\in B$ then
$\Theta(t,x)\in B$ for all $t$ when $\Theta(t,x)$ is defined.
\end{theorem}

\begin{theorem}\label{QQQ}
Let $x \in X$ be a point  of discontinuity  of the function
$\omega^0 (x)$.  Then there is such open in $\omega^0$ saturated
downwards set $W$ that $x \in \partial \mathcal{B}^0 (W)$.
\end{theorem}

The function $\omega^0 (x)$ is upper semicontinuous, hence, in any
point $x^*$ of its discontinuity the {\it lower semicontinuity} is
broken: there exist a point $y^*\in \omega^0(x^*)$, a number
$\eta>0$, and a sequence $x_i \to x^*$ such that
$$\rho(y^*,y)>\eta \; \mbox{for any} \; y\in \omega^0(x_i) \;
\mbox{and all} \; i.$$

The classical Smale order for hyperbolic systems was defined on a
finite totality $A_1, \ldots A_n$ of basic sets that are closed,
invariant, and transitive (i.e. containing a dense orbit). $A_i
\succ A_j$ if there exists such $x\in X$ that  $x$-trajectory is
whole, $\alpha(x) \subset A_i$, $\omega(x) \subset A_j$. Such
special trajectories exist in the general case of coarsened
dynamical system also.

\begin{theorem}\label{P4.22} Let $X$ be  connected, $\omega^0$   be  disconnected.
Then there is such $x \in X$ that $x$-motion is whole and $x \not
\in \omega^0$.  There  is also such partition of $\omega^0$ that
$$ \omega^0 = W_1 \cup W_2,\ W \cap W_2 = \varnothing, \ \alpha_f
(x) \subset W_1, \ \omega^0 (x) \subset W_2, $$ and $W_{1,2}$ are
open and, at the same time, closed subsets of $\omega^0$ (it means
that $W_{1,2}$ are preimages of open--closed subsets of the
quotient space $\omega^0/\sim_{\Theta} $).
\end{theorem}
This theorem can be applied, by descent, to connected closures of
coarsened basins of attraction $\mathcal{B}^0 (Y)$ (see Theorem
\ref{P4.21}).

Theorems \ref{P4.18}--\ref{P4.22} give us the picture of coarsened
phase portrait of a general dynamical system, and this portrait is
qualitatively close to phase portraits of structurally stable
systems: rough 2D systems, the Morse--Smale systems and the
hyperbolic Smale systems. For proofs and some applications we
address to \cite{Diss,SloRelMono}.

\paragraph{Stability of the coarsened phase portrait under smooth
perturbations of vector fields}

In order to analyze stability of this picture under the
perturbation of the vector field (or the diffeomorphism, for
discrete time dynamics) it is necessary to introduce $C^k$
$\varepsilon$-fattening in the space of smooth vector fields
instead of periodic $\varepsilon$-fattening of phase points. We
shall discuss a $C^k$-smooth dynamical system $\Theta$  on a
compact $C^m$-manifold $M (0\leq k\leq m)$. Let $\Theta_t$ be the
semigroup of phase flow transformations (shifts in time $t\geq 0$)
and $U_{\varepsilon}(\Theta)$ be the set of phase flows that
corresponds to a closed $\varepsilon$-neighborhood of system
$\Theta_t$ in the $C^k$-norm topology of vector fields. The
positive semi-trajectory of phase point $x$ is a set
$\Theta(x)=\{\Theta_t(x)\colon t\geq 0\}$. The $C^k$
$\varepsilon$-fattened semi-trajectory is
$\Theta^{\varepsilon}(x)=\bigcup_{\Phi \in
U_{\varepsilon}(\Theta)}\Phi(x)$. Let us take this set with all
limits for $t\rightarrow \infty$. It is the closure
$\overline{\Theta^{\varepsilon}(x)}$. After that, let us take the
limit $\varepsilon\rightarrow 0$: $P_x=\bigcap_{\varepsilon>0}
\overline{\Theta^{\varepsilon}(x)}$ (it is an analogue of
$\Theta(x)\cup \omega_0(x)$ from our previous consideration for
general dynamical systems). Following \cite{[69]} let us call this
set $P_x$ a {\it prolongation} of the semi-trajectory $\Theta(x)$.

A trajectory of a dynamical system is said to be {\it stable
under} $C^k$ {\it constantly-acting perturbations} if its
prolongation is equal to its closure: $P_x=\overline{\Theta(x)}$

For a given dynamical system let $L(\Theta)$ denote the union of
all trajectories that are stable in the above sense and let
$\mathbb{L}_1$ be the set of all dynamical systems $\Theta$ for
which $L(\Theta)$ is dense in phase space:
$\overline{L(\Theta)}=M$. All structurally stable systems belong
to $\mathbb{L}_1$. The main result of \cite{[69]} is as follows:

\begin{theorem} The set $\mathbb{L}_1$ is a dense $\mathbb{G_\delta}$ in the
space of $C^k$ dynamical systems with the $C^k$ norm.\footnote{In
a topological space a $\mathbb{G_\delta}$ set is a countable
intersection of open sets. A complement of a dense
$\mathbb{G_\delta}$ set is a countable union of nowhere dense
sets. It is a set of first category, or a meagre set. }
\end{theorem}

So, for almost all smooth dynamical systems almost all
trajectories are stable under smooth constantly-acting
perturbations: this type of stability is typical.

\section{Conclusion}

Two basic ideas of coarse-graining are presented. In the Ehrenfests'
inspired approach the dynamics of distributions with averaging is
studied. In the metric approach the starting point of analysis is
dynamics of sets with periodical $\varepsilon$-fattening.

The main question of the Ehrenfests' coarse-graining is: where
should we take the coarse-graining time $\tau$? There are two limit
cases: $\tau \to 0$ and $\tau \to \infty$ (physically, $\infty$ here
means the time that exceeds all microscopic time scales). The  first
limit, $\tau \to 0$, returns us to the quasi-equilibrium
approximation. The second limit is, in some sense, exact (if it
exists). Some preliminary steps in the study of this limit are made
in \cite{GKGeoNeo,Plenka,GorKar}. On this way, the question about
proper values of the Prandtl number, as well, as many other similar
questions about kinetic coefficients, has to be solved.

The constructed family of chains between conservative (with the
Karlin--Succi involution) and maximally dissipative (with
Ehrenfests' projection) ones give us a possibility to model
hydrodynamic systems with various dissipation (viscosity)
coefficients that are decoupled with time steps. The {\it collision
integral} is successfully substituted by combinations of the
involution and projection.

The direct descendant of the Ehrenfests' coarse-graining, the
kinetic approach to filtering of continuum equations, seems to be
promising and physically reasonable: if we need to include the
small eddies energy into internal energy, let us lift the
continuum mechanics to kinetics where all the energies live
together, make there the necessary filtering, and then come back.
Two main questions: when the obtained filtered continuum mechanics
is stable, and when there is way back from filtered kinetics to
continuum mechanics, have unexpectedly the similar answer: the
filter width $\Delta$ should be proportional to the square root of
the Knudsen number. The coefficient of this proportionality is
calculated from the entropic stability conditions.

The metric coarse-graining by $\varepsilon$-motions in the limit
$\varepsilon \to 0$ gives the stable picture with the totally
disconnected system of basic sets that form sources and sinks
structure in the phase space. Everything looks nice, but now we need
algorithms for effective computation and representation of the
coarsened phase portrait even in modest dimensions 3-5 (for discrete
time systems in dimensions 2-4).

It is necessary to build a bridge between theoretical topological
picture and applied computations. In some sense, it is the main
problem of modern theory of dynamical systems to develop language
and tools for constructive analysis of arbitrary dynamics. Of
course, the pure topological point of view is unsufficient, and we
need an interplay between measure and topology of dynamical systems,
perhaps, with inclusion of some physical and probabilistic ideas.

\acknowledgement A couple of years ago, Wm. Hoover asked me to
explain clearly the difference between various types of
coarse-graining. This paper is the first attempt to answer. I am
grateful to H.C. \"Ottinger, and L. Tatarinova for scientific
discussion and collaboration. Long joint work with I.V. Karlin was
very important for my understanding of the Ehrenfests'
coarse-graining and lattice Boltzmann models.


\begin{thebibliography}{999}
\bibitem{Ehrenfest}P. Ehrenfest, T.  Ehrenfest-Afanasyeva,
The Conceptual Foundations of the Statistical Approach in
Mechanics, In: {\it Mechanics Enziklop\"adie der Mathematischen
Wissenschaften}, Vol. 4., Leipzig, 1911. (Reprinted: P. Ehrenfest,
T.  Ehrenfest-Afanasyeva, The Conceptual Foundations of the
Statistical Approach in Mechanics, Dover Phoneix, 2002.)

\bibitem{Grabert}H. Grabert,  Projection operator techniques in nonequilibrium statistical
mechanics, Springer Verlag, Berlin, 1982.

\bibitem{CMIM}A.N. Gorban, I.V. Karlin, A.Yu. Zinovyev,
Constructive methods of invariant manifolds for kinetic problems,
{\it Phys. Reports} { 396}, 4-6 (2004), 197--403. Preprint online:
http://arxiv.org/abs/cond-mat/0311017.

\bibitem{GorKar}A.N. Gorban, I.V. Karlin, Invariant manifolds for
physical and chemical kinetics, {\it Lect. Notes Phys.}, Vol. 660,
Springer, Berlin, Heidelberg, New York, 2005.

\bibitem{GKIOeNONNEWT2001}A.N. Gorban, I.V. Karlin, P. Ilg,  and H.C. \"{O}ttinger,
Corrections and enhancements of quasi--equilibrium states, {\it J.
Non--Newtonian Fluid Mech.} { 96} (2001), 203--219.

\bibitem{Wilson}K.G. Wilson, and J. Kogut. The renormalization group and the
$\epsilon$-expansion, {\it  Phys. Reports} 12C (1974), 75-200.

\bibitem{RG}O. Pashko, Y.  Oono, The Boltzmann equation is a renormalization
group equation,  {\it Int. J. Mod. Phys. B} { 14} (2000),
555--561.

\bibitem{Kun3}Y. Hatta, T. Kunihiro, Renormalization group method applied to kinetic equations:
roles of initial values and time, {\it Annals Phys.} { 298} (2002),
24--57.

\bibitem{KevFree}I.G. Kevrekidis, C.W. Gear, J.M. Hyman,
P.G. Kevrekidis,  O. Runborg, C. Theodoropoulos,  Equation-free,
coarse-grained multiscale computation: enabling microscopic
simulators to perform system-level analysis, {\it Comm. Math.
Sci.} { 1} 4 (2003), 715--762.

\bibitem{Raz}A.J. Chorin, O.H. Hald, R. Kupferman,  Optimal
prediction with memory,  {\it Physica D} 166 (2002), 239--257.

\bibitem{GKOeTPRE2001}A.N. Gorban, I.V. Karlin, H.C. \"{O}ttinger,  and L.L. Tatarinova,
 Ehrenfests' argument extended to a formalism of nonequilibrium
thermodynamics, {\it Phys.Rev.E} { 63} (2001), 066124.

\bibitem{UNIMOLD}A.N. Gorban, I.V. Karlin, Uniqueness of thermodynamic projector and kinetic
basis of molecular individualism, {\it Physica A} { 336}, 3-4
(2004), 391--432.

\bibitem{Leray}J. Leray, Sur les movements d'un fluide visqueux remplaissant
l'espace, {\it Acta Mathematica}, 63 (1934), 193–-248.

\bibitem{Smag}J. Smagorinsky, General Circulation Experiments with the Primitive Equations:
I. The Basic Equations, {\it Mon. Weather Rev.} { 91} (1963),
99--164.

\bibitem{Germano}M. Germano, Turbulence: the filtering approach, {\it J. Fluid
Mech.}
238 (1992), 325--336.

\bibitem{Carati}D. Carati, G.S. Winckelmans, H. Jeanmart, On the modelling of the
subgrid-scale and filtered-scale stress tensors in large-eddy
simulation, {\it J. Fluid Mech.}, 441 (2001), 119--138.

\bibitem{LES2005}M. Lesieur,  O. M\'etais, P. Comte, Large-Eddy Simulations of
Turbulence, Cambridge University Press, 2005.

\bibitem{AnsKarlFiltr}S. Ansumali, I.V. Karlin, S. Succi, Kinetic Theory
of Turbulence Modeling: Smallness Parameter, Scaling and Microscopic
Derivation of Smagorinsky Model, Physica A, 338 (2004), 379--394.



\bibitem{LB2}S. Succi, The lattice Boltzmann equation
for fluid dynamics and beyond, Clarendon Press, Oxford, 2001.


\bibitem{Smeil}S. Smale, Structurally stable systems are not dense,
{\it Amer.  J. Math. }  { 88} (1966), 491--496.

\bibitem{[69]}V.A. Dobrynski\u\i, A.N. Sharkovski\u\i, Genericity of the
dynamical systems almost all orbits of which are stable under
sustained perturbations, {\it Soviet Math. Dokl.} 14 (1973),
997--1000.

\bibitem{Diss}A.N. Gorban,  Slow relaxations and bifurcations of
omega-limit sets of dynamical systems, {\it PhD Thesis in Physics
\& Math.} (Differential Equations \& Math.Phys), Kuibyshev,
Russia, 1980.

\bibitem{SloRelMono}A.N. Gorban, Singularities of Transition
Processes in Dynamical Systems: Qualitative Theory of Critical
Delays, {\it Electronic Journal of Differential Equations},
Monograph 05, 2004.
http://ejde.math.txstate.edu/Monographs/05/abstr.html (Includes
English translation of \cite{Diss}.)

\bibitem{Hub}B.A. Huberman, W.F. Wolff, Finite precision and transient behavior,
{\it Phys. Rev. A} 32, 6 (1985), 3768--3770.

\bibitem{Beck}C. Beck, G. Roepsorff, Effects of phase space discretization on
the long--time behavior of dynamical systems, {\it Physica D} 25
(1987), 173--180.

\bibitem{GrebYor}C. Grebogi, E. Ott, J.A. Yorke, Roundoff--induced periodicity and
the correlation dimension of chaotic attractors, {\it Phys. Rev.
A} 38, 7  (1988), 3688--3692.

\bibitem{Diamond}P. Diamond, P. Kloeden, A. Pokrovskii, A. Vladimirov, Collapsing
effect in numerical simulation of a class of chaotic dynamical
systems and random mappings with a single attracting centre, {\it
Physica D} 86 (1995), 559--571.

\bibitem{Longa}L. Longa, E.M.F. Curado, A. Oliveira, Rounoff--induced coalescence
of chaotic trajectories, {\it Phys. Rev. E} 54, 3 (1996),
R2201--R2204.

\bibitem{Binder}P.-M. Binder, J.C. Idrobo, Invertibility of dynamical systems in
granular phase space, {\it Phys. Rev. E} 58, 6 (1998), 7987--7989.

\bibitem{Hoower}C. Dellago, Wm. G. Hoover, Finite--precision stationary states at
and away from equilibrium, {\it Phys. Rev. E} 62, 5 (2000),
6275--6281.

\bibitem{BBoll}B. Bollobas, Random Graphs (Cambridge Studies in Advanced
Mathematics), Cambridge University Press, 2001.

\bibitem{[52]}M.I. Freidlin, A.D. Wentzell, Random Perturbations of Dynamical
Systems (Grundlehren der mathematischen Wissenschaften, V.260),
Springer-Verlag New York, Berlin, Heidelberg, 1998.


\bibitem{LArn}L. Arnold, Random Dynamical Systems, Springer Monographs in
Mathematics, 16, Springer, Berlin, Heidelberg, New York, 2002.


\bibitem{G1}A.N. Gorban, Equilibrium encircling. Equations of chemical kinetics and their
thermodynamic analysis, Nauka, Novosibirsk, 1984.

\bibitem{Ocherki}A.N. Gorban, V.I. Bykov, G.S. Yablonskii,
Essays on chemical relaxation,  Nauka, Novosibirsk, 1986.

\bibitem{YBGE}G.S.Yablonskii, V.I.Bykov, A.N. Gorban, and V.I.Elokhin,  Kinetic
Models of Catalytic Reactions (Series ``Comprehensive Chemical
Kinetics," V.32, ed. by R.G. Compton), Elsevier, Amsterdam, 1991,

\bibitem{Zeld}Y.B. Zeldovich, Proof of the Uniqueness of the Solution of the
Equations of the Law of Mass Action, In: {\it Selected Works of
Yakov Borisovich Zeldovich,} Vol. 1, J.P. Ostriker (Ed.),
Princeton University Press, Princeton, USA, 1996, 144-148.

\bibitem{BGK}Bhatnagar, P.L., Gross, E.P., Krook, M., A model for collision processes in
gases. I. Small amplitude processes in charged and neutral
one--component systems, {\it Phys. Rev.}, {\bf 94}, 3 (1954),
511--525.


\bibitem{GKMod}Gorban, A.N., Karlin, I.V.,  General approach to
constructing models of the Boltzmann equation, {\it Physica A}, {
206} (1994), 401--420.


\bibitem{Hilbert}D. Hilbert, Begr\"undung der kinetischen Gastheorie, {\it Math.
Annalen}
{ 72} (1912), 562--577.

\bibitem{Ruelle}D. Ruelle, Smooth Dynamics and New Theoretical Ideas in Nonequilibrium Statistical Mechanics,
{\it J. Stat. Phys.} { 95} (1-2) (1999), 393-468.

\bibitem{Kull}S. Kullback, Information theory and statistics, Wiley, New York, 1959.

\bibitem{ENTR1}A.N. Gorban, I.V.  Karlin,  Family of additive entropy functions out of
thermodynamic limit, {\it Phys. Rev. E} { 67} (2003), 016104.


\bibitem{ENTR3}P. Gorban, Monotonically equivalent entropies and solution of additivity equation,
{\it Physica A} { 328}  (2003), 380-390.

\bibitem{Abe}S. Abe,  Y. Okamoto (Eds.), Nonextensive statistical mechanics and its
applications, Springer, Heidelberg, 2001.

\bibitem{Boghos}B.M. Boghosian, P.J. Love, P.V. Coveney, I.V. Karlin, S. Succi, J.
Yepez, Galilean-invariant lattice-Boltzmann models with H theorem,
{\it Phys. Rev. E} 68,  2 (2003), 025103(R)

\bibitem{Gibb}G.W. Gibbs, Elementary Principles of Statistical Mechanics, Dover, 1960.

\bibitem{Janes1}E.T. Jaynes, Information theory and statistical mechanics, in:
 Statistical Physics. Brandeis Lectures, V.3, K. W. Ford, ed., New York: Benjamin, 1963, pp. 160--185.

\bibitem{Zubarev}D. Zubarev, V. Morozov, G. R\"opke, Statistical mechanics of
nonequilibrium processes, Akademie Verlag, Berlin,  V.1, 1996,
V.2, 1997.

\bibitem{Grad}H. Grad, On the kinetic theory of rarefied gases, {\it Comm. Pure and Appl.
Math.} { 2} 4, (1949), 331--407.

\bibitem{Garsia1}J.T. Alvarez-Romero, L.S. Garc\'{\i}a-Col\'{\i}n,  The foundations of informational statistical
thermodynamics revisited, {\it Physica A} { 232}, 1-2 (1996),
207--228

\bibitem{Nett}R.E. Nettleton, E.S. Freidkin,  Nonlinear reciprocity and the maximum entropy formalism, {\it Physica
A} { 158}, 2 (1989), 672--690.

\bibitem{Orlov84}N.N. Orlov, L.I. Rozonoer,
The macrodynamics of open systems and the variational principle of
the local potential, {\it J. Franklin Inst.}  { 318} (1984),
283--314 and 315--347.

\bibitem{KoRoz}A.M. Kogan, L.I. Rozonoer,  On the macroscopic description of kinetic processes, {\it Dokl.
AN SSSR} { 158}, 3 (1964), 566--569.

\bibitem{Ko}A.M. Kogan, Derivation of Grad--type equations and study of their properties by the method of
entropy maximization, {\it Prikl. Matem. Mech.} { 29} (1) (1965),
122--133.

\bibitem{Roz}L.I. Rozonoer, Thermodynamics of nonequilibrium processes far from equilibrium, in:
Thermodynamics and Kinetics of Biological Processes (Nauka,
Moscow, 1980), 169--186.

\bibitem{Kark}J. Karkheck, G.  Stell,  Maximization of entropy, kinetic equations, and irreversible thermodynamics
{\it Phys. Rev. A} { 25}, 6 (1984), 3302--3327.

\bibitem{BGKTMF}N.N. Bugaenko, A.N. Gorban, I.V. Karlin, Universal Expansion of
the Triplet Distribution Function, {\it Teoret. i Matem. Fisika} {
88}, 3 (1991), 430--441 (Transl.: Theoret. Math. Phys. (1992),
977--985).

\bibitem{MBCh}A.N. Gorban, I.V. Karlin, Quasi--equilibrium
approximation and non--standard expansions in the theory of the
Boltzmann kinetic equation, in: "Mathematical Modelling in Biology
and Chemistry. New Approaches", ed. R. G. Khlebopros, Nauka,
Novosibirsk, P. 69--117 (1991).[in Russian]

\bibitem{MBChLANL}A.N. Gorban, I.V. Karlin,
Quasi--equilibrium closure hierarchies for the Boltzmann equation,
{\it Physica A} 360, 2 (2006), 325--364. (Includes translation of
the first part of \cite{MBCh}).

\bibitem{Lever}C.D. Levermore, Moment Closure Hierarchies for Kinetic Theories,
{\it J. Stat. Phys.} { 83} (1996), 1021--1065.

\bibitem{Bal}R. Balian, Y. Alhassid, H. Reinhardt,  Dissipation in many--body systems:
A geometric approach based on information theory,  {\it Phys.
Reports} { 131}, 1 (1986), 1--146.

\bibitem{Degon}P. Degond, C. Ringhofer, Quantum moment hydrodynamics and the entropy principle, {\it J.
Stat. Phys.} { 112} (2003), 587--627.

\bibitem{IKOePhA02}P. Ilg, I.V. Karlin, H.C. {\"O}ttinger,  Canonical distribution functions in polymer dynamics: I.
Dilute solutions of flexible polymers, {\it Physica A} { 315}
(2002), 367--385.

\bibitem{IKOePhA03}P. Ilg, I.V. Karlin, M. Kr\"oger, H.C. {\"O}ttinger, Canonical distribution functions in polymer
dynamics: II Liquid--crystalline polymers, {\it Physica A} { 319}
(2003), 134--150.

\bibitem{IlKr}P. Ilg, M. Kr\"oger,  Magnetization dynamics, rheology, and an effective description of
ferromagnetic units in dilute suspension, {\it Phys. Rev. E} { 66}
(2002) 021501. Erratum, {\it Phys. Rev. E} { 67} (2003),
049901(E).

\bibitem{IKar2}P. Ilg, I.V. Karlin, Combined micro--macro integration scheme from an invariance principle:
application to ferrofluid dynamics, {\it J. Non--Newtonian Fluid
Mech.} 120, 1-3 (2004), 33--40 2004. Ppeprint online:
http://arxiv.org/abs/cond-mat/0401383.

\bibitem{Robertson}B. Robertson, Equations of motion in nonequilibrium statistical mechanics, {\it Phys.
Rev.}
{ 144} (1966), 151--161.

\bibitem{Morrison}P.J. Morrison, Hamiltonian description of the ideal fluid, {\it Rev.
Mod. Phys.} 70 (1998), 467–-521.


\bibitem{WIG}E. Wigner, On the quantum correction for thermodynamic equilibrium,
{\it Phys. Rev.} { 40} (1932), 749--759.

\bibitem{CAL}A.O. Caldeira, A.J. Leggett, Influence of damping on quantum interference: An exactly soluble
model, {\it Phys. Rev. A} { 31} (1985), 1059--1066.

\bibitem{GKMex2001}A.N. Gorban, I.V. Karlin, Reconstruction lemma and
fluctuation--dissipation theorem, {\it Revista Mexicana de Fisica}
{ 48}, Supl. 1 (2002), 238--242.

\bibitem{GKPRE02}A.N. Gorban, I.V. Karlin, Macroscopic dynamics through
coarse--graining: A solvable example, {\it Phys. Rev. E} { 56 }
(2002), 026116.

\bibitem{GKGeoNeo}A.N. Gorban, I.V. Karlin, Geometry of irreversibility, in: Recent
Developments in Mathematical and Experimental Physics, Vol. C, Ed.
F. Uribe, Kluwer, Dordrecht, 2002, 19--43.

\bibitem{Plenka}A.N. Gorban, I.V. Karlin, Geometry of irreversibility: The film of nonequilibrium
states, Preprint IHES/P/03/57, Institut des Hautes \'Etudes
Scientifiques in Bures-sur-Yvette (France), 2003. 69 p. Preprint
on-line: http://arXiv.org/abs/cond-mat/0308331.


\bibitem{KTGOePhA2003}I.V. Karlin, L.L. Tatarinova, A.N. Gorban, H.C. \"{O}ttinger, Irreversibility
in the short memory approximation, {\it Physica A} { 327}, 3-4
(2003), 399--424.

\bibitem{LebBloEnt}J.L. Lebowitz, Statistical Mechanics: A Selective Review of Two
Central Issues, {\it Rev. Mod. Phys.} 71 (1999), S346.

\bibitem{Bentr}S. Goldstein, J.L. Lebowitz, On the (Boltzmann) Entropy of
Nonequilibrium Systems, {\it Physica D} 193 (2004), 53--66.


\bibitem{Chapman}Chapman, S., Cowling, T., Mathematical theory of non-uniform gases, Third
edition, Cambridge University Press, Cambridge, 1970.


\bibitem{Cercignani}C. Cercignani,  The Boltzmann equation and its applications, Springer,
New York, 1988.

\bibitem{Struch}L. Mieussens, H. Struchtrup, Numerical Comparison
of Bhatnagar--Gross--Krook models with proper Prandtl number, {\it
Phys. Fluids} { 16}, 4 (2004), 2797--2813.

\bibitem{Lew}R.M. Lewis, A unifying principle in statistical mechanics,
 {\it J. Math. Phys.} { 8} (1967), 1448--1460.

\bibitem{Lya}Lyapunov A.M., The general problem of the stability of motion, Taylor \& Francis,
London, 1992.

\bibitem{Shnol}L.B. Ryashko, E.E. Shnol, On exponentially attracting invariant manifolds of ODEs,
{\it Nonlinearity} { 16} (2003), 147--160.

\bibitem{Kev}C. Foias, M.S. Jolly, I.G. Kevrekidis, G.R. Sell, E.S. Titi,
On the computation of inertial manifolds, {\it Phys. Lett. A} {
131}, 7--8 (1988), 433--436.

\bibitem{Sone}Y. Sone, Kinetic theory and fluid dynamics, Birkh\"auser, Boston, 2002.

\bibitem{GearKap}C.W. Gear, T.J. Kaper, I.G. Kevrekidis, A. Zagaris, Projecting
to a slow manifold: singularly perturbed systems and legacy codes,
{\it SIAM J. Appl. Dynamical Systems} { 4}, 3 (2005), 711--732.

\bibitem{Succi89}F. Higuera, S.  Succi, R. Benzi, Lattice  gas-dynamics with enhanced
collisions, {\it Europhys. Lett.} 9 (1989), 345--349.


\bibitem{ELB1}I.V. Karlin,   A.N.  Gorban, S. Succi, V. Boffi, V. Maximum
entropy principle for lattice kinetic equations, {\it Phys. Rev.
Lett.} 81 (1998), 6--9.

\bibitem{LBGK1}H. Chen, S. Chen, W. Matthaeus, Recovery of the Navier--Stokes
equation using a lattice--gas Boltzmann Method {\it Phys. Rev. A}
45 (1992), R5339--R5342.


\bibitem{LBGK2}Y.H. Qian, D. d'Humieres, P. Lallemand,
Lattice BGK models for Navier--Stokes equation, {\it Europhys.
Lett.} 17, (1992), 479--484.

\bibitem{LB3}S. Succi, I.V.  Karlin, H. Chen, Role of the $H$ theorem in lattice Boltzmann hydrodynamic
simulations, {\it Rev. Mod. Phys.} { 74} (2002), 1203--1220.

\bibitem{LBperfect}I.V. Karlin, A. Ferrante, H.C. \"Ottinger, Perfect entropy
functions of the Lattice Boltzmann method, {\it Europhys. Lett.} {
47}, 182--188 (1999).

\bibitem{LBentr}S. Ansumali, I.V. Karlin, Entropy function approach to the lattice
Boltzmann method, {\it J. Stat. Phys.} { 107} 291--308 (2002).

\bibitem{cubature}R. Cools, An encyclopaedia of cubature formulas, J. of
Complexity, Volume 19, Issue 3, June 2003, Pages 445-453.

\bibitem{Kagan}A. Gorban, B. Kaganovich, S. Filippov, A. Keiko, V. Shamansky, I.
Shirkalin, Thermodynamic Equilibria and Extrema: Analysis of
Attainability Regions and Partial Equilibrium, Springer, Berlin,
Heidelberg, New York,   2006 (in press).

\bibitem{AK4}Ansumali, S., Karlin, I.V., Kinetic Boundary condition for  the lattice Boltzmann
method, {\it Phys. Rev. E}, {\bf 66} (2002), 026311.

\bibitem{Higgins1}J.R. Higgins, Sampling Theory in Fourier and Signal Analysis:
Foundations, Clarendon, Oxford, 1996.

\bibitem{Higgins2}J.R. Higgins, R.L. Stens, Sampling Theory in Fourier and Signal
Analysis: Advanced Topics,  Clarendon, Oxford,  1999.

\bibitem{Germano2}J.G.M. Kuerten, B.J. Geurts, A.W. Vreman, M. Germano. Dynamic
inverse modeling and its testing in large-eddy simulations of the
mixing layer. {\it Phys. Fluids}, 11: 3778-3785, 1999.

\bibitem{Vreman}A.W. Vreman, The adjoint filter operator in large-eddy simulation of
turbulent flow, {\it Phys. Fluids}, 16 (2004), 2012--2022


\bibitem{Holm}B.J. Geurts, D.D. Holm, Nonlinear regularization for large-eddy
simulation, {\it Phys. Fluids}, 15, 1 (2003), L13--L16.

\bibitem{toyLeonard}P. Moeleker, A. Leonard, Lagrangian methods for the
tensor-diffusivity subgrid model, {\it J. Comp. Phys.}, 167 (2001),
1--21.

\bibitem{POD}G. Berkooz, P. Holmes, J.L. Lumley, The proper orthogonal
decomposition in the analysis of turbulent flows, {\it Annual Rev.
Fluid Mech.},  25 (1993), 539--575.

\bibitem{princ}I.T. Jolliffe,  Principal component analysis, Springer--Verlag, 1986.

\bibitem{PODGal}K. Kunisch, S. Volkwein, Galerkin Proper Orthogonal Decomposition
Methods for a General Equation in Fluid Dynamics, {\it SIAM J Numer.
Anal.} 40, 2 (2002), 492-–515.

\bibitem{GalTem}M. Marion, R.  Temam,  Nonlinear Galerkin methods, {\it SIAM J. Numer. Anal.},
26 (1989), 1139--1157.

\bibitem{GZComp2005}A. Gorban, A. Zinovyev, Elastic Principal Graphs and Manifolds
and their Practical Applications, {\it Computing} 75, 359–379
(2005),

\bibitem{[1]}G.D. Birkhoff, Dynamical systems, AMS Colloquium Publications, Providence, 1927. Online:
http://www.ams.org/online \_bks/coll9/

\bibitem{Katok}B. Hasselblatt, A. Katok, (Eds.), Handbook of Dynamical Systems, Elsevier, 2002.

\bibitem{Katok2}A. Katok, B. Hasselblat, Introduction to the Modern Theory of Dynamical Systems,
Encyclopedia of Math. and its Applications, Vol. 54, Cambridge
University Press, 1995.

\bibitem{Milnor}J. Milnor, On the concept of attractor,  {\it Comm. Math. Phys.} 99,
(1985), 177--195

\bibitem{Ashwin}P. Ashwin and J.R. Terry. On riddling and weak attractors. {\it Physica
D} 142 (2000), 87--100.

\bibitem{[53]}D.V. Anosov, About one  class  of  invariant  sets  of  smooth
     dynamical systems, {\it Proceedings of  International
     conference  on  non-linear oscillation,}  V.2.   Kiev,   1970,
     39--45.
\bibitem{[54]}P. Walters,  On  the  pseudoorbit  tracing  property  and  its
     relationship to stability, {\it Lect. Notes  Math.}   V. 668 (1978),
     231--244.

\bibitem{[55]}J.E. Franke, J.F. Selgrade,   Hyperbolicity   and   chain
     recurrence, {\it J. Different. Equat.}  { 26}, 1 (1977), 27--36.

\bibitem{[57]}H. Easton, Chain transitivity and the domain of  influence  of
     an invariant set, {\it Lect. Notes Math.,}  V.668 (1978), p.95--102.


\bibitem{sinai}Y. Sinai, Gibbs measures in ergodic theory, {\it Russ. Math.
Surveys} 166 (1972), pp. 21--69.

\bibitem{[41]}I.G. Malkin, On  the  stability  under  uniformly  influencing
     perturbations, {\it Prikl.  Matem.  Mech.}  { 8}, 3 (1944),
     241--245.

\bibitem{[42]}V.E. Germaidze, N.N. Krasovskii,  On  the  stability  under
     sustained perturbations, {\it Prikl. Matem.  Mech.}
     { 21}, 6  (1957), 769--775.

\bibitem{[44}I.G. Malkin, The motion stability theory. Moscow: Nauka, 1966.

\bibitem{[45]}A. Strauss, A.J. Yorke,  Identifying  perturbations  which
     preserved asymptotic stability, {\it Proc. Amer. Math. Soc.} { 22}, 2 (1969), 513--518.


\bibitem{Gromov99}M. Gromov, Metric structures for Riemannian and non-Riemannian
spaces.  Progress in Mathematics, 152. Birkhauser Boston, Inc.,
Boston, MA, 1999.



\end{thebibliography}
\end{document}